\documentclass[12pt]{article}

  \usepackage{graphicx}
  \usepackage{epsfig}
  \usepackage{amssymb}
  \usepackage{subfigure}
    \usepackage{feynmf}
\evensidemargin - 1.truecm
\begin{document}
\begin{fmffile}{Fact}

\newcommand{\be}{\begin{equation}}
\newcommand{\ee}{\end{equation}}
\newcommand{\nn}{\nonumber}
\newcommand{\bea}{\begin{eqnarray}}
\newcommand{\eea}{\end{eqnarray}}
\newcommand{\bfig}{\begin{figure}}
\newcommand{\efig}{\end{figure}}
\newcommand{\bc}{\begin{center}}
\newcommand{\ec}{\end{center}}

\begin{titlepage}
\nopagebreak
{\flushright{
        \begin{minipage}{5cm}
        Freiburg-THEP 03/11\\
        TTP03-20\\
        {\tt hep-ph/0307295}\\
        \end{minipage}        }

}
\renewcommand{\thefootnote}{\fnsymbol{footnote}}
\vspace*{-1.5cm}                        
\vskip 3.5cm
\begin{center}
\boldmath
{\Large \bf QED vertex form factors at two loops}\unboldmath
\vskip 1.cm
{\large  R.~Bonciani\footnote{Email: {\tt
Roberto.Bonciani@physik.uni-freiburg.de}}
\footnote{This work was supported by the European Union under
contract HPRN-CT-2000-00149},}
\vskip .2cm
{\it Facult\"at f\"ur Mathematik und Physik, Albert-Ludwigs-Universit\"at
Freiburg, \\ D-79104 Freiburg, Germany} 
\vskip .2cm
{\large P.~Mastrolia\footnote{Email: {\tt Pierpaolo.Mastrolia@bo.infn.it}}},
\vskip .2cm
{\it Dipartimento di Fisica dell'Universit\`a di Bologna, 
I-40126 Bologna, Italy} 
\vskip .1cm
{\it and Institut f\"ur Theoretische Teilchenphysik,
Universit\"at Karlsruhe, \\ D-76128 Karlsruhe, Germany}
\vskip .2cm
{\large E.~Remiddi\footnote{Email: {\tt Ettore.Remiddi@bo.infn.it}}}
\vskip .2cm
{\it Dipartimento di Fisica dell'Universit\`a di Bologna
and INFN Sezione di Bologna, I-40126 Bologna, Italy}
\end{center}
\vskip 1.2cm

\begin{abstract}
We present the closed analytic expression of the form factors of the 
two-loop QED vertex amplitude for on-shell electrons of finite mass $m$ 
and arbitrary momentum transfer $S=-Q^2$. The calculation is carried 
out within the continuous $D$-dimensional regularization scheme, with 
a single continuous parameter $D$, the dimension of the space-time, 
which regularizes at the same time UltraViolet (UV) and InfraRed (IR) 
divergences. The results are expressed in terms of 1-dimensional 
harmonic polylogarithms of maximum weight 4.

\end{abstract}
\vfill
\end{titlepage}

\section{Introduction}

This paper is devoted to the evaluation of the 2-loop contributions 
(second order in the expansion in terms of the fine-structure constant) 
of the form factors of the QED vertex amplitude, for arbitrary momentum
transfer $S=-Q^2$ and on-shell external fermion lines, of finite mass 
$m$, in the continuous $D$ regularization scheme. 

The analytic calculation of the imaginary parts of the 
form factors at two-loop level for arbitrary momentum transfer, 
together with the value of the charge slope of the electron, 
were obtained long ago 
in \cite{2loop1} using the Pauli-Villars regularization scheme for the 
ultraviolet (UV) divergences, and giving a small fictitious mass 
$\lambda$ to the photon for the regularization of the soft infrared 
(IR) divergences. The results were given in terms of Nielsen's 
polylogarithms \cite{Nielsen,Kolbig} of maximum weight 3. The properly 
subtracted dispersion relations for the evaluation of the corresponding 
real parts were also written, but their explicit analytic evaluation 
was not carried out in \cite{2loop1}, because the results could not be
expressed in terms of Nielsen's polylogarithms only. 

In \cite{Pie} that analytic integration of the dispersion relations 
could at last be performed, expressing the real parts 
of the 2-loop form factors in terms of the 
1-dimensional harmonic polylogarithms (HPLs), introduced in the 
meanwhile \cite{Polylog,Polylog3}, of maximum weight $w=4$.

In this paper we present the calculation of the real and imaginary parts
of the two-loop form factors within the framework of dimensional
regularization \cite{DimReg}. Both UV and soft IR divergences are 
regularized in terms of the same parameter $D$, the continuous 
dimension of the space-time. 

The Feynman diagrams involved are shown in Fig. \ref{fig1}. 
The fermion lines carry momenta $p_1$ and $p_2$ and are both incoming 
(as in the kinematical case of electron-positron annihilation), 
the outgoing photon has momentum $Q=p_1 + p_2$. 
The electron mass is $m$ and the mass-shell condition for the two 
fermions is $p_1^2 = p_2^2 = -m^2$~. 

\bfig
\bc
\subfigure[]{
\begin{fmfgraph*}(30,30)
\fmfleft{i1,i2}
\fmfright{o}
\fmf{fermion}{i1,v1}
\fmf{fermion}{i2,v2}
\fmf{photon,label=$Q$,l.s=left}{v5,o}
\fmflabel{$p_2$}{i1}
\fmflabel{$p_1$}{i2}
\fmf{fermion,tension=.3}{v2,v3}
\fmf{fermion,tension=.3}{v3,v5}
\fmf{fermion,tension=.3}{v1,v4}
\fmf{fermion,tension=.3}{v4,v5}
\fmf{photon,tension=0}{v2,v1}
\fmf{photon,tension=0}{v4,v3}
\end{fmfgraph*} } 
%
%
%
\subfigure[]{
\begin{fmfgraph*}(30,30)
\fmfleft{i1,i2}
\fmfright{o}
\fmfforce{0.3w,0.6h}{v10}
\fmfforce{0.3w,0.4h}{v11}
\fmf{fermion}{i1,v1}
\fmf{fermion}{i2,v2}
\fmf{photon,label=$Q$,l.s=left}{v5,o}
\fmflabel{$p_2$}{i1}
\fmflabel{$p_1$}{i2}
\fmf{fermion,tension=.3}{v2,v3}
\fmf{fermion,tension=.3}{v3,v5}
\fmf{fermion,tension=.3}{v1,v4}
\fmf{fermion,tension=.3}{v4,v5}
\fmf{photon,tension=0}{v2,v4}
\fmf{photon,tension=0}{v1,v3}
\end{fmfgraph*} }
%
%
\subfigure[]{
\begin{fmfgraph*}(30,30)
\fmfleft{i1,i2}
\fmfright{o}
\fmfforce{0.2w,0.93h}{v2}
\fmfforce{0.2w,0.07h}{v1}
\fmfforce{0.2w,0.5h}{v3}
\fmfforce{0.8w,0.5h}{v5}
\fmfforce{0.2w,0.3h}{v10}
\fmf{fermion}{i1,v1}
\fmf{fermion}{i2,v2}
\fmf{photon,label=$Q$,l.s=left}{v5,o}
\fmflabel{$p_2$}{i1}
\fmflabel{$p_1$}{i2}
\fmf{fermion,tension=0}{v2,v5}
\fmf{fermion,tension=0}{v3,v4}
\fmf{photon,tension=.4}{v1,v4}
\fmf{fermion,tension=.4}{v4,v5}
\fmf{fermion,tension=0}{v1,v3}
\fmf{photon,tension=0}{v2,v3}
\end{fmfgraph*} }
%
%
\subfigure[]{
\begin{fmfgraph*}(30,30)
\fmfleft{i1,i2}
\fmfright{o}
\fmfforce{0.2w,0.93h}{v1}
\fmfforce{0.2w,0.07h}{v2}
\fmfforce{0.2w,0.5h}{v3}
\fmfforce{0.8w,0.5h}{v5}
\fmf{fermion}{i1,v2}
\fmf{fermion}{i2,v1}
\fmf{photon,label=$Q$,l.s=left}{v5,o}
\fmflabel{$p_2$}{i1}
\fmflabel{$p_1$}{i2}
\fmf{fermion,tension=0}{v2,v5}
\fmf{fermion,tension=0}{v3,v4}
\fmf{photon,tension=.4}{v1,v4}
\fmf{fermion,tension=.4}{v4,v5}
\fmf{fermion,tension=0}{v1,v3}
\fmf{photon,tension=0}{v2,v3}
\end{fmfgraph*} } \\
%
%
\subfigure[]{
\begin{fmfgraph*}(30,30)
\fmfleft{i1,i2}
\fmfright{o}
\fmfforce{0.2w,0.93h}{v2}
\fmfforce{0.2w,0.07h}{v1}
\fmfforce{0.8w,0.5h}{v5}
\fmfforce{0.2w,0.4h}{v9}
\fmfforce{0.5w,0.45h}{v10}
\fmfforce{0.2w,0.5h}{v11}
\fmf{fermion}{i1,v1}
\fmf{fermion}{i2,v2}
\fmf{photon,label=$Q$,l.s=left}{v5,o}
\fmflabel{$p_2$}{i1}
\fmflabel{$p_1$}{i2}
\fmf{fermion}{v2,v3}
\fmf{photon,tension=.25,right}{v3,v4}
\fmf{fermion,tension=.25}{v3,v4}
\fmf{fermion}{v4,v5}
\fmf{fermion}{v1,v5}
\fmf{photon}{v1,v2}
\end{fmfgraph*} }
%
%
\hspace{5mm}
\subfigure[]{
\begin{fmfgraph*}(30,30)
\fmfleft{i1,i2}
\fmfright{o}
\fmfforce{0.2w,0.93h}{v1}
\fmfforce{0.2w,0.07h}{v2}
\fmfforce{0.8w,0.5h}{v5}
\fmf{fermion}{i1,v2}
\fmf{fermion}{i2,v1}
\fmf{photon,label=$Q$,l.s=left}{v5,o}
\fmflabel{$p_2$}{i1}
\fmflabel{$p_1$}{i2}
\fmf{fermion}{v2,v3}
\fmf{photon,tension=.25,left}{v3,v4}
\fmf{fermion,tension=.25}{v3,v4}
\fmf{fermion}{v4,v5}
\fmf{fermion}{v1,v5}
\fmf{photon}{v1,v2}
\end{fmfgraph*} }
%
%
\hspace{5mm}
\subfigure[]{
\begin{fmfgraph*}(30,30)
\fmfleft{i1,i2}
\fmfright{o}
\fmfforce{0.2w,0.93h}{v2}
\fmfforce{0.2w,0.07h}{v1}
\fmfforce{0.2w,0.3h}{v3}
\fmfforce{0.2w,0.7h}{v4}
\fmfforce{0.8w,0.5h}{v5}
\fmf{fermion}{i1,v1}
\fmf{fermion}{i2,v2}
\fmf{photon,label=$Q$,l.s=left}{v5,o}
\fmflabel{$p_2$}{i1}
\fmflabel{$p_1$}{i2}
\fmf{fermion}{v2,v5}
\fmf{photon}{v1,v3}
\fmf{photon}{v2,v4}
\fmf{fermion}{v1,v5}
\fmf{fermion,right}{v4,v3}
\fmf{fermion,right}{v3,v4}
\end{fmfgraph*} }
%
%
\vspace*{8mm}
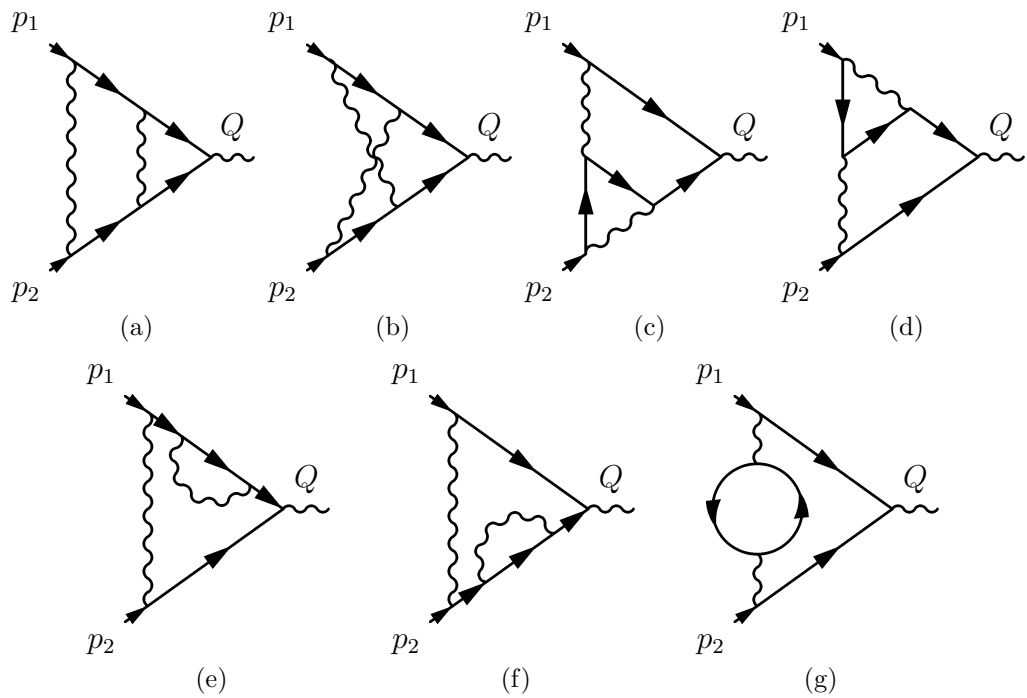
\caption{\label{fig1} 2-loop vertex diagrams for the QED form factors.
The fermionic external lines are on the mass-shell
$p_{1}^{2}=p_{2}^{2}=-m^2$; the wavy line on the r.h.s. carries 
momentum $Q=p_{1}+p_{2}$, with $Q^{2}=-S$. The arrows label the flow of
the momenta $p_1$ and $p_2$. }
\ec
\efig

On Lorentz-invariance grounds, all the vertex diagrams can be expressed 
in terms of at most three factors, corresponding to the three vectors 
proportional to the Dirac matrices $\gamma^{\mu}$, 
to $\sigma^{\mu \nu} Q_{\nu}$, with 
$ \sigma^{\mu \nu} = - \frac{1}{2}[ \gamma^\mu \gamma^\nu 
                                  - \gamma^\nu \gamma^\nu ] $ 
and to $Q^{\mu}$ (in QED the third form factor is of course vanishing, when
the contributions of the various graphs are summed). The on mass-shell 
form factors are functions of the momentum transfer $S=-Q^2$ and of the 
mass of the fermions $m$. Their value, when projected out from the 
Feynman graph amplitudes in the continuous $D$-dimensional regularization 
scheme, is a combination of the scalar integrals associated to the Feynman 
graphs. In \cite{Bon1} all those integrals were separately expressed 
in terms of 17 independent scalar integrals, called Master Integrals 
(MIs), by means of a reduction algorithm based on the integration by 
part identities (IBPs) \cite{Chet}, Lorentz invariance identities (LI) 
\cite{Rem3} and general symmetry relations, implemented for the computer
language FORM \cite{FORM}. The MIs were then calculated in ~\cite{Bon1} 
by the differential equations method \cite{Kot,Rem1,Rem2}.

In this paper we use those results in order to evaluate the explicit 
analytic value of the form factors for any of the (unrenormalized) 
diagrams of Fig. \ref{fig1} and for the full renormalized vertex amplitude 
as well.

The paper is structured as follows.

After a general introduction on the QED form factors, recalled in section 
\ref{Intro}, in section \ref{diagrams} we give the unrenormalized form 
factors for each of the 2-loop Feynman diagrams entering in the calculation
of the 2-loop QED vertex amplitude, within the $D$-dimensional 
regularization scheme. In section \ref{renormalization} the 
subtractions for the renormalization of UV divergences at the second 
order in the fine-structure constant are listed. In section 
\ref{fullyren} we present the full UV-renormalized form factors for the 
2-loop QED vertex amplitude in the space-like region $-S=Q^2>0$; we also 
discuss the analytic continuation to the physical region $S=-Q^2 > 4m^2$,
presenting the imaginary parts of the form factors. In sections 
\ref{Q2>>a} and \ref{Q2<<a} the behaviours of the form factors for large
and small momentum transfer are given (recovering in particular the 
two loop values of the electron $(g-2)$ and of the charge form factor 
slope). In the appendix \ref{app1} we list the definition of the 
propagators used in the explicit calculations, and in appendix 
\ref{app2} we give the 1-loop contributions to the QED vertex form 
factors up to first order in $(D-4)$.

\section{The QED form factors \label{Intro}}

Let us call $V^{\mu}(p_1,p_2)$ the QED vertex amplitude,
corresponding to the annihilation of an electron and a positron, of
momenta $p_1$ and $p_2$, with the two particles on the
mass-shell ($p_{1}^{2}=p_{2}^{2}=-m^2$). Let us define the following
two vectors:
\be
Q^{\mu} = p_{1}^{\mu}+p_{2}^{\mu} \, , \qquad 
\Delta^{\mu} = p_{1}^{\mu}-p_{2}^{\mu} \, , 
\label{b0001}
\ee
such that $Q^{2}=-S$, where $S$ is the c.m. energy squared; in the 
following we will also use the related dimensionless variables 
\be
q^2 = \frac{Q^2}{m^2} \ , \hskip 2cm s = \frac{S}{m^2} \ . 
\label{defq2ands} 
\ee 

In general $V^{\mu}(p_1,p_2)$ can be expressed in terms of three 
dimensionless scalar form factors $F_i(q^2), i=1,2,3$, depending 
only on the dimensionless variable $q^2$ of Eq. (\ref{defq2ands}), 
as follows: 
\begin{eqnarray} 
\hspace*{-0.5cm} 
 V^{\mu}(p_1,p_2) &=& \bar{v}(p_2) \Gamma^{\mu}(p_1,p_2) u(p_1) 
                     \\ 
\hspace*{-0.5cm}  \Gamma^{\mu}(p_1,p_2) &=& \left[ F_{1}(q^2) \ \gamma^{\mu} 
  + \frac{1}{2m} F_{2}(q^2) \ \sigma^{\mu \nu} Q_{\nu} 
  - \frac{i}{m} F_{3}(q^2) \ Q^{\mu} \right] \ ,  \label{b0002} 
\end{eqnarray} 
where $\bar{v}(p_2), u(p_1) $ are the spinor wave functions of the 
positron and the electron, 
$\sigma^{\mu \nu} = - \frac{i}{2}[ \gamma^{\mu},\gamma^{\nu}]$. 
Usually $F_1(q^2)$ is known as the {\it charge} (Dirac) {\it form factor} 
whereas $F_2(q^2)$ as the {\it magnetic} (Pauli) {\it form factor}.
If we consider single Feynman diagrams, any form factors in Eq.
(\ref{b0002}) can be in general different from zero. However, if
$\Gamma^{\mu}(p_1,p_2)$ is the full vertex amplitude, the conservation
of the electromagnetic current forces the third form factor to vanish,
$F_{3}(q^2) = 0$.

The extraction of each form factors $F_i(q^2)$ from Eq. (\ref{b0002}) 
can be carried out by the following general projector operators 
$P_{\mu}^{(i)}$:
\be
\! \! \! \! P_{\mu}^{(i)}(m,p_1,p_2) = 
         \frac{ -i \! \not{\! p_{1}} \! + \! m }{m} 
\biggl[   g_{1}^{(i)} \gamma_{\mu} 
        + \frac{i}{m} g_{2}^{(i)} \Delta_{\mu} 
        - \frac{i}{m} g_{3}^{(i)} Q_{\mu} \biggr] 
         \frac{ i \! \not{\! p_{2}} \! + \! m }{m} ,
\label{b0003} 
\ee 
where the constants $g_{j}^{(i)}, j=1,2,3$, are
properly chosen to have:
\be
{\rm Tr}\left( P_{\mu}^{(i)}(m,p_1,p_2) \Gamma^{\mu}(p_{1},p_{2}) \right) 
  = F_{i}(q^2) .
\label{b0004}
\ee
Let us observe that since we work in a $D$-dimensional space (to 
regularize the divergences arising in the computation) the {\it trace} 
operation is consistently  performed in $D$ dimensions as well.

The explicit values of the constants are:
\bea
g_{1}^{(1)} & = & - \frac{2}{(D-2)}\ \frac{1}{q^2+4} \, ,
\label{b0005} \\
g_{2}^{(1)} & = & - \frac{8(D-1)}{(D-2)}\ \frac{1}{(q^2+4)^2} \, ,
\label{b0006} \\
g_{1}^{(2)} & = & - \frac{8}{(D-2)}\ \frac{1}{q^{2}(q^2+4)} \, ,
\label{b0008} \\
g_{2}^{(2)} & = & - \frac{8}{(D-2)}\ \frac{1}{(q^2+4)^2} \left[
\frac{4}{q^2} + D-2 \right]  \, ,
\label{b0009} \\
g_{3}^{(3)} & = & - \frac{2}{q^{2}(q^2+4)}  \, ,
\label{b00013} 
\eea
and $g_{3}^{(1)} = g_{3}^{(2)} = g_{1}^{(3)} = g_{2}^{(3)} = 0$.

As the spinor traces are in $D$-dimensional space-time, 
in all the above formulas Eqs. (\ref{b0005}-\ref{b00013}), 
all the r.h.s., strictly speaking, should be multiplied by the overall 
constant $(1/4)\ {\rm Tr} {\bf 1}$, where ${\rm Tr} {\bf 1}$ is the trace 
of the unit 
Dirac matrix in $D$-continuous dimensions. The overall constant is in fact 
undetermined for arbitrary $D$, except for its limiting value at $D=4$, 
which is $1$. For simplicity, we will therefore omit systematically 
that overall factor. 

In QED, the form factors are given as an expansion in powers 
of $(\alpha/\pi)$, $\alpha=e^2/4\pi$ being the fine-structure 
constant. Showing explicitly the dependence on the regularizing 
dimension, we write the expansion as 
\bea
F_{1}(D,q^2) & = & 
   1 + \left( \frac{\alpha}{\pi} \right) F^{(1l)}_{1}(D,q^2) 
     + \left( \frac{\alpha}{\pi} \right)^{2} F^{(2l)}_{1}(D,q^2) 
     + {\mathcal O} \Bigg( \left( \frac{\alpha}{\pi} \right) ^{3} \Bigg)
     \, , 
\label{b00020} 
\\
F_{2}(D,q^2) & = & \quad \; \; \,
       \left( \frac{\alpha}{\pi} \right) F^{(1l)}_{2}(D,q^2) 
     + \left( \frac{\alpha}{\pi} \right) ^{2} F^{(2l)}_{2}(D,q^2) 
     + {\mathcal O} \Bigg( \left( \frac{\alpha}{\pi} \right) ^{3} \Bigg) 
     \, , 
\label{b00021} 
\\
F_{3}(D,q^2) & = & \quad \; \; \, 
       \left( \frac{\alpha}{\pi} \right) F^{(1l)}_{3}(D,q^2) 
     + \left( \frac{\alpha}{\pi} \right) ^{2} F^{(2l)}_{3}(D,q^2) 
     + {\mathcal O} \Bigg( \left( \frac{\alpha}{\pi} \right) ^{3} \Bigg)
     \, ,
\label{b00022} 
\eea
where the superscripts ``$1l$'' and ``$2l$'' stand for 1- and 2-loop
contributions, the first term $1$ in $F_{1}(D,q^2)$ is the tree 
approximation and $F_{3}(D,q^2)$ vanishes, as already said, when 
all the graphs are summed up. 

In the following we will focus on the detailed 2-loop computations of  
$F^{(2l)}_1(D,q^2)$, $F^{(2l)}_2(D,q^2)$ and $F^{(2l)}_{3}(D,q^2)$. We 
will present the unrenormalized form factors for each Feynman diagram 
and we will discuss the renormalization procedure, giving the fully 
UV-renormalized result for the QED vertex amplitude. The 1-loop
contributions to the form factors, $F^{(1l)}_{1}(D,q^2)$  and 
$F^{(1l)}_{2}(D,q^2)$, 
will be recalled in appendix \ref{app2} for completeness. 

\section{ Unrenormalized contributions \label{diagrams} }

The diagrams which contribute to the order $(\alpha/\pi)^2$ are shown 
in Fig. \ref{fig1}. Following Eq. (\ref{b0002}), we will indicate by 
${\mathcal F}^{(2l,{\tt graph})}_{i}(D,q^2) \ (i=1,2,3),\ 
{\tt graph} \in \{ {\tt a,..,g} \}$ the contribution of each 
unrenormalized graph to the unrenormalized form factors: 
\bea
\! \! \! \! \! \! \left( \underbrace{
\parbox{20mm}{
\begin{fmfgraph*}(15,15)\fmfkeep{generic}
\fmfforce{0.5w,0.5h}{c1}
\fmfleft{i1,i2}
\fmfright{o1}
\fmfblob{.7w}{c1}
\fmf{plain}{i1,c1}
\fmf{plain}{i2,c1}
\fmf{photon}{c1,o1}
\end{fmfgraph*}
} \hspace*{-.5cm}
}_{{\tt graph} \ \in \ \{ {\tt a,...,g} \} }
\right)^{\mu}
& = & \ \bar{v}(p_2) \Bigg[ {\mathcal F}^{(2l,{\tt graph})}_{1}(D,q^2) 
\, \gamma^{\mu} + \frac{1}{2m} 
{\mathcal F}^{(2l,{\tt graph})}_{2}(D,q^2) \, 
\sigma^{\mu \nu} Q_{\nu} \nn\\
\! \! \! \! \! \! & & \qquad \quad - \frac{i}{m} 
    {\mathcal F}^{(2l,{\tt graph})}_{3}(D,q^2) \ Q^{\mu} \Bigg]u(p_1) 
    \ . 
\label{eq:unrenFF}
\eea

In so doing, each ${\mathcal F}^{(2l,{\tt graph})}_{i}(D,q^2)$ can be 
extracted by the $D$-dimensional projection defined in Eq. 
(\ref{b0004}). After the computation of the {\it trace}, it turns out
that each form factor is expressed in terms of several hundreds of
scalar integrals. According to 
\cite{Bon1}, one can express all those integrals in terms of only 17 
Master Integrals (MIs) {\it via} integration-by-parts identities, 
Lorentz invariance and general symmetry relations (exactly in $D$). 

As an example consider the {\it magnetic} form factor of the diagram 
(g) in Fig. \ref{fig1}. It can be written as a linear combination of 5 
MIs, the coefficients being  ratios of simple polynomials in $D$ 
and $q^2$, the fermion mass squared $m^2$ appearing as a dimensional 
scale factor. (It is to be noted that in the formulas which follow, 
and which are exact in $D$, the dimensionless variables $q^2$ and 
$D$ are never entangled in a same non factorisable polynomial in the 
denominators). 
\bea
\hspace*{-6mm} {\mathcal F}^{(2l,{\tt g})}_{2}(D,q^2) & = & 
\frac{1}{(q^2+4)} 
\Biggl\{
\frac{32(D-4)}{(D-5)} \nn\\
\hspace*{-6mm} & & - \frac{64(D^2-10D+19)}{(D-1)(D-5)} 
\frac{1}{(q^2+4)} 
\Biggr\}
\, \, \parbox{15mm}{\begin{fmfgraph*}(15,15)
\fmfleft{i1,i2}
\fmfright{o}
\fmf{plain}{i1,v1}
\fmf{plain}{i2,v2}
\fmf{photon}{v3,o}
\fmf{plain,tension=.3}{v2,v3}
\fmf{plain,tension=.3}{v1,v3}
\fmf{plain,tension=0,right=.5}{v2,v1}
\fmf{plain,tension=0,right=.5}{v1,v2}
\end{fmfgraph*} } \nn\\
\hspace*{-6mm} & & - \frac{64(D^3 \! - \! 8D^2 \! + \! 23D \! - \! 
26)}{(D-1)(D-5)} \frac{1}{(q^2 \! + \! 4)^2} \left( \frac{1}{m^2}
\, \, \parbox{15mm}{\begin{fmfgraph*}(15,15)
\fmfleft{i1,i2}
\fmfright{o}
\fmf{plain}{i1,v1}
\fmf{plain}{i2,v2}
\fmf{photon}{v3,o}
\fmflabel{$\! (p_{2} \! \cdot \! k_{1})$ }{o}
\fmf{plain,tension=.3}{v2,v3}
\fmf{plain,tension=.3}{v1,v3}
\fmf{plain,tension=0,right=.5}{v2,v1}
\fmf{plain,tension=0,right=.5}{v1,v2}
\end{fmfgraph*} } \hspace*{14mm} \right) \nn\\
\hspace*{-6mm} & & - \frac{1}{(q^2+4)} 
\Biggl\{
\frac{4(D^4-14D^3+59D^2-82D+16)}{(D-1)(D-4)(D-6)} \nn\\
\hspace*{-6mm} & & + \frac{16(3D-8)(D^2-10D+19)}{(D-1)(D-4)(D-5)} 
\frac{1}{(q^2+4)} 
\Biggr\} \left( \frac{1}{m^2} \, \, 
\parbox{15mm}{\begin{fmfgraph*}(15,15)
\fmfleft{i}
\fmfright{o}
\fmf{plain}{i,v1}
\fmf{plain}{v2,o}
\fmf{plain,tension=.15,left}{v1,v2}
\fmf{plain,tension=.15}{v1,v2}
\fmf{plain,tension=.15,right}{v1,v2}
\end{fmfgraph*} } \right) \nn\\
\hspace*{-6mm} & & - \frac{32(D-2)(D^2-10D+19)}{(D-1)(D-5)} 
\frac{1}{(q^2+4)^2}
\, \, \left( \frac{1}{m^2} \, \, 
\parbox{15mm}{\begin{fmfgraph*}(15,15)
\fmfleft{i}
\fmfright{o}
\fmf{photon}{i,v1}
\fmf{photon}{v2,o}
\fmf{plain,tension=.22,left}{v1,v2}
\fmf{plain,tension=.22,right}{v1,v2}
\fmf{plain,right=45}{v2,v2}
\end{fmfgraph*} } \right) \nn\\
\hspace*{-6mm} & & - \frac{2(D - 2)^2}{(q^2 + 4)} 
\Biggl\{
\frac{(3D^4- 53D^3 + 323D^2-795D+642)}{(D-1)(D-3)(D-4)(D-5)(D-6)}
\nn\\
\hspace*{-6mm} & & + 
\frac{4(2D-5)(D^2-10D+19)}{(D-1)(D-3)(D-4)(D-5)} \frac{1}{(q^2+4)} 
\Biggr\} \! \! 
\left( \frac{1}{m^4} \! \! \! \! 
\parbox{15mm}{\begin{fmfgraph*}(15,15)
\fmfleft{i}
\fmfright{o}
\fmf{phantom}{i,v1}
\fmf{phantom}{v1,o}
\fmf{plain,right=45}{v1,v1}
\fmf{plain,left=45}{v1,v1}
\end{fmfgraph*} } \hspace*{-5mm} \right) \! ,
\label{b00023}
\eea
where the MIs depicted on the r.h.s. are those of Fig. 7 of \cite{Bon1}.

Similar formulas hold for the other form factors and graphs, but are too 
lengthy to be reported here. 

Once the form factors are expressed in terms of MIs, one expands the
result around $D=4$ and inserts the values of the MIs, also given in 
\cite{Bon1} as an expansion around $D=4$. In this way one finally obtains 
the required analytic result where 
both UV and soft IR divergences, regulated by the same parameter
$D$, appear as poles in $(D-4)$. In the case of Eq. (\ref{b00023}),
using the Eqs. (88,93,123,125,B.1) of \cite{Bon1}, one will get the 
expression given later in this paper in Eq. (\ref{ex2}).

In this section we will give the contributions 
${\mathcal F}^{(2l,{\tt graph})}_{i}(D,q^2)$ to the 
{\it unrenormalized} form factors from each of the 2-loop 
still unrenormalized graphs (i.e. where the renormalization of the 
inserted 1-loop subgraphs has not yet been carried out). 

The propagators of the graphs are considered in the Feynman gauge and 
the corresponding denominators ${\mathcal D}$'s, which will appear in 
the following formulas, are listed in the appendix \ref{app1}.

The resulting {\it unrenormalized} form factors are given for space-like 
$Q$ ($Q^2>0$ or $S = -Q^2 <0$) and are expressed in terms of 
1-dimensional harmonic polylogarithms \cite{Polylog,Polylog3} of argument 
\be
x = \frac{\sqrt{q^2+4} - \sqrt{q^2} }{\sqrt{q^2+4} + \sqrt{q^2} } 
  = \frac{\sqrt{Q^2+4m^2} - \sqrt{Q^2} }{\sqrt{Q^2+4m^2} + \sqrt{Q^2} } 
\label{b00015}
\ .
\ee

Let us comment shortly here the normalization of the $D$-dimensional 
integrals, or, which is the same, the choice of the loop integration 
measure in $D$ continuous dimensions. With the natural choice 
\be 
\mu_{0}^{(4-D)} \int \frac{d^D k}{(2\pi)^{(D-2)}} \, ,
\label{b00016}
\ee 
where $\mu_0$ is the mass scale, 
one obtains for the simplest loop integral, the 1-loop massive tadpole, 
\be
\mu_{0}^{(4-D)} \int \frac{d^D k}{(2\pi)^{(D-2)}} \frac{1}{k^2+m^2} =
C(D) \, \left( \frac{m^2}{\mu_0^{2}} \right)^{\frac{D-4}{2}} \,
\frac{m^2}{(D-2)(D-4)} \, ,
\label{b00017}
\ee
where $C(D)$ is the following function of the space-time dimension $D$:
\be
C(D) = (4 \pi)^{\frac{(4-D)}{2}} \Gamma \left( 3 - \frac{D}{2} \right) 
\label{b00018} \, ,
\ee
with the limiting value $C(4)=1$ for $D=4$. Note that the explicit 
form of $C(D)$ is essentially irrelevant, as in any physical quantity, 
finite for $D=4$, $C(D)$ can be replaced by 1. The explicit expression 
in Eq. (\ref{b00018}) can matter only in detailed comparisons with 
calculations using a different integration measure. 

To simplify the writing of all the subsequent formulae, we will use 
the following $D$-continuous integration measure 
\be 
\int{\mathfrak D}^Dk = \frac{1}{C(D)} 
\left( \frac{m^2}{\mu_0^{2}} \right)^{\frac{4-D}{2}} 
\int \frac{d^D k}{(2\pi)^{(D-2)}} \, ,
\label{b00016a} 
\ee 
using $m$ as mass scale. With that choice, the 1-loop tadpole simply reads 
\be 
\int{\mathfrak D}^Dk \ \frac{1}{k^2+m^2} = \frac {m^2}{(D-2)(D-4)} \ . 
\label{b00017a} 
\ee 

By using for each loop the integration measure ${\mathcal D}^Dk$ of 
Eq. (\ref{b00016a}) we find the results which follow for the contributions 
to the unrenormalized form factors from the various 2-loop graphs, with 
still unrenormalized 1-loop insertions. 

$\bullet$ The {\it Ladder} graph {\tt(a)} of Fig. \ref{fig1}, defined as
\bea
\parbox{20mm}{\begin{fmfgraph*}(15,15)
\fmfleft{i1,i2}
\fmfright{o}
\fmf{plain}{i1,v1}
\fmf{plain}{i2,v2}
\fmf{photon}{v5,o}
\fmf{plain,tension=.3}{v2,v3}
\fmf{plain,tension=.3}{v3,v5}
\fmf{plain,tension=.3}{v1,v4}
\fmf{plain,tension=.3}{v4,v5}
\fmf{photon,tension=0}{v2,v1}
\fmf{photon,tension=0}{v4,v3}
\end{fmfgraph*} } & = & 
\int {\mathfrak{D}}^Dk_1\;{\mathfrak{D}}^Dk_2 \ 
\frac{{\mathcal N}_{({\tt a})}^{\mu}}{{\mathcal D}_{1} {\mathcal D}_{2} 
{\mathcal D}_{9} {\mathcal D}_{10} {\mathcal D}_{12} 
{\mathcal D}_{13} } \ , 
\label{b1} 
\eea
where 
\bea
\! \! \! \! \! \! {\mathcal N}_{({\tt a})}^{\mu} & = & \bar{v}(p_2)  \gamma_{\sigma}
[i ( \not{\! p_{2}} + \! \! \not{\! k_{1}}) \! + \! m] 
\gamma_{\lambda} 
[i ( \not{\! p_{2}} + \!  \! \not{\! k_{1}} - \!  \! \not{\! k_{2}}) 
\! + \! m] \gamma^{\mu}
[-i ( \not{ \! p_{1}} -  \! \! \not{\! k_{1}} +  \! \! \not{\! k_{2}})
\! + \! m] \times
\nn\\
\! \! \! \! \! \! & & \times
\gamma^{\lambda} 
[-i ( \not{\! p_{1}} -  \! \! \not{\! k_{1}} ) \!  + \! m]
\gamma^{\sigma} u(p_1)  ,
\eea
gives: 
\bea
\hspace*{-5mm} {\mathcal F}^{(2l,{\tt a})}_{1}(D,q^2) & = & 
            \frac{1}{(D-4)^{2}} \Biggl\{ 
            \frac{1}{8} \! 
          + \! \frac{1}{2} \biggl[ 1 \!  
          - \frac{1}{(1\! -\! x)} \! 
   - \frac{1}{(1\! +\! x)} \biggr] H(0;x) \! 
          + \biggl[ 1 \! + \! \frac{1}{(1\! -\! x)^2} 
          \nn\\
\hspace*{-5mm} & &  \hspace*{20mm}     
          - \frac{1}{(1-x)}
          + \frac{1}{(1+x)^2}
          - \frac{1}{(1+x)} \biggr] H(0,0;x)
\Biggr\} \nn\\
\hspace*{-5mm} & & \nn\\
\hspace*{-5mm} & + & \frac{1}{(D-4)} \Biggl\{
          - \frac{5}{32} 
   - \frac{1}{8} \biggl[ 1 - \frac{2}{(1-x)}
   \biggr] H(0;x) - \frac{1}{2} \biggl[  1 - \frac{2}{(1-x)} \nn\\
\hspace*{-5mm} & & \hspace*{18mm}        + \frac{2}{(1 \!  - \!  x)^2}
   \biggr] H(0,0;x) 
          +  \!  \frac{1}{4} \biggl[ 1 \!  
          -  \!  \frac{1}{(1 \!  - \!  x)} \! 
   -  \!  \frac{1}{(1 \!  + \!  x)} \biggr] \bigl[ 
     \zeta(2) \nn\\
\hspace*{-5mm} & & \hspace*{18mm}     
          - H(0;x) - H(0,0;x) + 2 H(-1,0;x) 
\bigr] \nn\\
\hspace*{-5mm} & &  \hspace*{18mm}       
          + \frac{1}{2} \biggl[ 1  \! +  \! \frac{1}{(1 \! - \! x)^2} 
          -  \! \frac{1}{(1 \! - \! x)}
          +  \! \frac{1}{(1 \! + \! x)^2}  
   -  \! \frac{1}{(1 \! + \! x)} \biggr] \bigl[ \zeta(3)\nn\\
\hspace*{-5mm} & & \hspace*{18mm}      
          - 2 H(0,0;x) - 2 H(0,0,0;x) + 4 H(-1,0,0;x)\nn\\
\hspace*{-5mm} & & \hspace*{18mm}        
          + 4 H(0,1,0;x) \bigr]
\Biggr\}   \nn\\
\hspace*{-5mm} & & \nn\\
\hspace*{-5mm} & & 
          + \frac{55}{128}
          + \biggl[ \frac{5}{4(1-x)} 
          - \frac{81}{2(1+x)^4} 
          + \frac{177}{2(1+x)^3} 
          - \frac{123}{2(1+x)^2} \nn\\
\hspace*{-5mm} & &           + \frac{55}{4(1+x)} \! 
          -  \! \frac{23}{8} \biggr] \zeta(2) \! 
          +  \! \biggl[ \frac{3}{5(1-x)^2} \! 
          -  \! \frac{31}{80(1-x)} \! 
          -  \! \frac{51}{2(1+x)^5} \nn\\
\hspace*{-5mm} & &           +  \! \frac{255}{4(1 \! + \! x)^4} \! 
          -  \! \frac{1071}{20(1 \! + \! x)^3} \! 
          +  \! \frac{687}{40(1 \! + \! x)^2} \!
          -  \! \frac{303}{80(1 \! + \! x)} \!
          +  \! \frac{29}{20} \biggr] \zeta^{2}(2) \nn\\
\hspace*{-5mm} & &           - \biggl[ \frac{1}{(1-x)^2}
          - \frac{5}{4(1-x)} 
          - \frac{60}{(1+x)^4}
          + \frac{120}{(1+x)^3}
          - \frac{141}{2(1+x)^2} \nn\\
\hspace*{-5mm} & &           + \frac{41}{4(1+x)} 
   - 2 \zeta(3) \biggr] \zeta(3)
   - \frac{3}{4(1+x)} \biggl[ 1 
   - \frac{1}{(1+x)}  \biggr] \nn\\
\hspace*{-5mm} & &      
       +  \biggl[
            \frac{19}{32} 
 - \biggl( \frac{2}{(1-x)^2}
          - \frac{33}{32(1-x)} \!
          + \frac{81}{2(1+x)^5} \!
          - \frac{465}{4(1+x)^4} \nn\\  
\hspace*{-5mm} & &      \hspace*{5mm}
          + \frac{979}{8(1+x)^3} 
   - \frac{865}{16(1+x)^2}
          + \frac{327}{32(1+x)}
          - \frac{9}{8} \biggr) \zeta(2) \nn\\  
\hspace*{-5mm} & &      \hspace*{5mm}
        - \biggl( \frac{1}{4(1-x)^2}  \!
          +  \!\frac{1}{4(1-x)} \!
          -  \!\frac{60}{(1+x)^5} \!
          +  \!\frac{150}{(1+x)^4} \!
          -  \!\frac{126)}{(1+x)^3} \nn\\  
\hspace*{-5mm} & &      \hspace*{5mm}
          + \frac{157}{4(1+x)^2}  
          - \frac{31}{4(1+x)}
          + \frac{9}{4} \biggr) \zeta(3)
          - \frac{1}{(1-x)}
          + \frac{9}{4(1+x)^3} \nn\\  
\hspace*{-5mm} & &      \hspace*{5mm}
          - \frac{27}{8(1+x)^2}
          + \frac{15}{16(1+x)}
          \biggr] H(0;x)  \nn\\
\hspace*{-5mm} & &        
     -  \biggl[
           \biggl( \frac{1}{4(1-x)}
          + \frac{1}{4(1+x)} 
          - \frac{1}{4} \biggr) \zeta(2)
          - \biggl( 1
   +  \! \frac{1}{(1 \! - \! x)^2}
          - \frac{1}{(1 \! - \! x)} \nn\\
\hspace*{-5mm} & &   \hspace*{5mm}        
          + \frac{1}{(1+x)^2} \!
          - \frac{1}{(1+x)}
          \biggr) \zeta(3)
          \biggr] H(-1;x) \nn\\
& &        +  \biggl[
            \frac{27}{8}   \! 
   + \! \biggl( \frac{7}{4(1 \! - \! x)^2}  
   -  \! \frac{13}{8(1 \! - \! x)}  \!
          -  \! \frac{15}{(1 \! + \! x)^5} \!
          + \! \frac{75}{2(1 \! + \! x)^4}   \!
   -  \! \frac{63}{2(1 \!+ \!x)^3}   \nn\\
\hspace*{-5mm} & & \hspace*{5mm}         
          + \frac{23}{2(1+x)^2} 
          - \frac{29}{8(1+x)}  
          + \frac{9}{4} \biggr)  \zeta(2)
          + \frac{17}{8(1-x)^2} 
          - \frac{27}{8(1-x)}  \nn\\
\hspace*{-5mm} & & \hspace*{5mm}     
          + \frac{39}{2(1+x)^4}    
   - \frac{93}{2(1+x)^3}  \!
          + \frac{77}{2(1+x)^2}  \! 
          - \frac{47}{4(1+x)}  \!
          \biggr] H(0,0;x)  \nn\\
\hspace*{-5mm} & &  
       +  \biggl[ \biggl(
            \frac{1}{2(1-x)}
   - \frac{60}{(1+x)^5}
          + \frac{150}{(1+x)^4}
          - \frac{126}{(1+x)^3}
          + \frac{39}{(1+x)^2} \nn\\
\hspace*{-5mm} & & \hspace*{5mm}      
          - \frac{15}{2(1+x)}  
   + 2 \biggr) \zeta(2)
          \biggr] H(1,0;x) \nn\\
\hspace*{-5mm} & &  
       +  \biggl[ \biggl( 1  \! 
          +  \! \frac{1}{2(1 \! - \! x)^2} 
          -  \! \frac{1}{2(1 \! - \! x)}  
          +  \! \frac{1}{2(1 \! + \! x)^2}  
          -  \! \frac{1}{2(1 \! + \! x)}  \biggr) \zeta(2)
          \biggr] \times \nn\\
\hspace*{-5mm} & &  \hspace*{5mm} \times \bigl[ H(0,1;x) 
+ 2 H(-1,0;x) + H(0,-1;x) \bigr] \nn\\
\hspace*{-5mm} & &  
       -  \biggl[
            \frac{3}{2} \! 
          -  \! \frac{5}{2(1 \! - \! x)}   \! 
   -  \! \frac{15}{(1 \! + \! x)^3} \! 
          +  \! \frac{45}{2(1 \! + \! x)^2}  \! 
          -  \! \frac{8}{(1 \! + \! x)}
          \biggr] H(-1,0;x) \nn\\
\hspace*{-5mm} & &        
       +  \biggl[
          \frac{35}{8} \! 
          +  \! \frac{1}{(1 \! - \! x)^2}  \! 
          -  \! \frac{71}{32(1 \! - \! x)} \! 
          - \! \frac{81}{2(1 \! + \! x)^5} \! 
          + \! \frac{525}{4(1 \! + \! x)^4} \!
          - \! \frac{1219}{8(1 \! + \! x)^3} \nn\\
\hspace*{-5mm} & &  \hspace*{5mm}    
          + \frac{1173}{16(1 \! + \! x)^2}
          - \frac{471}{32(1 \! + \! x)} 
          \biggr] H(0,0,0;x) \nn\\
& &        -  \biggl[
            \frac{25}{4} \!
          + \! \frac{4}{(1-x)^2}  \!
          - \! \frac{17}{4(1-x)} \!
          + \! \frac{60}{(1+x)^4} \!
          - \! \frac{120}{(1+x)^3} \!
          + \! \frac{73}{(1+x)^2} \nn\\
\hspace*{-5mm} & &  \hspace*{5mm}    
          - \frac{53}{4(1+x)}
          \biggr] H(-1,0,0;x) \nn\\
\hspace*{-5mm} & &        +  \frac{1}{2} \biggl[
            1 
          - \frac{1}{(1-x)}
          - \frac{1}{(1+x)}
          \biggr] H( -1, -1, 0;x) \nn\\
\hspace*{-5mm} & &        +  \biggl[
          \frac{5}{4}
          + \frac{1}{4(1-x)} 
          + \frac{30}{(1+x)^4}
          - \frac{60}{(1+x)^3}
          + \frac{71}{2(1+x)^2} \nn\\
\hspace*{-5mm} & &     \hspace*{5mm}    
          - \frac{21}{4(1+x)}
          \biggr] H(0,-1,0;x) \nn\\
\hspace*{-5mm} & &        -  \biggl[ 
            \frac{3}{2} \!
          + \frac{2}{(1 \! - \! x)^2} \!
          - \frac{2}{(1 \! - \! x)} \!
          + \frac{1}{(1 \! + \! x)^2} \!
          - \frac{1}{(1 \! + \! x)} \!
          \biggr] H(0,1,0;x)  \nn\\
\hspace*{-5mm} & &        +  \biggl[ 
          \frac{3}{2} \!
          + \! \frac{30}{(1 \! + \! x)^4} \!
          - \! \frac{60}{(1 \! + \! x)^3} \!
          + \! \frac{71}{2(1 \! + \! x)^2} \!
          - \! \frac{11}{2(1 \! + \! x)} \!
          \biggr] H(1,0,0;x) \nn\\
\hspace*{-5mm} & &        +  \frac{1}{4} \biggl[
          1 
          +  \frac{1}{(1-x)^2} 
          -  \frac{1}{(1-x)} 
          +  \frac{1}{(1+x)^2} 
          -  \frac{1}{(1+x)}  
          \biggr] \times \nn\\
\hspace*{-5mm} & &   \hspace*{5mm}  
             \times \bigl[ 5 H(0,0,0,0;x)
          + 16 H(-1, \! -1, \! 0, \! 0;x) 
          - 4  H(-1, \! 0, \! 0, \! 0;x)\nn\\
\hspace*{-5mm} & &   \hspace*{5mm}  
   + 2  H(-1, \! 0, \! 1, \! 0;x)  \!
   - \! 16 H(0, \! -1, \! -1, \! 0;x)  \!
   + \! 3  H(0, \! -1, \! 1, \! 0;x) \nn\\
\hspace*{-5mm} & &   \hspace*{5mm}     
          - \! 12 H(0,0,1,0;x) \! 
   + \! 12 H(0,1, \! -1,0;x) \! 
   - \! 4 H(0,1,1,0;x) \bigr] \nn\\  
\hspace*{-5mm} & &        +  \biggl[
            \frac{7}{2} \! 
          +  \! \frac{3}{2(1-x)^2}  \! 
          -  \! \frac{1}{(1-x)} \! 
          -  \! \frac{60}{(1+x)^5} \! 
          +  \! \frac{150}{(1+x)^4}    \!        
          -  \! \frac{126}{(1+x)^3}\nn\\
\hspace*{-5mm} & &   \hspace*{5mm} 
          + \frac{81}{2(1+x)^2} 
          - \frac{9}{(1+x)}
          \biggr] H(0,-1,0,0;x)  \nn\\
\hspace*{-5mm} & &        +  \biggl[
            \frac{5}{2} \! 
          +  \! \frac{7}{2(1-x)^2}  \! 
          -  \! \frac{15}{4(1-x)}  \! 
          +  \! \frac{30}{(1+x)^5} \! 
          -  \! \frac{75}{(1+x)^4}   \!         
          +  \! \frac{63}{(1+x)^3} \nn\\
\hspace*{-5mm} & &   \hspace*{5mm} 
          - \frac{16}{(1+x)^2}
          + \frac{1}{4(1+x)} 
          \biggr] H(0,0,-1,0;x) 
 \nn\\
\hspace*{-5mm} & &        -  \biggl[
            3
          + \frac{2}{(1-x)^2}
          - \frac{7}{4(1-x)} 
          - \frac{30}{(1+x)^5}
          + \frac{75}{(1+x)^4}           
          - \frac{63}{(1+x)^3} \nn\\
\hspace*{-5mm} & &  \hspace*{5mm} 
          + \frac{43}{2(1+x)^2} 
          - \frac{23}{4(1+x)} 
          \biggr] H(0,1,0,0;x)  \nn\\
\hspace*{-5mm} & &        +  \biggl[
            2 
          + \frac{1}{2(1-x)} 
          - \frac{60}{(1+x)^5}
          + \frac{150}{(1+x)^4}
          - \frac{126}{(1+x)^3}          
          + \frac{39}{(1+x)^2} \nn\\
\hspace*{-5mm} & &   \hspace*{5mm} 
          - \frac{15}{2(1+x)} 
          \biggr] H(1,0,0,0;x)  \nn\\
\hspace*{-5mm} & + & 
  {\mathcal O} (D-4) \, , \\
\hspace*{-5mm} {\mathcal F}^{(2l,{\tt a})}_{2}(D,q^2) & = & 
          \frac{1}{(D-4)} \Biggl\{ 
        \frac{1}{4} \biggl[
            \frac{1}{(1-x)} \! 
          - \! \frac{1}{(1+x)}
          \biggr] H(0;x) \! 
       -  \biggl[
            \frac{1}{(1-x)^2}  \! 
   - \! \frac{1}{(1-x)}  
         \nn\\
\hspace*{-5mm} & &  \hspace*{18mm}          
          - \frac{1}{(1+x)^2}
          + \frac{1}{(1+x)}
          \biggr] H(0,0;x)
\Biggr\}   \nn\\
\hspace*{-5mm} & & \nn\\
\hspace*{-5mm} & & 
          +  \biggl[ \frac{7}{8(1-x)^2}  \! 
          -  \! \frac{5}{4(1-x)}  \! 
          +  \! \frac{81}{2(1+x)^4}  \! 
          -  \! \frac{177}{2(1+x)^3}  \! 
   +  \! \frac{465}{8(1+x)^2} \nn\\
\hspace*{-5mm} & &     
          - \frac{39}{4(1+x)} \biggr]  \zeta(2)
          + \biggl[ \frac{17}{160(1-x)} 
          + \frac{51}{2(1+x)^5} 
   - \frac{255}{4(1+x)^4}  \nn\\
\hspace*{-5mm} & &          
          + \frac{2057}{40(1+x)^3} 
          - \frac{1071}{80(1+x)^2} 
          + \frac{17}{160(1+x)}  \biggr]  \zeta^{2}(2)           
          - \biggl[ \frac{5}{2(1-x)^2} \nn\\
\hspace*{-5mm} & & 
          - \frac{5}{2(1-x)} \! 
          + \frac{60}{(1+x)^4} \! 
          - \frac{120}{(1+x)^3} \!
   + \frac{131}{2(1+x)^2}          
          - \frac{11}{2(1 \! + \! x)} \biggr]  \zeta(3) \nn\\
\hspace*{-5mm} & &  
          + \frac{3}{4(1+x)} \biggl[ 
     1
   - \frac{1}{(1+x)} \biggr] \nn\\
\hspace*{-5mm} & &        +  \biggl[ \biggl(
            \frac{7}{8(1 \! - \! x)^3} \! 
          -  \! \frac{45}{16(1 \! - \! x)^2}  \! 
          +  \! \frac{87}{32(1 \! - \! x)} \!  
          +  \! \frac{81}{2(1 \! + \! x)^5}  \! 
          -  \! \frac{465}{4(1 \! + \! x)^4}  \nn\\
\hspace*{-5mm} & & \hspace*{5mm}      
          +   \frac{119}{(1+x)^3} 
          -   \frac{191}{4(1  +  x)^2}  
          +   \frac{119}{32(1  +  x)}  \biggr) \zeta(2) 
          -   \biggl( \frac{1}{4(1  -  x)}  \nn\\ 
\hspace*{-5mm} & & \hspace*{5mm}   
          +   \frac{60}{(1  +   x)^5}  
          - \frac{150}{(1 \! + \! x)^4} \! 
          + \frac{121}{(1 \! + \! x)^3} \!
          - \frac{63}{2(1 \! + \! x)^2}  \! 
          +  \! \frac{1}{4(1 \! + \! x)}  \biggr) \zeta(3) \nn\\
\hspace*{-5mm} & & \hspace*{5mm} 
          -  \! \frac{3}{8(1-x)}        
          -  \! \frac{9}{4(1+x)^3} \! 
          +  \! \frac{27}{8(1+x)^2}
   - \! \frac{3}{4(1+x)} 
          \biggr] H(0;x)   \nn\\
\hspace*{-5mm} & &    
       +  \biggl[ \biggl(
            \frac{1}{16(1-x)}  \! 
          +  \! \frac{15}{(1+x)^5} \! 
          -  \! \frac{75}{2(1+x)^4} \! 
          +  \! \frac{121}{4(1+x)^3}  \!   
          -  \! \frac{63}{8(1+x)^2}  \nn\\
\hspace*{-5mm} & & \hspace*{5mm}       
          + \frac{1}{16(1+x)} \biggr)  \zeta(2)
          + \frac{17}{8(1-x)^2}
          - \frac{7}{4(1-x)} 
          - \frac{39}{2(1+x)^4} \nn\\
\hspace*{-5mm} & & \hspace*{5mm}    
          + \frac{93}{2(1+x)^3} 
          - \frac{295}{8(1+x)^2}
          + \frac{19}{2(1+x)}
          \biggr] H(0,0;x) \nn\\
\hspace*{-5mm} & &        
       -  \biggl[ 
            \frac{3}{4(1-x)}
          + \frac{15}{(1+x)^3}
          - \frac{45}{2(1+x)^2}
          + \frac{27}{4(1+x)}
          \biggr] H(-1,0;x) \nn\\
\hspace*{-5mm} & &        
       +  \biggl[ \biggl(
            \frac{1}{4(1-x)}
          + \frac{60}{(1+x)^5}
          - \frac{150}{(1+x)^4}
          + \frac{121}{(1+x)^3}
          - \frac{63}{2(1+x)^2}  \nn\\
\hspace*{-5mm} & & \hspace*{5mm} 
          + \frac{1}{4(1+x)} \biggr)  \zeta(2)
          \biggr] H(1,0;x) \nn\\ 
\hspace*{-5mm} & &     
       +  \biggl[
            \frac{7}{8(1-x)^3} \! 
          -  \! \frac{29}{16(1-x)^2}  \!  
          +  \! \frac{55}{32(1-x)}   \! 
   +  \! \frac{81}{2(1+x)^5}  \! 
          -  \! \frac{525}{4(1+x)^4}   \nn\\
\hspace*{-5mm} & & \hspace*{5mm} 
          + \frac{149}{(1+x)^3}
          - \frac{263}{4(1+x)^2}
   + \frac{215}{32(1+x)} 
          \biggr] H(0,0,0;x) \nn\\ 
\hspace*{-5mm} & &      
       +  \biggl[
            \frac{60}{(1+x)^4} \!
          - \frac{120}{(1+x)^3} \!
          + \frac{68}{(1+x)^2} 
          - \frac{8}{(1+x)} \! 
          \biggr] H(-1,0,0;x)  \nn\\ 
\hspace*{-5mm} & &  
       -  \biggl[
            \frac{1}{(1-x)^2}
          - \frac{1}{(1-x)}
          + \frac{30}{(1+x)^4} 
          - \frac{60}{(1+x)^3}
          + \frac{33}{(1+x)^2} \nn\\
\hspace*{-5mm} & & \hspace*{5mm} 
          - \frac{3}{(1+x)}
          \biggr] \bigl[ H(0,-1,0;x) +  H(1,0,0;x) \bigr] \nn\\ 
\hspace*{-5mm} & &  
       -  \biggl[
            \frac{1}{(1-x)^2}
          - \frac{1}{(1-x)}
          - \frac{1}{(1+x)^2}
          + \frac{1}{(1+x)}
          \biggr] H(0,1,0;x)   \nn\\ 
\hspace*{-5mm} & &        
       -  \biggl[
            \frac{1}{8(1-x)}   
          + \frac{30}{(1+x)^5}  
          - \frac{75}{(1+x)^4}
          + \frac{121}{2(1+x)^3} 
          - \frac{63}{4(1+x)^2} \nn\\
\hspace*{-5mm} & & \hspace*{5mm} 
          + \frac{1}{8(1+x)} 
          \biggr] \bigl[ H(0,0,-1,0;x) + H(0,1,0,0;x) \nn\\
\hspace*{-5mm} & & \hspace*{25mm} 
   - 2 H(1,0,0,0;x) - 2 H(0,-1,0,0;x) \bigr] \nn\\
\hspace*{-5mm} & +&  {\mathcal O} (D-4) \, , \\
\hspace*{-5mm} {\mathcal F}^{(2l,{\tt a})}_{3}(D,q^2) & = &  0 \, . \\
\nonumber
\eea

$\bullet$ The {\it Cross} graph {\tt(b)} of Fig. \ref{fig1}, defined as
\bea
\parbox{20mm}{\begin{fmfgraph*}(15,15)
\fmfleft{i1,i2}
\fmfright{o}
\fmf{plain}{i1,v1}
\fmf{plain}{i2,v2}
\fmf{photon}{v5,o}
\fmf{plain,tension=.3}{v2,v3}
\fmf{plain,tension=.3}{v3,v5}
\fmf{plain,tension=.3}{v1,v4}
\fmf{plain,tension=.3}{v4,v5}
\fmf{photon,tension=0}{v2,v4}
\fmf{photon,tension=0}{v1,v3}
\end{fmfgraph*} 
}  & = & 
\int {\mathfrak{D}}^Dk_1\;{\mathfrak{D}}^Dk_2 \ 
\frac{{\mathcal N}_{({\tt b})}^{\mu}}{{\mathcal D}_{1} {\mathcal D}_{2} 
{\mathcal D}_{9} {\mathcal D}_{11} {\mathcal D}_{12} 
{\mathcal D}_{13} } \, , 
\label{b2} 
\eea
where 
\bea
\! \! \! \! \! \! {\mathcal N}_{({\tt b})}^{\mu} & = &  \bar{v}(p_2)  \gamma_{\sigma}
[i ( \not{\! p_{2}} - \!  \! \not{\! k_{2}})  \! + \! m] 
\gamma_{\lambda} 
[i ( \not{\! p_{2}} + \!  \! \not{\! k_{1}} - \! \! \not{\! k_{2}})  \! 
+ \! m]
\gamma^{\mu}
[-i ( \not{\! p_{1}} -  \!  \! \not{\! k_{1}} +  \! \! \not{\! k_{2}})  
\! + \! m] \times
\nn\\
\! \! \! \! \! \! & & \times
\gamma^{\sigma} 
[-i ( \not{\! p_{1}} - \!  \! \not{\! k_{1}} )  \! + \! m]
\gamma^{\lambda} u(p_1) ,
\eea
gives: 
\bea
\hspace*{-5mm} {\mathcal F}^{(2l,{\tt b})}_{1}(D,q^2) & = & \frac{1}{(D-4)} 
\Biggl\{   \frac{1}{4}  
          -  \frac{1}{2} \biggl[ 
     1 
   + \frac{1}{(1-x)^2}  
          - \frac{1}{(1-x)}
          + \frac{1}{(1+x)^2}  \nn\\
\hspace*{-5mm} & &   \hspace*{18mm}     
          - \frac{1}{(1+x)} \biggr] \bigl[ \zeta(3)
   - \zeta(2) H(0;x) 
   + 2 H(0,0,0;x) \nn\\
\hspace*{-5mm} & &   \hspace*{38mm} 
   - 2 H(0,-1,0;x) 
   + 2 H(0,1,0;x) \bigr]
\Biggr\}   \nn\\
\hspace*{-5mm} & &
          - \frac{1}{2}
          + \frac{1}{2} \biggl[ \frac{1}{(1-x)} 
          - \frac{45}{(1+x)^4} 
          + \frac{99}{(1+x)^3} 
          - \frac{53}{(1+x)^2} \nn\\
\hspace*{-5mm} & &           - \frac{3}{(1+x)} \! 
          +  \! \frac{27}{4} \biggr] \zeta(2) \! 
          +  \! \biggl[ \frac{37}{40(1-x)^2}  \! 
          -  \! \frac{971}{320(1-x)}  \! 
          -  \! \frac{321}{20(1+x)^5} \nn\\
\hspace*{-5mm} & &           
          + \frac{321}{8(1+x)^4} 
          - \frac{2299}{80(1+x)^3} 
          + \frac{125}{32(1+x)^2} 
          - \frac{79}{64(1+x)} \nn\\
\hspace*{-5mm} & &           
          + \frac{119}{40} \biggr]  \zeta^{2}(2)
          -  \biggl[ \frac{6}{(1+x)^4}
          - \frac{12}{(1+x)^3}
          + \frac{7}{(1+x)^2}
          - \frac{1}{(1+x)} \nn\\
\hspace*{-5mm} & &           + 2 \biggr]  \zeta(3)
          - \frac{3}{4 (1+x)} \biggl[ 
     1 
   - \frac{1}{(1+x)} \biggr] \nn\\
\hspace*{-5mm} & &         
          -   \biggl[
            \frac{1}{4}
          - \biggl( \frac{25}{32(1-x)} 
          + \frac{45}{2(1+x)^5}
          - \frac{279}{4(1+x)^4} 
          + \frac{597}{8(1+x)^3}  \nn\\
\hspace*{-5mm} & &      \hspace*{5mm}   
          - \frac{463}{16(1+x)^2}
          + \frac{121}{32(1+x)} 
          - \frac{3}{4} \biggr) \zeta(2)
          - \biggl( \frac{1}{4(1-x)^2} \nn\\
\hspace*{-5mm} & &       \hspace*{5mm}  
          - \frac{7}{8(1-x)} \! 
          - \frac{6}{(1+x)^5} \! 
          + \!  \frac{15}{(1+x)^4} \! 
          - \frac{25}{2(1+x)^3} \! 
          + \!  \frac{4}{(1+x)^2} \nn\\
\hspace*{-5mm} & &       \hspace*{5mm}  
          -  \! \frac{47}{8(1+x)}  \! 
          +  \! \frac{13}{4} \biggr) \zeta(3) \! 
          -  \! \frac{5}{8(1-x)}  \! 
          -  \! \frac{9}{4(1+x)^3}  \! 
          +  \! \frac{27}{8(1+x)^2}  \nn\\
\hspace*{-5mm} & &    \hspace*{5mm}  
          - \frac{1}{(1+x)}
          \biggr] H(0;x)  \nn\\
\hspace*{-5mm} & &        
          + \biggl[
            \frac{45}{(1 \! + \! x)^4} \! 
          -  \! \frac{90}{(1 \! + \! x)^3} \! 
          +  \! \frac{93}{2(1 \! + \! x)^2}  \! 
          -  \! \frac{3}{2(1 \! + \! x)}  \! 
          -  \! \frac{3}{2} 
          \biggr] \zeta(2) H(-1;x) \nn\\
\hspace*{-5mm} & &      
          -  \biggl[
            \frac{13}{4}
          + \biggl( \frac{1}{2(1-x)^2} 
          + \frac{49}{32(1-x)} 
          + \frac{15}{2(1+x)^5} 
          - \frac{75}{4(1+x)^4} \nn\\
\hspace*{-5mm} & &     \hspace*{5mm}          
          +  \! \frac{105}{8(1+x)^3}  \! 
          -  \! \frac{7}{16(1+x)^2}  \! 
          +  \! \frac{17}{32(1+x)}  \! 
          -  \! \frac{3}{2} \biggr)   \zeta(2) \! 
          +  \! \frac{3}{8(1-x)^2} \nn\\
\hspace*{-5mm} & &     \hspace*{5mm}    
          - \frac{23}{8(1-x)} 
          - \frac{51}{2(1+x)^4} 
          + \frac{129}{2(1+x)^3} 
          - \frac{103}{2(1+x)^2} \nn\\
\hspace*{-5mm} & &     \hspace*{5mm}
          + \frac{43}{4(1+x)}  \! 
          \biggr] H(0,0;x) \nn\\
\hspace*{-5mm} & &     
       -  \biggl[
            \frac{1}{2(1-x)^2}  \! 
          -  \! \frac{23}{16(1-x)} \! 
          -  \! \frac{45}{(1+x)^5}    \! 
   +  \! \frac{225}{2(1+x)^4}  \! 
          -  \! \frac{351}{4(1+x)^3}  \nn\\
\hspace*{-5mm} & &     \hspace*{5mm}     
          + \frac{157}{8(1+x)^2} 
          - \frac{71}{16(1+x)} 
   + \frac{7}{2}
          \biggr] \zeta(2) H(0,-1;x) \nn\\
\hspace*{-5mm} & &       
       +  \biggl[
            3
          -  \! \frac{3}{(1 \! - \! x)} \! 
          +  \! \frac{21}{(1 \! + \! x)^3} \! 
          -  \! \frac{63}{2(1 \! + \! x)^2}  \! 
          +  \! \frac{15}{2(1 \! + \! x)}
          \biggr]  \! H(-1,0;x)    \nn\\
\hspace*{-5mm} & &  
       -  \biggl[
            \frac{5}{4} \! 
          -  \! \biggl( \frac{1}{(1 \! - \! x)^2} \! 
          -  \! \frac{5}{2(1 \! - \! x)} \! 
          +  \! \frac{12}{(1 \! + \! x)^5} \! 
          -  \! \frac{30}{(1 \! + \! x)^4} \! 
          +  \! \frac{26}{(1 \! + \! x)^3} \nn\\
\hspace*{-5mm} & &          \hspace*{5mm}     
          -  \! \frac{8}{(1+x)^2} \! 
          +  \! \frac{7}{2(1 \! + \! x)} \biggr) \zeta(2)  \! 
          -  \! \frac{2}{(1-x)} \! 
          +  \! \frac{6}{(1 \! + \! x)^3} \! 
          -  \! \frac{9}{(1 \! + \! x)^2}\nn\\
\hspace*{-5mm} & &         \hspace*{5mm}     
          + \frac{5}{2(1+x)} 
          \biggr] H(1,0;x) \nn\\
\hspace*{-5mm} & &        
          -  \biggl[
            1
          - \frac{25}{32(1-x)} 
          - \frac{45}{2(1+x)^5} 
          + \frac{255}{4(1+x)^4} 
          - \frac{501}{8(1+x)^3} \nn\\
\hspace*{-5mm} & &     \hspace*{5mm}          
          + \frac{359}{16(1+x)^2} 
          - \frac{105}{32(1+x)} 
          \biggr] H(0,0,0;x)  \nn\\
\hspace*{-5mm} & &   
       -   \! \biggl[
            3 \! 
          -  \! \frac{42}{(1+x)^4} \! 
   +  \! \frac{84}{(1+x)^3} \! 
          -  \! \frac{41}{(1+x)^2} \! 
          -  \! \frac{1}{(1+x)}
          \biggr] H(0,-1,0;x)  \nn\\
\hspace*{-5mm} & &    
       +  \!  \biggl[
            \frac{5}{2} \! 
   -  \! \frac{3}{(1 \! + \! x)^4} \! 
          +  \! \frac{6}{(1 \! + \! x)^3} \! 
          -  \! \frac{3}{2(1 \! + \! x)^2}  \! 
          -  \! \frac{3}{2(1 \! + \! x)} 
          \biggr] H(-1, 0, 0;x) \nn\\
\hspace*{-5mm} & &        +   \biggl[
            \frac{3}{2} \! 
          -  \! \frac{12}{(1+x)^4} \! 
          +  \! \frac{24}{(1+x)^3} \! 
          -  \! \frac{11}{(1+x)^2} \! 
          -  \! \frac{1}{(1+x)} \! 
          \biggr]  \! H(0,1,0;x) \nn\\
\hspace*{-5mm} & &        -   \biggl[
            1
          - \frac{36}{(1+x)^4}
          + \frac{72}{(1+x)^3}
          - \frac{36}{(1+x)^2}
          \biggr] H(1,0,0;x) \nn\\
\hspace*{-5mm} & &        +  \biggl[
            \frac{11}{2}
          + \frac{3}{(1-x)^2}
          - \frac{169}{32(1-x)} 
          - \frac{3}{2(1+x)^5} 
          + \frac{15}{4(1+x)^4} \nn\\
\hspace*{-5mm} & &     \hspace*{5mm}
          - \frac{9}{8(1+x)^3} 
          + \frac{15}{16(1+x)^2} 
          - \frac{153}{32(1+x)} 
          \biggr] H(0,0,0,0;x) \nn\\
\hspace*{-5mm} & &        
        -  \biggl[
            7
          + \frac{4}{(1-x)^2}
          - \frac{79}{16(1-x)} 
          + \frac{3}{(1+x)^5}
          - \frac{15}{2(1+x)^4} \nn\\
\hspace*{-5mm} & &     \hspace*{5mm}    
          + \frac{17}{4(1+x)^3} 
          + \frac{41}{8(1+x)^2} 
          - \frac{159}{16(1+x)} 
          \biggr] H(0,-1,0,0;x) \nn\\
\hspace*{-5mm} & &        -  \biggl[
            6 \! 
          +  \! \frac{6}{(1-x)^2} \! 
          -  \! \frac{43}{8(1-x)} \! 
          -  \! \frac{42}{(1+x)^5} \! 
          +  \! \frac{105}{(1+x)^4} \! 
          -  \! \frac{159}{2(1+x)^3} \nn\\
\hspace*{-5mm} & &     \hspace*{5mm}      
          + \frac{81}{4(1+x)^2}
          - \frac{35}{8(1+x)}
          \biggr] H(0,0,-1,0;x) \nn\\
\hspace*{-5mm} & &          
       +  \biggl[
            5
          +  \! \frac{4}{(1-x)^2}   \! 
   -  \! \frac{9}{2(1-x)} \! 
          -  \! \frac{12}{(1+x)^5} \! 
          +  \! \frac{30}{(1+x)^4} \! 
          -  \! \frac{22}{(1+x)^3} \nn\\
\hspace*{-5mm} & &     \hspace*{5mm} 
          + \frac{7}{(1+x)^2}
   - \frac{9}{2(1+x)} \! 
          \biggr] H(0, \! 0, \! 1, \! 0, \! x)   \nn\\
\hspace*{-5mm} & & 
       +   \! \biggl[
            4 \! 
          +  \! \frac{2}{(1-x)^2} \! 
          -  \! \frac{17}{4(1-x)} \! 
          + \frac{36}{(1+x)^5}
          - \frac{90}{(1+x)^4} \! 
          +  \! \frac{69}{(1+x)^3} \nn\\
\hspace*{-5mm} & &     \hspace*{5mm}        
          -  \! \frac{23}{2(1+x)^2} \! 
          -  \! \frac{21}{4(1+x)} \! 
          \biggr] H(0,1,0,0;x) \nn\\
\hspace*{-5mm} & &        
       +  \biggl[
            1 \! 
          +  \! \frac{2}{(1-x)^2} \! 
          -  \! \frac{7}{2(1-x)} \! 
          +  \! \frac{12}{(1+x)^5} \! 
          -  \! \frac{30}{(1+x)^4} \! 
          +  \! \frac{26}{(1+x)^3} \nn\\
\hspace*{-5mm} & &     \hspace*{5mm}
          - \frac{7}{(1+x)^2}
          + \frac{5}{2(1+x)}
          \biggr] H(1,0,0,0;x)  \nn\\
\hspace*{-5mm} & &        
          + \frac{1}{2} \biggl[
             1
          +     \frac{1}{(1 \! - \! x)^2}  \! 
          -  \! \frac{1}{(1 \! - \! x)} \! 
          +  \! \frac{1}{(1 \! + \! x)^2} \! 
          -  \! \frac{1}{(1 \! + \! x)} 
          \biggr] \bigl[ 2 \zeta(3) ( H(1;x) \nn\\
\hspace*{-5mm} & &       \hspace*{5mm}
          - H(-1;x) )  
   - 2 \zeta(2) H(-1,0;x) 
   + \zeta(2) H(0,1;x) \nn\\
\hspace*{-5mm} & &       \hspace*{5mm}
   + 4 H( \! -1, \! 0, \! -1, \! 0;x)
   - 4 H( \! -1, \! 0, \! 0, \! 0;x)
   - 4 H( \! -1, \! 0,1, \! 0;x)\nn\\
\hspace*{-5mm} & &       \hspace*{5mm}
   +  \! 10 H(0, \! -1, \! -1, \! 0, \! x) \! 
   -  \! 6 H(0, \! -1, \! 1, \! 0, \! x) \! 
   -  \! 6 H(0, \! 1, \! -1, \! 0, \! x)\nn\\
\hspace*{-5mm} & &       \hspace*{5mm}
   + 2 H(0,1,1,0;x)
   - 4 H(1,0,-1,0;x)
   + 4 H(1,0,1,0;x) \bigr]
   \nn\\
\hspace*{-5mm} &+ &  {\mathcal O} (D-4) \, , \\
\hspace*{-5mm} {\mathcal F}^{(2l,{\tt b})}_{2}(D,q^2) & = & 
        - \biggl[ \frac{1}{8(1-x)^2}  
          + \frac{5}{4(1-x)} 
          - \frac{45}{2(1+x)^4} 
          + \frac{99}{2(1+x)^3} \nn\\
\hspace*{-5mm} & &      
          - \frac{197}{8(1+x)^2} 
          - \frac{15}{4(1+x)} \biggr] \zeta(2) 
        + \biggl[ \frac{39}{80(1-x)^3} 
          - \frac{117}{160(1-x)^2} \nn\\
\hspace*{-5mm} & &           
          - \frac{337}{320(1-x)} 
          + \frac{321}{20(1+x)^5} 
          - \frac{321}{8(1+x)^4} 
          + \frac{137}{5(1+x)^3} \nn\\
\hspace*{-5mm} & &           
          - \frac{39}{40(1+x)^2} 
          - \frac{337}{320(1+x)}  \biggr]  \zeta^{2}(2)
        + \biggl[ \frac{1}{2(1-x)^2}  
          - \frac{1}{2(1-x)}   \nn\\
\hspace*{-5mm} & &           
          + \frac{6}{(1+x)^4}
          - \frac{12}{(1+x)^3}
          + \frac{13}{2(1+x)^2}  
          - \frac{1}{2(1+x)}  \biggr]  \zeta(3) \nn\\
\hspace*{-5mm} & &           
          - \frac{3}{4(1+x)^2}
          + \frac{3}{4(1+x)} \nn\\
\hspace*{-5mm} & &           
       -  \biggl[ \biggl(
            \frac{1}{8(1 \! - \! x)^3}  \!
          -  \! \frac{13}{16(1 \! - \! x)^2}  \! 
          +  \! \frac{19}{32(1 \! - \! x)}  \! 
   +  \! \frac{45}{2(1 \! + \! x)^5} \! 
          -  \! \frac{279}{4(1 \! + \! x)^4}  \nn\\
\hspace*{-5mm} & &     \hspace*{5mm}
          +  \frac{291}{4(1+x)^3}
          -  \frac{25}{(1+x)^2}
          -  \frac{13}{32(1+x)}  \biggr)  \zeta(2) 
          -  \biggl( \frac{1}{2(1-x)^3}   \nn\\
\hspace*{-5mm} & &     \hspace*{5mm}
          -  \! \frac{3}{4(1-x)^2} 
          + \frac{1}{8(1-x)}  
          + \frac{6}{(1+x)^5}
          - \frac{15}{(1+x)^4} 
          + \frac{12}{(1+x)^3} \nn\\
\hspace*{-5mm} & &     \hspace*{5mm}
          - \frac{3}{(1+x)^2}
          + \frac{1}{8(1+x)}  \biggr) \zeta(3) 
          + \frac{5}{16(1-x)}
          + \frac{9}{4(1+x)^3}  \nn\\
\hspace*{-5mm} & &     \hspace*{5mm}
          - \frac{27}{8(1+x)^2}
          + \frac{13}{16(1+x)}
          \biggr] H(0;x) \nn\\
\hspace*{-5mm} & &          
       -  \biggl[
            \frac{3}{4(1-x)^2} 
          - \frac{3}{4(1-x)}
          + \frac{45}{(1+x)^4}
          - \frac{90}{(1+x)^3}
          + \frac{171}{4(1+x)^2}  \nn\\
\hspace*{-5mm} & &     \hspace*{5mm}
          + \frac{9}{4(1+x)} 
          \biggr] \zeta(2) H(-1;x) \nn\\
\hspace*{-5mm} & &        
       +  \biggl[ \biggl(
            \frac{5}{8(1 \! - \! x)^3} \! 
          -  \! \frac{15}{16(1 \! - \! x)^2}  \! 
          -  \! \frac{15}{32(1 \! - \! x)}  \! 
          +  \! \frac{15}{2(1 \! + \! x)^5} \! 
          -  \! \frac{75}{4(1 \! + \! x)^4} \nn\\
\hspace*{-5mm} & &     \hspace*{5mm}
          + \frac{25}{2(1+x)^3} 
          - \frac{15}{32(1+x)} \biggr)  \zeta(2)
          - \frac{19}{8(1-x)^2}
          + \frac{21}{4(1-x)} \nn\\
\hspace*{-5mm} & &     \hspace*{5mm}
          -  \! \frac{51}{2(1+x)^4} \! 
          +  \! \frac{129}{2(1+x)^3} \! 
          -  \! \frac{395}{8(1+x)^2} \! 
          +  \! \frac{15}{2(1+x)} \! 
          \biggr] H(0,0;x) \nn\\
\hspace*{-5mm} & &        
        -  \biggl[ \biggl(
            \frac{1}{4(1-x)} 
          + \frac{12}{(1+x)^5}
          - \frac{30}{(1+x)^4}
          + \frac{25}{(1+x)^3}
          - \frac{15}{2(1+x)^2} \nn\\
\hspace*{-5mm} & &     \hspace*{5mm}
          + \frac{1}{4(1+x)}   \biggr)  \zeta(2)
          - \frac{1}{(1-x)}
          - \frac{6}{(1+x)^3}
          + \frac{9}{(1+x)^2}\nn\\
\hspace*{-5mm} & &     \hspace*{5mm}
          - \frac{2}{(1+x)}
          \biggr] H(1,0;x) \nn\\
\hspace*{-5mm} & &       
       -  \biggl[
             \! \frac{19}{4(1-x)} \! 
          +  \! \frac{21}{(1+x)^3} \! 
          -  \! \frac{63}{2(1+x)^2} \! 
          +  \! \frac{23}{4(1+x)} \! 
          \biggr] \!  H(-1,0;x) \nn\\
\hspace*{-5mm} & &        
       -  \biggl[
            \frac{3}{4(1-x)^3} \! 
          - \frac{9}{8(1-x)^2}  \! 
          -  \! \frac{21}{16(1-x)}  \! 
          +  \! \frac{45}{(1 \! + \! x)^5} \! 
          -  \! \frac{225}{2(1 \! + \! x)^4} \nn\\
\hspace*{-5mm} & &      \hspace*{5mm}     
          +  \! \frac{84}{(1+x)^3} \! 
          -  \! \frac{27}{2(1+x)^2}  \! 
          -  \! \frac{21}{16(1+x)}  \! 
          \biggr]  \zeta(2) H(0,-1;x) \nn\\
\hspace*{-5mm} & &        
      -  \biggl[
            \frac{1}{8(1-x)^3}
          - \frac{13}{16(1-x)^2}
          + \frac{19}{32(1-x)}
          + \frac{45}{2(1+x)^5} \nn\\
\hspace*{-5mm} & &      \hspace*{5mm}    
          - \frac{255}{4(1 \! + \! x)^4} \! 
          +  \! \frac{243}{4(1 \! + \! x)^3} \! 
          -  \! \frac{19}{(1 \! + \! x)^2} \! 
          -  \! \frac{13}{32(1 \! + \! x)} \! 
          \biggr]  \! H(0,0,0;x) \nn\\
\hspace*{-5mm} & &        
       +  \biggl[
            \frac{12}{(1+x)^4}
          - \frac{24}{(1+x)^3}
          + \frac{10}{(1+x)^2}
          + \frac{2}{(1+x)}
          \biggr] H(0,1,0;x) \nn\\
\hspace*{-5mm} & &        
       +  \biggl[
            \frac{1}{(1-x)^2}
          - \frac{1}{(1-x)}
          - \frac{36}{(1+x)^4}
          + \frac{72}{(1+x)^3}
          - \frac{33}{(1+x)^2} \nn\\
\hspace*{-5mm} & &     \hspace*{5mm}    
          - \frac{3}{(1+x)}
          \biggr]  H(1,0,0;x)  \nn\\
\hspace*{-5mm} & &        
       -  \biggl[
            \frac{3}{4(1-x)^2}
          - \frac{3}{4(1-x)}
          - \frac{3}{(1+x)^4} 
          + \frac{6}{(1+x)^3}
          - \frac{5}{4(1+x)^2} \nn\\
\hspace*{-5mm} & &     \hspace*{5mm}   
          - \frac{7}{4(1+x)}
          \biggr] H(-1,0,0;x) \nn\\
\hspace*{-5mm} & &        
       +   \! \biggl[
            \frac{1}{2(1-x)^2} \! 
          -  \! \frac{1}{2(1-x)} \! 
          -  \! \frac{42}{(1+x)^4} \! 
          +  \! \frac{84}{(1+x)^3} \! 
          -  \! \frac{75}{2(1+x)^2} \nn\\
\hspace*{-5mm} & &           
          - \frac{9}{2(1+x)}
          \biggr] H(0,-1,0;x)  \nn\\
\hspace*{-5mm} & &           
       +  \biggl[
            \frac{5}{8(1 \! - \! x)^3}
          - \frac{15}{16(1 \! - \! x)^2}
          - \frac{11}{32(1 \! - \! x)} \! 
          +  \! \frac{3}{2(1 \! + \! x)^5} \! 
          -  \! \frac{15}{4(1 \! + \! x)^4} \nn\\
\hspace*{-5mm} & &     \hspace*{5mm} 
          +  \! \frac{1}{(1+x)^3} \! 
          +  \! \frac{9}{4(1+x)^2}
          - \frac{11}{32(1+x)}
          \biggr] H(0,0,0,0;x)  \nn\\
\hspace*{-5mm} & &          
       -  \biggl[
            \frac{3}{4(1-x)^3}
          -  \! \frac{9}{8(1-x)^2} 
   +  \! \frac{11}{16(1-x)} \! 
          -  \! \frac{3}{(1 \! + \! x)^5} \! 
          +  \! \frac{15}{2(1 \! + \! x)^4} \nn\\
\hspace*{-5mm} & &     \hspace*{5mm} 
          -  \! \frac{4}{(1+x)^3} \! 
          -  \! \frac{3}{2(1 \! + \! x)^2}
          + \frac{11}{16(1 \! + \! x)}
          \biggr] H(0,-1,0,0;x) \nn\\
\hspace*{-5mm} & &          
       +  \biggl[
            \frac{1}{2(1-x)^3}
          - \frac{3}{4(1-x)^2} 
   + \frac{17}{8(1-x)}
          -  \frac{42}{(1+x)^5}
          +  \frac{105}{(1+x)^4}  \nn\\
\hspace*{-5mm} & &     \hspace*{5mm} 
          -  \! \frac{76}{(1+x)^3} \! 
          +  \! \frac{9}{(1+x)^2}
   + \frac{17}{8(1+x)}
          \biggr] H(0,0,-1,0;x) \nn\\
\hspace*{-5mm} & &            
       -  \biggl[
            \frac{3}{4(1-x)}
          - \frac{12}{(1+x)^5}
   + \frac{30}{(1+x)^4}
          -  \frac{21}{(1+x)^3}
          +  \frac{3}{2(1+x)^2} \nn\\
\hspace*{-5mm} & &     \hspace*{5mm} 
          +  \frac{3}{4(1+x)}
          \biggr]  \! H(0,0,1,0;x) \nn\\
\hspace*{-5mm} & &        
       +  \biggl[
            \frac{1}{(1-x)^3}
          - \frac{3}{2(1-x)^2}
          + \frac{7}{4(1-x)}
          - \frac{36}{(1+x)^5}
          + \frac{90}{(1+x)^4} \nn\\
\hspace*{-5mm} & &     \hspace*{5mm}        
          - \frac{66}{(1+x)^3}
          + \frac{9}{(1+x)^2}
          + \frac{7}{4(1+x)}
          \biggr] H(0,1,0,0;x) \nn\\
\hspace*{-5mm} & &        
       -  \biggl[
            \frac{1}{4(1-x)} \! 
          + \frac{12}{(1+x)^5} \! 
          - \frac{30}{(1+x)^4} \! 
          + \frac{25}{(1+x)^3} \! 
          - \frac{15}{2(1+x)^2} \nn\\
\hspace*{-5mm} & &      \hspace*{5mm}       
          + \frac{1}{4(1+x)}
          \biggr] H(1,0,0,0;x)  \nn\\
\hspace*{-5mm} & + & {\mathcal O} (D-4) \, , \\
\hspace*{-5mm} {\mathcal F}^{(2l,{\tt b})}_{3}(D,q^2) & = & 0 \, .
\nonumber
\eea

$\bullet$ The {\it Down-Corner} graph {\tt(c)} of Fig. \ref{fig1}, defined as
\bea
\parbox{20mm}{\begin{fmfgraph*}(15,15)
\fmfleft{i1,i2}
\fmfright{o}
\fmfforce{0.2w,0.93h}{v2}
\fmfforce{0.2w,0.07h}{v1}
\fmfforce{0.2w,0.5h}{v3}
\fmfforce{0.8w,0.5h}{v5}
\fmf{plain}{i1,v1}
\fmf{plain}{i2,v2}
\fmf{photon}{v5,o}
\fmf{plain,tension=0}{v2,v5}
\fmf{plain,tension=0}{v3,v4}
\fmf{photon,tension=.4}{v1,v4}
\fmf{plain,tension=.4}{v4,v5}
\fmf{plain,tension=0}{v1,v3}
\fmf{photon,tension=0}{v2,v3}
\end{fmfgraph*} }  & = & 
\int {\mathfrak{D}}^Dk_1\;{\mathfrak{D}}^Dk_2\ 
\frac{{\mathcal N}_{({\tt c})}^{\mu}}{{\mathcal D}_{1} {\mathcal D}_{5} 
{\mathcal D}_{7} {\mathcal D}_{8} {\mathcal D}_{9} 
{\mathcal D}_{10} } \, , 
\label{b3} 
\eea
where 
\bea
{\mathcal N}_{({\tt c})}^{\mu} & = &  \bar{v}(p_2) \gamma_{\sigma} 
[i ( \not{\! k_{2}}) +m] 
\gamma_{\lambda} 
[i ( \not{\! k_{1}} + \! \not{\! k_{2}}) +m] 
\gamma_{\sigma} 
[i ( \not{\! p_{2}} + \! \not{\! k_{1}}) +m]
\gamma^{\mu} \times
\nn\\
& & \times
[-i ( \not{\! p_{1}} - \! \not{\! k_{1}}) +m] 
\gamma^{\lambda} u(p_1)  ,
\eea
gives: 
\bea
\hspace*{-5mm} {\mathcal F}^{(2l,{\tt c})}_{1}(D,q^2) & = & \frac{1}{(D-4)^{2}} 
\Biggl\{
         \frac{1}{8}
       +  \biggl[
            1
          - \frac{1}{(1-x)}
          - \frac{1}{(1+x)}
          \biggr] H(0;x) 
\Biggr\}   \nn\\
\hspace*{-5mm} & - & \frac{1}{(D-4)} \Biggl\{
            \frac{3}{32}
   - \frac{7 \zeta(2)}{2}
   - \frac{1}{8} \biggl[
     1
   - \frac{2}{(1+x)} \biggr] H(0;x) \nn\\
\hspace*{-5mm} & &    \hspace*{18mm}  
   - \frac{1}{4} \biggl[
     1 
          - \frac{1}{(1-x)}
          - \frac{1}{(1+x)}
    \biggr] \bigl[ 
           2 \zeta(2)
   - 4 H(0;x)  \nn\\
\hspace*{-5mm} & &    \hspace*{36mm}  
   + H(0,0;x)
   + 2 H(1,0;x) \bigr]
\Biggr\} \nn\\
\hspace*{-5mm} & & 
          + \frac{17}{128} \! 
       + \biggl[ 
            \frac{23}{16}  \! 
   + \frac{1}{2(1-x)} \! 
          - \frac{15}{4(1+x)^4} \! 
          + \frac{15}{2 (1+x)^3} \! 
          - \frac{23}{8 (1+x)^2}\nn\\
\hspace*{-5mm} & &           
          - \frac{1}{2(1+x)} \biggr] \zeta(2)
       - \biggl[ \frac{9}{(1+x)^2}
          - \frac{9}{(1+x)}
          + \frac{9}{2}  \biggr] \zeta(2) \ln{2}           
       + \biggl[ 
            \frac{1}{20} \nn\\
\hspace*{-5mm} & & 
   -  \! \frac{31}{160(1-x)} \! 
          -  \! \frac{3}{5 (1 \! + \! x)^5} \! 
          +  \! \frac{3}{2  (1 \! + \! x)^4} \! 
          -  \! \frac{71}{40 (1 \! + \! x)^3} \! 
          +  \! \frac{93}{80 (1 \! + \! x)^2} \nn\\
\hspace*{-5mm} & &           
          - \frac{31}{160 (1 \! + \! x)} \biggr] \zeta^{2}(2)
       - \biggl[ \frac{5}{8 (1 \! - \! x)} \! 
          -  \! \frac{6}{ (1 \! + \! x)^4} \! 
          +  \! \frac{12}{ (1 \! + \! x)^3} \! 
          -  \! \frac{8}{ (1 \! + \! x)^2} \nn\\
\hspace*{-5mm} & &           
          + \frac{11}{8 (1+x)}
          + \frac{1}{8}  \biggr] \zeta(3)
          + \frac{13}{8 (1+x)} \biggr[ 
     1  
          - \frac{1}{(1+x)} \biggr] \nn\\
\hspace*{-5mm} & &           
       +  \biggl[
            \frac{49}{32}
       - \biggl( \frac{85}{64 (1-x)}
          + \frac{15}{4 (1+x)^5}
          - \frac{75}{8 (1+x)^4}           
          + \frac{117}{16 (1+x)^3} \nn\\
\hspace*{-5mm} & &  \hspace*{5mm} 
          +  \! \frac{117}{32 (1+x)^2}  \! 
          -  \! \frac{331}{64 (1+x)} \! 
          + \!  \frac{3}{8} \biggr)  \zeta(2) \! 
       - \biggl( \frac{5}{16(1-x)}   \!    
          +  \! \frac{6}{(1+x)^5} \nn\\
\hspace*{-5mm} & &  \hspace*{5mm} 
          - \frac{15}{(1+x)^4} \! 
          + \frac{45}{4(1+x)^3} \! 
          - \frac{15}{8(1+x)^2} \! 
          + \frac{5}{16(1+x)}   \!   
          - \frac{1}{2}  \biggr) \zeta(3) \nn\\
\hspace*{-5mm} & &  \hspace*{5mm}
          -  \frac{21}{16 (1-x)}   \! 
          -   \!  \frac{39}{8 (1+x)^3}   \! 
          +   \!  \frac{117}{16 (1+x)^2} \! 
          -   \!  \frac{67}{16 (1+x)}  \!  
          \biggr]   H(0;x) \nn\\
\hspace*{-5mm} & &           
       -  \biggl[
            \frac{1}{4 (1-x)}
          + \frac{1}{4 (1+x)}
          - \frac{1}{4}
          \biggr] \zeta(2) H(1;x) \nn\\
\hspace*{-5mm} & &        +  \biggl[
            \frac{21}{2 (1+x)^2}
          - \frac{21}{2 (1+x)}
          + \frac{9}{4}
          \biggr] \zeta(2) H(-1;x) \nn\\
\hspace*{-5mm} & &        
   +  \biggl[
            \frac{35}{8} 
       - \biggl( \frac{1}{16 (1-x)}
          + \frac{3}{ (1+x)^5}
          - \frac{15}{2 (1+x)^4}
          + \frac{19}{4 (1+x)^3} \nn\\
\hspace*{-5mm} & &   \hspace*{5mm}        
          +  \frac{3}{8 (1+x)^2} \! 
          -  \! \frac{1}{16 (1+x)} \! 
          -  \! \frac{1}{4} \biggr) \zeta(2) \! 
          +  \! \frac{3}{16 (1-x)^2}  \!          
          -  \! \frac{47}{16 (1-x)} \nn\\
\hspace*{-5mm} & &   \hspace*{5mm}        
          -  \! \frac{15}{4 (1+x)^4} \! 
          +  \! \frac{3}{(1+x)^3} \! 
          +  \! \frac{31}{8 (1+x)^2}  \! 
          -  \! \frac{43}{8 (1+x)} \! 
          \biggr] H(0,0;x)  \nn\\
\hspace*{-5mm} & &           
       -  \biggl[
            \frac{29}{8} \! 
          -  \! \frac{4}{ (1-x)} \! 
          -  \! \frac{6}{ (1+x)^3}  \! 
          +  \! \frac{9 }{(1+x)^2} \! 
          -  \! \frac{25}{4 (1+x)} \! 
          \biggr] H(-1,0;x) \nn\\
\hspace*{-5mm} & &           
       + \frac{1}{9} \biggl[
            \frac{3}{(1-x)} \! 
          +  \! \frac{12}{(1+x)^3} \! 
          - \frac{18}{(1+x)^2} \! 
          +  \! \frac{3}{(1+x)} \! 
          \biggr] \zeta(2) H(0,-1;x)  \nn\\
\hspace*{-5mm} & &      
       +  \biggl[
            \frac{11}{8}   \!    
       + \biggl( \frac{1}{4(1-x)} \! 
          +  \! \frac{6}{(1 \! + \! x)^5} \! 
          - \frac{15}{(1 \! + \! x)^4} \! 
          +  \! \frac{11}{(1 \! + \! x)^3} \! 
          - \frac{3}{2 (1 \! + \! x)^2} \nn\\
\hspace*{-5mm} & &      \hspace*{5mm}   
          + \frac{1}{4 (1+x)}
          - \frac{1}{2} \biggr) \zeta(2)
          - \frac{3}{2 (1-x)}
          - \frac{3}{ (1+x)^3}
          + \frac{9}{2 (1+x)^2} \nn\\
\hspace*{-5mm} & &     \hspace*{5mm}        
          - \frac{11}{4 (1+x)}
          \biggr] H(1,0;x)  \nn\\
\hspace*{-5mm} & &      
       -  \biggl[
            \frac{13}{8}
          + \frac{21}{64 (1-x)}
          + \frac{15}{4 (1+x)^5}
          - \frac{51}{8 (1+x)^4}
          + \frac{21}{16 (1+x)^3} \nn\\
\hspace*{-5mm} & &     \hspace*{5mm}  
          + \frac{197}{32 (1+x)^2}
          - \frac{363}{64 (1+x)}
          \biggr] H(0,0,0;x) \nn\\
\hspace*{-5mm} & &        
       - \frac{1}{2} \biggl[
            1
          - \frac{1}{(1-x)}
          - \frac{1}{(1+x)}
          \biggr] \bigl[
            H(1,1,0;x)
   + 6 H(-1,-1,0;x)  \nn\\
\hspace*{-5mm} & &     \hspace*{5mm}  
          - 3 H(-1,1,0;x)
          - 3 H(1,-1,0;x) \bigr] \nn\\
\hspace*{-5mm} & &        
       +  \biggl[
            3
          - \frac{9}{4 (1-x)}
          + \frac{7}{2 (1+x)^2}
          - \frac{23}{4 (1+x)}
          \biggr] H(-1,0,0;x)  \nn\\
\hspace*{-5mm} & &        
       +  \biggl[
            \frac{19}{4}
          - \frac{9}{4(1-x)}
          + \frac{12}{(1+x)^4}       
          - \frac{24}{(1+x)^3}
          + \frac{17}{ (1+x)^2} \nn\\
\hspace*{-5mm} & &      \hspace*{5mm} 
          - \frac{29}{4 (1+x)}
          \biggr] H(0,-1,0;x)  \nn\\
\hspace*{-5mm} & &     
       -  \biggl[
            \frac{9}{4}      
   - \frac{1}{(1-x)}
          + \frac{6}{ (1+x)^4}
          - \frac{12}{ (1+x)^3}
          + \frac{17}{2 (1+x)^2}  \nn\\
\hspace*{-5mm} & &    \hspace*{5mm} 
   - \frac{7}{2 (1+x)}
          \biggr] H(0,1,0;x)  \nn\\
\hspace*{-5mm} & &        
       +  \biggl[
            \frac{3}{4}
          + \frac{1}{(1-x)}
          + \frac{12}{ (1+x)^4}
          - \frac{24}{ (1+x)^3}
          + \frac{27}{2 (1+x)^2} \nn\\
\hspace*{-5mm} & &     \hspace*{5mm}      
          - \frac{1}{2 (1+x)}
          \biggr]  H(1,0,0;x)   \nn\\
\hspace*{-5mm} & &        
       - \frac{1}{8} \biggl[
            \frac{1}{(1-x)} \! 
          +  \! \frac{4}{(1+x)^3} \! 
          - \frac{6}{(1+x)^2} \! 
          +  \! \frac{1}{(1+x)} \! 
          \biggr]  \bigl[ H(0,0,0,0;x) \nn\\
\hspace*{-5mm} & &  \hspace*{5mm}       
          - H(0,-1,0,0;x) \bigr] \nn\\
\hspace*{-5mm} & &        
       -  \biggl[
            1
          - \frac{1}{2 (1-x)} \! 
          - \frac{12}{ (1+x)^5} \! 
          + \frac{30}{ (1+x)^4} \! 
          - \frac{22}{ (1+x)^3} \! 
          + \frac{3}{ (1+x)^2} \nn\\
\hspace*{-5mm} & &  \hspace*{5mm}           
          - \frac{1}{2(1+x)}
          \biggr] H(0,0,-1,0;x)  \nn\\
\hspace*{-5mm} & &        
       +  \biggl[
            \frac{1}{2} \! 
          - \frac{1}{4(1-x)} \! 
          - \frac{6}{(1+x)^5}  \! 
          +  \! \frac{15}{(1+x)^4} \! 
          - \frac{11}{(1+x)^3} \! 
          +  \! \frac{3}{2(1+x)^2}\nn\\
\hspace*{-5mm} & &   \hspace*{5mm}     
          - \frac{1}{4(1+x)}
          \biggr] H(0,0,1,0;x) \nn\\
\hspace*{-5mm} & &        
        -  \biggl[
            1 \! 
          -  \! \frac{3}{8(1-x)} \! 
          -  \! \frac{12}{(1+x)^5} \! 
          +  \! \frac{30}{(1+x)^4} \! 
          -  \! \frac{43}{2(1+x)^3}    \!      
          +  \! \frac{9}{4(1+x)^2} \nn\\
\hspace*{-5mm} & &   \hspace*{5mm}    
          - \frac{3}{8(1+x)}
          \biggr] H(0,1,0,0;x)  \nn\\
\hspace*{-5mm} & &           
       -  \biggl[
            \frac{1}{2} \! 
          -  \! \frac{1}{4(1-x)} \! 
          -  \! \frac{6}{(1+x)^5} \! 
          +  \! \frac{15}{(1+x)^4} \! 
          -  \! \frac{11}{(1+x)^3} \! 
          +  \! \frac{3}{2(1+x)^2} \nn\\
\hspace*{-5mm} & &           - \frac{1}{4(1+x)}
          \biggr]  H(1,0,0,0;x)  \nn\\
\hspace*{-5mm} & + & {\mathcal O} (D-4) \, , \\
\hspace*{-5mm} {\mathcal F}^{(2l,{\tt c})}_{2}(D,q^2) & = & \frac{1}{(D-4)} 
\Biggl\{ 
       \frac{1}{4} \biggl[
          \frac{1}{(1-x)}
          - \frac{1}{(1+x)}
          \biggr] H(0;x) 
\Biggr\}   \nn\\
\hspace*{-5mm} & & 
       + \biggl[ \frac{1}{16 (1-x)^2} \! 
          +  \! \frac{1}{16 (1-x)}  \! 
          +  \! \frac{15}{4 (1 \! + \! x)^4} \! 
          -  \! \frac{15}{2 (1 \! + \! x)^3} \! 
          +  \! \frac{89}{16 (1 \! + \! x)^2} \nn\\
\hspace*{-5mm} & &           
          - \frac{31}{16 (1+x)} \biggr]  \zeta(2)
   - \frac{6 \zeta(2) \ln{2}}{ (1+x)} \biggl[
     1 
   - \frac{1}{ (1+x)} \biggr] 
       - \biggl[ \frac{27}{40 (1-x)^3}  \nn\\
\hspace*{-5mm} & &           
          -  \! \frac{81}{80 (1-x)^2} \! 
          +  \! \frac{3}{80 (1-x)} \! 
          -  \! \frac{3}{5 (1 \! + \! x)^5} \! 
          +  \! \frac{3}{2 (1 \! + \! x)^4} \! 
          -  \! \frac{69}{40 (1 \! + \! x)^3} \nn\\
\hspace*{-5mm} & &           
          +  \! \frac{87}{80 (1 \! + \! x)^2} \! 
          +  \! \frac{3}{80 (1 \! + \! x)} \biggr]  \zeta^{2}(2) \! 
       -  \! \biggl[ \frac{1}{4 (1 \! - \! x)^2} \! 
          -  \! \frac{1}{4 (1 \! - \! x)} \! 
          -  \! \frac{6}{ (1 \! + \! x)^4} \nn\\
\hspace*{-5mm} & &           
          +  \! \frac{12}{ (1+x)^3} \! 
          - \frac{27}{4 (1+x)^2} \! 
          +  \! \frac{3}{4 (1 \! + \! x)} \biggr]  \zeta(3) \! 
          - \frac{13}{8 (1 \! + \! x)} \biggl[
     1  \! 
   - \frac{1}{(1 \! + \! x)} \biggr] 
           \nn\\
\hspace*{-5mm} & &        
       +  \biggl[ \biggl(
            \frac{1}{16  (1 \! - \! x)^3} \! 
          -  \! \frac{27}{32 (1 \! - \! x)^2} \! 
          +  \! \frac{57}{64 (1 \! - \! x)} \! 
          +  \! \frac{15}{4 (1 \! + \! x)^5} \! 
          -  \! \frac{75}{8 (1 \! + \! x)^4} \nn\\
\hspace*{-5mm} & &      \hspace*{5mm}       
          + \frac{7}{(1+x)^3}
          + \frac{21}{8 (1+x)^2}
          - \frac{263}{64 (1+x)} \biggr) \zeta(2) 
          - \biggl( \frac{1}{4  (1-x)^3} \nn\\
\hspace*{-5mm} & &      \hspace*{5mm} 
          + \frac{3}{8 (1-x)^2}
          - \frac{3}{8 (1-x)}
          + \frac{6}{ (1+x)^5}
          - \frac{15}{ (1+x)^4}
          + \frac{43}{4 (1+x)^3} \nn\\
\hspace*{-5mm} & &      \hspace*{5mm} 
          - \frac{9}{8 (1+x)^2}
          - \frac{3}{8 (1+x)} \biggr) \zeta(3) 
          - \frac{27}{32 (1-x)}
          + \frac{39}{8 (1+x)^3} \nn\\
\hspace*{-5mm} & &      \hspace*{5mm} 
          - \frac{117}{16 (1+x)^2}
          + \frac{105}{32 (1+x)}
          \biggr] H(0;x) \nn\\
\hspace*{-5mm} & &        
       + \frac{3}{2} \biggl[
            \frac{1}{(1-x)^2}
          - \frac{1}{(1-x)}
          - \frac{5}{(1+x)^2}
          + \frac{5}{(1+x)}
          \biggr] \zeta(2) H(-1;x) \nn\\
\hspace*{-5mm} & &        
       -  \biggl[ \biggl(
            \frac{3}{4 (1-x)^3} \! 
          -  \! \frac{9}{8 (1-x)^2} \! 
          -  \! \frac{3}{16 (1-x)} \! 
          +  \! \frac{3}{(1 \! + \! x)^5}   \!      
          -  \! \frac{15}{2 (1 \! + \! x)^4} \nn\\
\hspace*{-5mm} & &    \hspace*{5mm}
          + \frac{9}{2 (1+x)^3}
          + \frac{3}{4 (1+x)^2}
          - \frac{3}{16 (1+x)}  \biggr) \zeta(2)     
          + \frac{13}{16 (1-x)^2} \nn\\
\hspace*{-5mm} & &    \hspace*{5mm}
          + \frac{1}{16 (1-x)}
          - \frac{15}{4 (1+x)^4}
          + \frac{3}{(1+x)^3}
          + \frac{51}{16 (1+x)^2} \nn\\
\hspace*{-5mm} & &    \hspace*{5mm}
          - \frac{53}{16 (1+x)}
          \biggr] H(0,0;x)  \nn\\
\hspace*{-5mm} & &       
       +  \biggl[
            \frac{5}{4 (1-x)}
          - \frac{6}{ (1+x)^3}   
          + \frac{9}{ (1+x)^2}
          - \frac{17}{4 (1+x)}
          \biggr] H(-1,0;x) \nn\\
\hspace*{-5mm} & &    
       +  \biggl[
            \frac{3 \zeta(2)}{8 (1-x)}
          - \frac{6 \zeta(2)}{ (1+x)^5}
          + \frac{15 \zeta(2)}{ (1+x)^4}
          - \frac{21 \zeta(2)}{2 (1+x)^3}
          + \frac{3 \zeta(2)}{4 (1+x)^2} \nn\\
\hspace*{-5mm} & &       \hspace*{5mm}    
          + \frac{3 \zeta(2)}{8 (1+x)}
          - \frac{1}{2(1-x)}
          + \frac{3}{(1+x)^3}
          - \frac{9}{2 (1+x)^2} \nn\\
\hspace*{-5mm} & &       \hspace*{5mm}   
          + \frac{2}{(1+x)}
          \biggr] H(1,0;x)  \nn\\
\hspace*{-5mm} & &       
       +  \biggl[
            \frac{1}{16 (1 \! - \! x)^3} \! 
          -  \! \frac{19}{32 (1 \! - \! x)^2} \! 
          +  \! \frac{41}{64 (1 \! - \! x)} \! 
          +  \! \frac{15}{4 (1 \! + \! x)^5} \! 
          -  \! \frac{51}{8 (1 \! + \! x)^4} \nn\\
\hspace*{-5mm} & &       \hspace*{5mm}
          + \frac{1}{(1+x)^3}
          + \frac{39}{8 (1+x)^2}
          - \frac{215}{64 (1+x)}
          \biggr] H(0,0,0;x) \nn\\
\hspace*{-5mm} & &        
       +  \biggl[
            \frac{6}{ (1+x)^4} \! 
          -  \! \frac{12}{ (1+x)^3} \! 
          +  \! \frac{7}{ (1+x)^2} \! 
          -  \! \frac{1}{(1+x)} \! 
          \biggr] H(0,1,0;x)  \nn\\
\hspace*{-5mm} & &        
          +   \! \frac{1}{2} \! \biggl[
            \frac{1}{(1-x)^2} \! 
          -  \! \frac{1}{(1-x)} \! 
          -  \! \frac{5}{(1+x)^2} \! 
          +  \! \frac{5}{(1+x)} \! 
          \biggr]  \! H(-1,0,0;x) \nn\\
\hspace*{-5mm} & &        
       -  \!  \biggl[
            \frac{12}{(1+x)^4} \! 
          -  \! \frac{24}{(1+x)^3} \! 
          +  \! \frac{14}{(1+x)^2} \! 
          -  \! \frac{2}{(1+x)} \! 
          \biggr]  \! H(0, \! -1,0;x)  \nn\\
\hspace*{-5mm} & &        
       -  \! \biggl[
            \frac{1}{2 (1-x)^2} \! 
          -  \! \frac{1}{2 (1-x)} \! 
          +  \! \frac{12}{ (1+x)^4} \! 
          -  \! \frac{24}{ (1+x)^3} \! 
          +  \! \frac{23}{2 (1+x)^2} \nn\\
\hspace*{-5mm} & &      \hspace*{5mm}  
          +  \! \frac{1}{2 (1 \! + \! x)} \! 
          \biggr] \!  H(1,0,0;x)   \nn\\
\hspace*{-5mm} & &   
       +  \frac{1}{2} \biggl[
            \frac{1}{(1-x)^3}
          - \frac{3}{2 (1-x)^2}
          - \frac{1}{(1+x)^3}
          + \frac{3}{2 (1+x)^2}
          \biggr] \times \nn\\
\hspace*{-5mm} & &      \hspace*{8mm}
    \times \bigl[ 3 \zeta(2) H(0,-1;x) 
   - H(0,0,0,0;x) 
   + H(0,-1,0,0;x) \bigr] \nn\\
\hspace*{-5mm} & &           
       +  \biggl[
            \frac{3}{4 (1-x)}
   - \frac{12}{(1+x)^5}
          + \frac{30}{ (1+x)^4}
          - \frac{21}{ (1+x)^3}
          + \frac{3}{2 (1+x)^2}  \nn\\
\hspace*{-5mm} & &      \hspace*{5mm}
   + \frac{3}{4 (1+x)}
          \biggr]  H(0,0,-1,0;x)  \nn\\
\hspace*{-5mm} & &           
       -  \biggl[
            \frac{3}{8 (1-x)}
          - \frac{6}{(1+x)^5} 
   + \frac{15}{(1+x)^4}
   - \frac{21}{2 (1+x)^3}
          + \frac{3}{4 (1+x)^2} \nn\\
\hspace*{-5mm} & &      \hspace*{5mm}
          + \frac{3}{8 (1+x)}
          \biggr] H(0,0,1,0;x) \nn\\
\hspace*{-5mm} & &        -  \biggl[
            \frac{1}{2 (1-x)^3}
          - \frac{3}{4 (1-x)^2}
          - \frac{3}{4 (1-x)}
          + \frac{12}{ (1+x)^5}
          - \frac{30}{ (1+x)^4} \nn\\
\hspace*{-5mm} & &      \hspace*{5mm}       
          + \frac{41}{2 (1+x)^3}
          - \frac{3}{4 (1+x)^2}
          - \frac{3}{4 (1+x)} 
          \biggr] H(0,1,0,0;x) \nn\\
\hspace*{-5mm} & & 
       +  \biggl[
            \frac{3}{8 (1-x)} \! 
          -  \! \frac{6}{ (1+x)^5} \! 
          +  \! \frac{15}{ (1+x)^4} \! 
          -  \! \frac{21}{2 (1+x)^3} \! 
          +  \! \frac{3}{4 (1+x)^2} \nn\\
\hspace*{-5mm} & &      \hspace*{5mm}    
          + \frac{3}{8 (1+x)}
          \biggr] H(1,0,0,0;x)  \nn\\
\hspace*{-5mm} & + & 
{\mathcal O} (D-4) , \\
\hspace*{-5mm} {\mathcal F}^{(2l,{\tt c})}_{3}(D,q^2) & = & 
      - \biggl[ \frac{3}{16 (1-x)^2}
          - \frac{3}{16 (1-x)}
          + \frac{21}{16 (1+x)^2}
          - \frac{21}{16 (1+x)} \biggr]  \zeta(2) \nn\\
\hspace*{-5mm} & &           
          - \frac{3 \zeta(2) \ln{2}}{ (1-x)}
 \biggl[ 
     1
   - \frac{1}{ (1-x)} \biggr] 
      + \biggl[ \frac{27}{80 (1-x)^3}
          - \frac{81}{160 (1-x)^2} \nn\\
\hspace*{-5mm} & &        
          - \frac{27}{80 (1+x)^3}
          + \frac{81}{160 (1+x)^2} \biggr] \zeta^{2}(2)
      - \biggl[ \frac{1}{8 (1-x)^2}
          - \frac{1}{8 (1-x)} \nn\\
\hspace*{-5mm} & &           
          + \frac{1}{8 (1+x)^2}
          - \frac{1}{8 (1+x)} \biggr] \zeta(3)
          - \frac{1}{16 (1-x)^2}
          + \frac{1}{16 (1-x)} \nn\\
\hspace*{-5mm} & &        
       -  \biggl[ \biggl( 
            \frac{3}{16 (1-x)^3}
          - \frac{69}{32 (1-x)^2}
          + \frac{33}{16 (1-x)}
          + \frac{9}{16 (1+x)^3} \nn\\
\hspace*{-5mm} & &           \hspace*{5mm}   
          - \frac{15}{32 (1+x)^2}
          - \frac{3}{16 (1+x)} \biggl)  \zeta(2)
       - \biggl( \frac{1}{8 (1-x)^3}
          + \frac{3}{16 (1-x)^2} \nn\\
\hspace*{-5mm} & &           \hspace*{5mm}   
          + \frac{1}{8 (1+x)^3}
          - \frac{3}{16 (1+x)^2}  \biggl)  \zeta(3)
          + \frac{1}{8 (1-x)^3}
          - \frac{3}{16 (1-x)^2} \nn\\
\hspace*{-5mm} & &          \hspace*{5mm}    
          - \frac{11}{32 (1-x)} \! 
          + \frac{13}{32 (1+x)} \! 
          \biggr] H(0;x)  \nn\\
\hspace*{-5mm} & &       
       +   \biggl[  \biggl(
            \frac{3}{8 (1-x)^3}
          - \frac{9}{16 (1-x)^2}
          -  \frac{3}{8 (1+x)^3}
          + \frac{9}{16 (1+x)^2} \biggr)  \zeta(2) \nn\\
\hspace*{-5mm} & &          \hspace*{5mm} 
          -  \! \frac{1}{8 (1-x)^4} \! 
          +  \! \frac{1}{4 (1-x)^3} \! 
          +  \! \frac{5}{8 (1-x)^2} \! 
          -  \! \frac{3}{4 (1-x)} \! 
          -  \! \frac{13}{16 (1 \! + \! x)^2}\nn\\
\hspace*{-5mm} & &          \hspace*{5mm} 
          + \frac{13}{16 (1+x)}
          \biggr] H(0,0;x) \nn\\
\hspace*{-5mm} & &    
      -  \biggl[
            \frac{3}{16 (1-x)^3}
          - \frac{49}{32 (1-x)^2}
          + \frac{23}{16 (1-x)}
          + \frac{9}{16 (1+x)^3} \nn\\
\hspace*{-5mm} & &         \hspace*{5mm}  
          - \frac{19}{32 (1+x)^2}
          - \frac{1}{16 (1+x)}
          \biggr] H(0,0,0;x) \nn\\
\hspace*{-5mm} & &        
       - \frac{1}{4}  \biggl[
            \frac{5}{(1 \! - \! x)^2}  \! 
          -  \! \frac{5}{(1 \! - \! x)} \! 
          -  \! \frac{1}{(1 \! + \! x)^2} \! 
          +  \! \frac{1}{(1 \! + \! x)} \! 
          \biggr] \bigl[  3 \zeta(2) H(-1;x)   \nn\\
\hspace*{-5mm} & &         \hspace*{5mm}  
          + H(-1,0,0;x) 
   - H(1,0,0;x) \bigr] \nn\\
\hspace*{-5mm} & &        
       - \frac{1}{4}  \biggl[
            \frac{1}{(1-x)^3}
          - \frac{3}{2 (1-x)^2}
          - \frac{1}{(1+x)^3}
          + \frac{3}{2 (1+x)^2}
          \biggr]  \times  \nn\\
\hspace*{-5mm} & &         \hspace*{8mm}  
        \times \bigl[ 3 \zeta(2) H(0,-1;x)
   + H(0,0,0,0;x)
   - H(0,-1,0,0;x)  \nn\\
\hspace*{-5mm} & &         \hspace*{13mm}  
   + H(0,1,0,0;x) \bigr] \nn\\
\hspace*{-5mm} & + & {\mathcal O} (D-4) \, .
\eea

$\bullet$ The {\it Up-Corner} graph {\tt(d)} of Fig. \ref{fig1}, defined as
\bea
\parbox{20mm}{\begin{fmfgraph*}(15,15)
\fmfleft{i1,i2}
\fmfright{o}
\fmfforce{0.2w,0.93h}{v1}
\fmfforce{0.2w,0.07h}{v2}
\fmfforce{0.2w,0.5h}{v3}
\fmfforce{0.8w,0.5h}{v5}
\fmf{plain}{i1,v2}
\fmf{plain}{i2,v1}
\fmf{photon}{v5,o}
\fmf{plain,tension=0}{v2,v5}
\fmf{plain,tension=0}{v3,v4}
\fmf{photon,tension=.4}{v1,v4}
\fmf{plain,tension=.4}{v4,v5}
\fmf{plain,tension=0}{v1,v3}
\fmf{photon,tension=0}{v2,v3}
\end{fmfgraph*} }  & = & 
\int {\mathfrak{D}}^Dk_1\;{\mathfrak{D}}^Dk_2\ 
\frac{{\mathcal N}_{({\tt d})}^{\mu}}{{\mathcal D}_{3} {\mathcal D}_{4} 
{\mathcal D}_{6} {\mathcal D}_{7} {\mathcal D}_{12} 
{\mathcal D}_{13} } \, , 
\label{b4} 
\eea
where 
\bea
{\mathcal N}_{({\tt d})}^{\mu} & = & \bar{v}(p_2) \gamma_{\sigma} 
[i ( \not{\! p_{2}} + \! \not{\! k_{1}} - \not{\! k_{2}}) +m] 
\gamma^{\mu} 
[-i ( \not{\! p_{1}} - \! \not{\! k_{1}} + \not{\! k_{2}}) +m] 
\gamma_{\lambda}   \times \nn\\
& & \times [-i ( \not{\! k_{2}}) +m] 
\gamma_{\sigma} 
[-i (\not{\! k_{1}}) +m]
\gamma^{\lambda} u(p_1) ,
\eea
due to reasons of symmetry gives:
\bea
{\mathcal F}^{(2l,{\tt d})}_{1}(D,q^2) & = &  {\mathcal F}^{(2l,{\tt c})}_{1}(D,q^2) 
\, , \\
{\mathcal F}^{(2l,{\tt d})}_{2}(D,q^2) & = &  {\mathcal F}^{(2l,{\tt c})}_{2}(D,q^2) 
\, , \\
{\mathcal F}^{(2l,{\tt d})}_{3}(D,q^2) & = & - {\mathcal F}^{(2l,{\tt c})}_{3}(D,q^2) 
\, .
\eea

$\bullet$ The {\it Up-Self-Mass-insertion} graph {\tt(e)} of Fig. \ref{fig1}, 
defined as 
\bea
\parbox{20mm}{\begin{fmfgraph*}(15,15)
\fmfleft{i1,i2}
\fmfright{o}
\fmfforce{0.2w,0.93h}{v2}
\fmfforce{0.2w,0.07h}{v1}
\fmfforce{0.8w,0.5h}{v5}
\fmf{plain}{i1,v1}
\fmf{plain}{i2,v2}
\fmf{photon}{v5,o}
\fmf{plain}{v2,v3}
\fmf{photon,tension=.25,right}{v3,v4}
\fmf{plain,tension=.25}{v3,v4}
\fmf{plain}{v4,v5}
\fmf{plain}{v1,v5}
\fmf{photon}{v1,v2}
\end{fmfgraph*} }  & = & 
\int {\mathfrak{D}}^Dk_1\;{\mathfrak{D}}^Dk_2\ 
\frac{{\mathcal N}_{({\tt e})}^{\mu}}{{\mathcal D}_{1} {\mathcal D}_{2} 
{\mathcal D}^{2}_{9} {\mathcal D}_{10} {\mathcal D}_{12}} 
\label{b5} 
\eea
where 
\bea
\! \! \! \! \! \! {\mathcal N}_{({\tt e})}^{\mu} & = &  \bar{v}(p_2)  \gamma_{\sigma}
[i ( \not{\! p_{2}} +  \!  \! \not{\! k_{1}}) \!  + \! m] 
\gamma_{\mu} 
[-i ( \not{\! p_{1}} -  \! \! \not{\! k_{1}})  \! + \! m] 
\gamma_{\lambda} 
[-i ( \not{\! p_{1}} - \!  \! \not{\! k_{1}} + \!  \! \not{\! k_{2}})  
\! + \! m]
\gamma^{\lambda} \times
\nn\\
\! \! \! \! \! \! & & \times
[-i ( \not{\! p_{1}} - \! \! \not{\! k_{1}}) \! +\! m] 
\gamma^{\sigma} u(p_1),
\eea
gives: 
\bea
\hspace*{-5mm} {\mathcal F}^{(2l,{\tt e})}_{1}(D,q^2) & = & \frac{1}{(D-4)^{2}} 
\Biggl\{ 
            \frac{11}{8}
          + \frac{3}{(1+x)^2}
          - \frac{3}{(1+x)}
        -  \biggl[
            1
          + \frac{1}{2(1-x)}  \nn\\
\hspace*{-5mm} & &  \hspace*{20mm}     
          - \frac{3}{(1+x)^3}
          + \frac{9}{2(1+x)^2}
          - \frac{4}{(1+x)}
          \biggr] H(0;x) 
\Biggr\}   \nn\\
\hspace*{-5mm} & + & \frac{1}{(D-4)} 
\Biggl\{ 
            \frac{19}{32}
          - \frac{2}{(1+x)^2}
          + \frac{2}{(1+x)}
       - \biggl( \frac{1}{4(1-x)}
          + \frac{3}{2(1+x)^3} \nn\\
\hspace*{-5mm} & &    \hspace*{18mm}         
          -  \! \frac{9}{4(1 \! + \! x)^2} \! 
          +  \! \frac{2}{(1 \! + \! x)}  \! 
          -  \! \frac{1}{2} \biggr) \zeta(2)  \nn\\
\hspace*{-5mm} & &    \hspace*{18mm}     
       +  \biggl[
            \frac{7}{8}
          + \frac{3}{4(1-x)}
          - \frac{5}{(1+x)^3}       
          + \frac{15}{2(1+x)^2}\nn\\
\hspace*{-5mm} & &    \hspace*{23mm} 
          - \frac{5}{(1+x)}
          \biggr] H(0;x) \nn\\
\hspace*{-5mm} & &    \hspace*{18mm}     
       -  \biggl[
            \frac{1}{4}
          - \frac{1}{(1-x)}
          + \frac{3}{2(1+x)^3}   
          - \frac{9}{4(1+x)^2}\nn\\
\hspace*{-5mm} & &    \hspace*{23mm} 
          + \frac{5}{4(1+x)}
          \biggr] H(0,0;x)  \nn\\
\hspace*{-5mm} & &    \hspace*{18mm}     
       - \frac{3}{2} \biggl[
            \frac{1}{(1 \! - \! x)} \! 
          -  \! \frac{2}{(1 \! + \! x)^3}    \!  
          +  \! \frac{3}{(1 \! + \! x)^2} \! 
          -  \! \frac{2}{(1 \! + \! x)} \! 
          \biggr] H( \! -1, \! 0, \! x)  \nn\\
\hspace*{-5mm} & &    \hspace*{18mm}     
       -  \biggl[
            \frac{1}{2}
          - \frac{1}{2(1-x)}        
          - \frac{1}{2(1+x)}
          \biggr] H(1,0;x) 
\Biggr\}   \nn\\
\hspace*{-5mm} & & 
          + \frac{179}{128}
       + \biggl[ \frac{3}{8(1-x)}
          + \frac{21}{4(1+x)^4}
          - \frac{13}{(1+x)^3}
          + \frac{105}{8(1+x)^2} \nn\\
\hspace*{-5mm} & &           
          - \frac{53}{8(1+x)}
          + \frac{31}{16} \biggr]  \zeta(2)
       + \biggl[ \frac{11}{8(1-x)}
          - \frac{3}{2(1+x)^3}
          + \frac{9}{4(1+x)^2} \nn\\
\hspace*{-5mm} & &           
          -  \! \frac{7}{8(1 \! + \! x)} \! 
          -  \! \frac{5}{8} \biggr]  \zeta(3)
          +  \! \frac{59}{8(1 \! + \! x)^2} \! 
          -  \! \frac{59}{8(1 \! + \! x)} \nn\\
\hspace*{-5mm} & &   
       -  \biggl[
            \frac{45}{32} \! 
       + \biggl( \frac{17}{8} \! 
          -  \! \frac{75}{64(1-x)}  \! 
   -  \! \frac{21}{4(1 \! + \! x)^5} \! 
          +  \! \frac{105}{8(1 \! + \! x)^4} \! 
          -  \! \frac{215}{16(1 \! + \! x)^3} \nn\\
\hspace*{-5mm} & &     \hspace*{5mm}     
          + \frac{225}{32(1+x)^2}
          - \frac{291}{64(1+x)} \biggr)  \zeta(2)
          + \frac{21}{16(1-x)}
          - \frac{89}{8(1+x)^3} \nn\\
\hspace*{-5mm} & &     \hspace*{5mm}  
          + \frac{267}{16(1+x)^2}
          - \frac{155}{16(1+x)}
          \biggr] H(0;x)  \nn\\
\hspace*{-5mm} & &    
       -  \biggl[
             \frac{3}{4(1 \! - \! x)} \! 
          -  \! \frac{3}{2(1 \! + \! x)^3} \! 
          +  \! \frac{9}{4(1 \! + \! x)^2} \! 
          -  \! \frac{3}{2(1 \! + \! x)} \! 
          \biggr] \zeta(2) H(-1;x)  \nn\\
\hspace*{-5mm} & &           
       -  \frac{1}{4} \biggl[
            1
          - \frac{1}{(1-x)}
          - \frac{1}{(1+x)}
          \biggr] \zeta(2) H(1;x) \nn\\
\hspace*{-5mm} & &         
       +  \biggl[
            \frac{13}{4} \! 
          + \frac{15}{16(1-x)^2} \! 
          - \frac{57}{16(1-x)} \! 
          + \frac{21}{4(1+x)^4} \! 
          - \frac{8}{(1+x)^3} \nn\\
\hspace*{-5mm} & &     \hspace*{5mm}        
          + \frac{45}{8(1+x)^2}
          - \frac{25}{8(1+x)}
          \biggr] H(0,0;x)  \nn\\
\hspace*{-5mm} & &         
       -  \biggl[
            \frac{13}{8} \! 
          -  \! \frac{15}{4(1-x)}  \! 
          +  \! \frac{5}{(1+x)^3} \! 
          -  \! \frac{15}{2(1+x)^2} \! 
          +  \! \frac{3}{(1+x)} \! 
          \biggr] H(-1,0;x)  \nn\\
\hspace*{-5mm} & &         
       +  \biggl[
            \frac{5}{4} 
          - \frac{3}{2(1-x)}
          - \frac{1}{(1+x)}
          \biggr] H(1,0;x)  \nn\\
\hspace*{-5mm} & &         
       -  \biggl[
            \frac{9}{8}
          + \frac{37}{64(1-x)} 
          - \frac{21}{4(1+x)^5}
          + \frac{105}{8(1+x)^4}
          - \frac{239}{16(1+x)^3} \nn\\
\hspace*{-5mm} & &     \hspace*{5mm}     
          + \frac{297}{32(1+x)^2} 
          - \frac{323}{64(1+x)}
          \biggr] H(0,0,0;x)  \nn\\
\hspace*{-5mm} & &         
       +  \biggl[
            3
          - \frac{9}{2(1-x)}
          + \frac{3}{(1+x)^3} 
          - \frac{9}{2(1+x)^2}
          \biggr] H(-1,-1,0;x)  \nn\\
\hspace*{-5mm} & &         
       -   \biggl[
             \frac{9}{4} \! 
          -  \! \frac{3}{(1 \!- \!x)} \! 
          +  \! \frac{3}{2(1 \!+ \!x)^3}  \! 
   -  \!\frac{9}{4(1 \!+ \!x)^2} \!
          -  \!\frac{3}{4(1 \!+ \!x)} \!
          \biggr] H(-1,0,0;x) \nn\\
\hspace*{-5mm} & &         
       -  \biggl[
            \frac{9}{4} \!
          -  \! \frac{3}{(1 \! - \! x)} \!
          +  \! \frac{3}{2(1 \! + \! x)^3}  \!
          -  \! \frac{9}{4(1 \! + \! x)^2} \!
          -  \! \frac{3}{4(1 \! + \! x)} \!
          \biggr] H(0,-1,0;x)\nn\\
\hspace*{-5mm} & &            
       - \frac{1}{4}  \biggl[
            1
          - \frac{1}{(1-x)}
   - \frac{1}{(1+x)}
          \biggr] \bigl[   \zeta(2) H(1;x)   
   + 6 H(-1,1,0;x) \nn\\
\hspace*{-5mm} & &    \hspace*{10mm}     
   - 4 H(0,1,0;x)  
          + 6 H(1,-1,0;x) 
   - 4 H(1,0,0;x)  \nn\\
\hspace*{-5mm} & &    \hspace*{10mm}   
   - 4 H(1,1,0;x) \bigr]\nn\\
\hspace*{-5mm} & + & 
{\mathcal O} (D-4) \, , \\
\hspace*{-5mm} {\mathcal F}^{(2l,{\tt e})}_{2}(D,q^2) & = & \frac{1}{(D-4)^{2}} 
\Biggl\{ 
            \frac{3}{(1+x)} \biggl[ 
     1
   - \frac{1}{(1+x)} \biggr]
       + \frac{3}{4} \biggl[
            \frac{1}{(1-x)}
          - \frac{4}{(1+x)^3} \nn\\
\hspace*{-5mm} & &    \hspace*{20mm}             
          + \frac{6}{(1+x)^2}
          - \frac{3}{(1+x)}
          \biggr] H(0;x) 
\Biggr\}   \nn\\
\hspace*{-5mm} & + & \frac{1}{(D-4)} 
\Biggl\{        
          - \frac{5}{(1+x)} \biggl[ 
     1
   - \frac{1}{(1+x)} \biggr] \nn\\
\hspace*{-5mm} & &     \hspace*{18mm}  
       -  \biggl[
            \frac{3}{4(1-x)} \! 
          -  \! \frac{8}{(1 \! + \! x)^3} \! 
          +  \! \frac{12}{(1 \! + \! x)^2}  \! 
          -  \! \frac{19}{4(1 \! + \! x)} \! 
          \biggr] H(0;x) \nn\\
\hspace*{-5mm} & &     \hspace*{18mm}
       + \frac{3}{8} \biggl[
            \frac{1}{(1-x)} \! 
          -  \! \frac{4}{(1+x)^3}   \!          
          +  \! \frac{6}{(1+x)^2} \! 
          -  \! \frac{3}{(1+x)} \! 
          \biggr] \bigl[ \zeta(2) \nn\\
\hspace*{-5mm} & &     \hspace*{25mm}
         -  H(0,0;x) 
  + 2 H(-1,0;x) \bigr]
\Biggr\}   \nn\\
\hspace*{-5mm} & & 
       - \biggl[ \frac{7}{16(1-x)^2}
          - \frac{1}{16(1-x)}
          + \frac{21}{4(1+x)^4}
          - \frac{29}{2(1+x)^3} \nn\\
\hspace*{-5mm} & &           
          + \frac{239}{16(1+x)^2}
          - \frac{97}{16(1+x)} \biggr]  \zeta(2)
          - \frac{51}{8(1+x)^2}
          + \frac{51}{8(1+x)} \nn\\
\hspace*{-5mm} & &           
       -  \biggl[ \biggl(
            \frac{7}{16(1-x)^3} 
          - \frac{21}{32(1-x)^2}
          - \frac{25}{64(1-x)}
          + \frac{21}{4(1+x)^5} \nn\\
\hspace*{-5mm} & &   \hspace*{5mm}         
          - \frac{105}{8(1+x)^4}
          + \frac{13}{(1+x)^3}
          - \frac{51}{8(1+x)^2}
          + \frac{119}{64(1+x)} \biggl)  \zeta(2) \nn\\
\hspace*{-5mm} & &   \hspace*{5mm} 
          -  \! \frac{53}{32(1-x)}   \! 
          +  \! \frac{105}{8(1+x)^3} \! 
          -  \! \frac{315}{16(1+x)^2} \! 
          +  \! \frac{263}{32(1+x)} \! 
          \biggr] H(0;x) \nn\\
\hspace*{-5mm} & &        
       +  \biggl[
            \frac{19}{16(1-x)^2}
          - \frac{25}{16(1-x)} 
          - \frac{21}{4(1+x)^4}
          + \frac{13}{2(1+x)^3} \nn\\
\hspace*{-5mm} & &   \hspace*{5mm} 
          - \frac{47}{16(1+x)^2}
          + \frac{33}{16(1+x)}
          \biggr] H(0,0;x) \nn\\
\hspace*{-5mm} & &        
          - \frac{1}{2} \biggl[
            \frac{1}{(1-x)}
          - \frac{1}{(1+x)}
          \biggr] H(1,0;x)  \nn\\
\hspace*{-5mm} & &        
       +  \!  \biggl[
            \frac{1}{4(1-x)}
          + \frac{8}{(1+x)^3} 
          - \frac{12}{(1+x)^2}
          + \frac{15}{4(1+x)}
          \biggr] H(-1,0;x)  \nn\\
\hspace*{-5mm} & &  
       -  \biggl[
            \frac{7}{16(1 \! - \! x)^3} \! 
          -  \! \frac{21}{32(1 \! - \! x)^2}  \! 
          -  \! \frac{49}{64(1 \! - \! x)} \! 
          +  \! \frac{21}{4(1 \! + \! x)^5} \! 
          -  \! \frac{105}{8(1 \! + \! x)^4} \nn\\
\hspace*{-5mm} & &   \hspace*{5mm} 
          + \frac{29}{2(1+x)^3}
          - \frac{69}{8(1+x)^2}
          + \frac{191}{64(1+x)}
          \biggr] H(0,0,0;x)  \nn\\
\hspace*{-5mm} & &  
       - \frac{3}{8} \biggl[
            \frac{1}{(1-x)} \! 
          -  \! \frac{12}{(1+x)^3} \! 
          +  \! \frac{6}{(1+x)^2}  \! 
          -  \! \frac{3}{(1+x)} \! 
          \biggr] \bigl[ \zeta(3) \! 
   - H(-1;x) \nn\\
\hspace*{-5mm} & &   \hspace*{10mm} 
   - 2 H(-1,-1,0;x) 
   + H(-1,0,0;x) 
   + H(0,-1,0;x) \bigr] \nn\\
\hspace*{-5mm} & + &  
{\mathcal O} (D-4) \, , \\
\hspace*{-5mm} {\mathcal F}^{(2l,{\tt e})}_{3}(D,q^2) & = & \frac{1}{(D-4)^{2}} 
\Biggl\{ 
          -  \frac{3}{2 (1-x)} \biggl[ 
     1
   -  \! \frac{1}{(1-x)} \biggr] \! 
       +  \! \frac{3}{2} \biggl[
            \frac{1}{(1-x)^3} \! 
          -  \! \frac{3}{2 (1-x)^2} \nn\\
\hspace*{-5mm} & &        \hspace*{20mm}     
          + \frac{3}{4 (1-x)}
          - \frac{1}{4 (1+x)}
          \biggr] H(0;x) 
\Biggr\}   \nn\\
\hspace*{-5mm} & + & \frac{1}{(D-4)} 
\Biggl\{          
          -  \frac{1}{2 (1-x)} \biggl[ 
     1
   - \frac{1}{(1-x)} \biggr]
       + \frac{3}{4} \biggl[  
            \frac{1}{(1-x)^3} \! 
          - \frac{3}{2(1-x)^2} \nn\\
\hspace*{-5mm} & &        \hspace*{18mm}   
          +  \! \frac{3}{4(1-x)} \! 
          - \frac{1}{4(1+x)} \biggr] \biggl[ \zeta(2) \! 
   +  \! \frac{2}{3} H(0;x) \! 
   - H(0,0;x) \nn\\
\hspace*{-5mm} & &        \hspace*{18mm}  
   + 2 H(-1,0;x) \biggr]
\Biggr\}   \nn\\
\hspace*{-5mm} & & 
       + \biggl[ \frac{1}{4 (1-x)^3}
          + \frac{9}{16 (1-x)^2}
          - \frac{9}{8 (1-x)}
          - \frac{7}{16 (1+x)^2} \nn\\
\hspace*{-5mm} & & 
          + \frac{3}{4 (1+x)} \biggr] \zeta(2)
          + \frac{19}{16 (1-x)} \biggl[ 
     1
   - \frac{1}{(1-x)} \biggr]\nn\\
\hspace*{-5mm} & & 
       +  \biggl[ \biggl(
            \frac{9}{16 (1-x)^3}
          - \frac{27}{32 (1-x)^2}
          + \frac{19}{32 (1-x)}
          - \frac{7}{16 (1+x)^3} \nn\\
\hspace*{-5mm} & &      \hspace*{5mm} 
          + \frac{21}{32 (1+x)^2}  
          - \frac{17}{32 (1+x)} \biggr) \zeta(2)
          - \frac{13}{8 (1-x)^3}
          + \frac{39}{16 (1-x)^2} \nn\\
\hspace*{-5mm} & &      \hspace*{5mm} 
          + \frac{21}{32 (1+x)^2} 
          + \frac{9}{32 (1-x)}
          - \frac{35}{32 (1+x)}
          \biggr] H(0;x) \nn\\
\hspace*{-5mm} & &         
       -  \biggl[
            \frac{7}{8 (1-x)^4}
          - \frac{3}{2 (1-x)^3}
          + \frac{3}{8 (1-x)^2}
          - \frac{1}{16 (1-x)} \nn\\
\hspace*{-5mm} & &      \hspace*{5mm} 
          + \frac{7}{16 (1+x)^2}
          - \frac{1}{8 (1+x)}
          \biggr] H(0,0;x) \nn\\
\hspace*{-5mm} & &        
      + \frac{1}{2}  \biggl[
            \frac{1}{(1-x)^3} \! 
          -  \! \frac{3}{2(1-x)^2} \! 
          -  \! \frac{3}{4(1-x)} \! 
          +  \! \frac{5}{4(1+x)} \! 
          \biggr] H(-1,0;x) \nn\\
\hspace*{-5mm} & &        
      +  \biggl[
            \frac{21}{16 (1-x)^3}
          - \frac{63}{32 (1-x)^2}
          + \frac{37}{32 (1-x)}
          - \frac{7}{16 (1+x)^3} \nn\\
\hspace*{-5mm} & &       \hspace*{5mm}     
      + \frac{21}{32 (1+x)^2}
          - \frac{23}{32 (1+x)}
          \biggr] H(0,0,0;x) \nn\\
\hspace*{-5mm} & &   
      - \frac{3}{4}  \biggl[
            \frac{1}{(1-x)^3} \! 
          -  \! \frac{3}{2(1-x)^2} \! 
          +  \! \frac{3}{4(1-x)} \! 
          -  \! \frac{1}{4(1+x)} \! 
          \biggr] \bigr[ \zeta(3) \nn\\
\hspace*{-5mm} & &       \hspace*{5mm}    
   - \zeta(2) H(-1;x)
   - 2 H(-1,-1,0;x) 
   + H(-1,0,0;x) \nn\\
\hspace*{-5mm} & &       \hspace*{5mm}    
   + H(0,-1,0;x) \bigr] \nn\\
\hspace*{-5mm} & + &  {\mathcal O} (D-4) \, .
\eea

$\bullet$ The {\it Down-Self-Mass-insertion} graph {\tt(f)} of 
Fig. \ref{fig1}, defined as 
\bea
\parbox{20mm}{\begin{fmfgraph*}(15,15)
\fmfleft{i1,i2}
\fmfright{o}
\fmfforce{0.2w,0.93h}{v1}
\fmfforce{0.2w,0.07h}{v2}
\fmfforce{0.8w,0.5h}{v5}
\fmf{plain}{i1,v2}
\fmf{plain}{i2,v1}
\fmf{photon}{v5,o}
\fmf{plain}{v2,v3}
\fmf{photon,tension=.25,left}{v3,v4}
\fmf{plain,tension=.25}{v3,v4}
\fmf{plain}{v4,v5}
\fmf{plain}{v1,v5}
\fmf{photon}{v1,v2}
\end{fmfgraph*} }  & = & 
\int {\mathfrak{D}}^Dk_1\;{\mathfrak{D}}^Dk_2\ 
\frac{{\mathcal N}_{({\tt f})}^{\mu}}{{\mathcal D}_{1} {\mathcal D}_{2} 
{\mathcal D}_{9} {\mathcal D}^{2}_{10} {\mathcal D}_{13}} \, , 
\label{b6} 
\eea
where 
\bea
\! \! \! \! \! \! {\mathcal N}_{({\tt f})}^{\mu} & = & \bar{v}(p_2) \gamma_{\sigma}
[i ( \not{\! p_{2}} +  \!  \! \not{\! k_{1}}) \!  + \! m] 
\gamma_{\lambda} 
[i ( \not{\! p_{2}} + \!  \! \not{\! k_{1}} - \!  \! \not{\! k_{2}}) \! 
 + \! m] 
\gamma^{\lambda} 
[i ( \not{\! p_{2}} +   \! \! \not{\! k_{1}})  \! + \! m] 
\gamma_{\mu} \times
\nn\\
\! \! \! \! \! \! & & \times
[-i ( \not{\! p_{1}} -  \! \! \not{\! k_{1}})  \! + \! m] 
\gamma^{\sigma} u(p_1) ,
\eea
due to reasons of symmetry, gives: 
\bea
{\mathcal F}^{(2l,{\tt f})}_{1}(D,q^2) & = & 
{\mathcal F}^{(2l,{\tt e})}_{1}(D,q^2) \, ,
\\
{\mathcal F}^{(2l,{\tt f})}_{2}(D,q^2) & = & 
{\mathcal F}^{(2l,{\tt e})}_{2}(D,q^2) \, ,
\\
{\mathcal F}^{(2l,{\tt f})}_{3}(D,q^2) & = & 
- {\mathcal F}^{(2l,{\tt e})}_{3}(D,q^2) \, .
\eea

$\bullet$ The {\it Vacuum-Polarization-insertion} graph {\tt(g)} 
of Fig. \ref{fig1}, defined as 
\bea
\parbox{20mm}{\begin{fmfgraph*}(15,15)
\fmfleft{i1,i2}
\fmfright{o}
\fmfforce{0.2w,0.93h}{v2}
\fmfforce{0.2w,0.07h}{v1}
\fmfforce{0.2w,0.3h}{v3}
\fmfforce{0.2w,0.7h}{v4}
\fmfforce{0.8w,0.5h}{v5}
\fmf{plain}{i1,v1}
\fmf{plain}{i2,v2}
\fmf{photon}{v5,o}
\fmf{plain}{v2,v5}
\fmf{photon}{v1,v3}
\fmf{photon}{v2,v4}
\fmf{plain}{v1,v5}
\fmf{plain,left}{v3,v4}
\fmf{plain,right}{v3,v4}
\end{fmfgraph*} }  & = & 
\int {\mathfrak{D}}^Dk_1\;{\mathfrak{D}}^Dk_2\ 
\frac{{\mathcal N}_{({\tt g})}^{\mu}}{{\mathcal D}^{2}_{1} {\mathcal D}_{7} 
{\mathcal D}_{8} {\mathcal D}_{9} {\mathcal D}_{10}} \, , 
\label{b7} 
\eea
where 
\bea
{\mathcal N}_{({\tt g})}^{\mu} & = & - \bar{v}(p_2) \gamma_{\sigma} 
[i ( \not{\! p_{2}} +  \!  \! \not{\! k_{1}})  \! + \! m] 
\gamma_{\mu} 
[-i ( \not{\! p_{1}} - \!  \! \not{\! k_{1}})  \! + \! m] 
\gamma_{\lambda} 
\gamma_{\lambda} 
[i ( \not{\! k_{1}} + \!  \! \not{\! k_{2}})  \! + \! m] \times
\nn\\
& & \times
\gamma^{\sigma} 
[i ( \not{\! k_{2}})  \! + \! m] u(p_1),
\eea
gives: 
\bea
\hspace*{-5mm} {\mathcal F}^{(2l,{\tt g})}_{1}(D,q^2) & = &\frac{1}{(D-4)^{2}} 
\Biggl\{ 
          - \frac{1}{3}
       - \frac{2}{3} \biggl[
            1
          - \frac{1}{(1-x)}
          - \frac{1}{(1+x)}
          \biggr] H(0;x)  
\Biggr\}   \nn\\
\hspace*{-5mm} & + & 
\frac{1}{(D-4)} \Biggl\{ 
            \frac{1}{8}
   + \frac{1}{6} \biggl[
     1
   - \frac{2}{(1-x)} \biggr] H(0;x) \nn\\
\hspace*{-5mm} & &      \hspace*{18mm}
       - \frac{1}{3} \biggl[
            1
          - \frac{1}{(1-x)}
          - \frac{1}{(1+x)}
          \biggr] \bigl[ \zeta(2)
   - H(0;x) \nn\\
\hspace*{-5mm} & &      \hspace*{26mm}
   - H(0,0;x)
   + 2 H(-1,0;x) \bigr]
\Biggr\}   \nn\\
\hspace*{-5mm} & & 
          + \frac{223}{864}
      + \biggl[ 2 \! 
          - \frac{1}{3(1-x)} \! 
          +  \! \frac{98}{3(1+x)^4} \! 
          - \frac{196}{3(1+x)^3} \! 
          +  \! \frac{229}{6(1+x)^2} \nn\\
\hspace*{-5mm} & &           
          - \frac{17}{3(1+x)} \biggr]  \zeta(2)
          - \frac{49}{9(1+x)} \biggl[ 
     1
   - \frac{1}{(1+x)}  \biggr]\nn\\
\hspace*{-5mm} & &           
       -  \biggl[
            \frac{409}{216} \! 
       - \biggl( \frac{1}{4(1-x)} \! 
          +  \! \frac{6}{(1 \! + \! x)^5} \! 
          -  \! \frac{15}{(1 \! + \! x)^4} \! 
          +  \! \frac{11}{(1 \! + \! x)^3} \! 
          -  \! \frac{3}{2(1 \! + \! x)^2} \nn\\
\hspace*{-5mm} & &     \hspace*{5mm}
          - \frac{1}{4(1+x)} \biggr)  \zeta(2)
          - \frac{77}{27(1-x)}
          - \frac{89}{9(1+x)^3}
          + \frac{89}{6(1+x)^2} \nn\\
\hspace*{-5mm} & &     \hspace*{5mm}       
          -  \! \frac{635}{108(1+x)}
          \biggr] H(0;x)  \nn\\
\hspace*{-5mm} & &   
       +  \biggl[
            \frac{5}{18}
          + \frac{1}{3(1-x)} 
          + \frac{62}{9(1+x)^4}
          - \frac{124}{9(1+x)^3}
          + \frac{163}{18(1+x)^2} \nn\\
\hspace*{-5mm} & &     \hspace*{5mm}
          - \frac{2}{(1+x)}
          \biggr] H(0,0;x) \nn\\
\hspace*{-5mm} & &        
       + \frac{1}{6} \biggl[
            1
          - \frac{2}{(1-x)}
          \biggr] H(-1,0;x)  \nn\\
\hspace*{-5mm} & &        
       -  \biggl[
            \frac{1}{3}
          - \frac{7}{12(1-x)}
          + \frac{6}{(1+x)^5}
          - \frac{15}{(1+x)^4}
          + \frac{11}{(1+x)^3}  \nn\\
\hspace*{-5mm} & &   \hspace*{5mm} 
          - \frac{3}{2(1+x)^2}       
   - \frac{7}{12(1+x)}
          \biggr] H(0,0,0;x) \nn\\
\hspace*{-5mm} & & 
       + \frac{1}{3} \biggl[
            1 \! 
          - \frac{1}{(1-x)} \! 
          - \frac{1}{(1+x)} \! 
          \biggr]  \bigl[
     \zeta(3) \! 
   -  \! \zeta(2) H(-1;x)  \! 
   + \!  H(-1,0;x) \nn\\
\hspace*{-5mm} & &   \hspace*{10mm} 
   - 2 H(-1,-1,0;x) 
   + H(-1,0,0;x)
   + H(0,-1,0;x) \bigr] \nn\\
\hspace*{-5mm} & + &  {\mathcal O} (D-4) \, , 
\label{ex1} \\
\hspace*{-5mm} {\mathcal F}^{(2l,{\tt g})}_{2}(D,q^2) & = & \frac{1}{(D-4)} \Biggl\{ 
       \frac{1}{3} \biggl[
            \frac{1}{(1+x)}
          - \frac{1}{(1-x)}
          \biggr] H(0;x) 
\Biggr\}   \nn\\
\hspace*{-5mm} & & 
       - \biggl[ \frac{1}{6(1-x)} \! 
          -  \! \frac{34}{(1+x)^4} \! 
          +  \! \frac{68}{(1+x)^3} \! 
          -  \! \frac{33}{(1+x)^2} \! 
          -  \! \frac{5}{6(1+x)} \biggr] \zeta(2) \nn\\
\hspace*{-5mm} & &  
          + \frac{17}{3(1+x)} \biggl[ 
     1
   - \frac{1}{(1+x)} \biggr] \nn\\
\hspace*{-5mm} & &           
       -  \biggl[ \biggl(
            \frac{3}{8(1-x)} \! 
          - \frac{6}{(1+x)^5} \! 
          +  \! \frac{15}{(1+x)^4} \! 
          - \frac{21}{2(1+x)^3} \! 
          +  \! \frac{3}{4(1+x)^2} \nn\\
\hspace*{-5mm} & &    \hspace*{5mm}       
          + \frac{3}{8(1+x)} \biggr)  \zeta(2)
          + \frac{11}{9(1-x)}
          + \frac{31}{3(1+x)^3}
          - \frac{31}{2(1+x)^2} \nn\\
\hspace*{-5mm} & &    \hspace*{5mm}     
          + \frac{71}{18(1+x)}
          \biggr] H(0;x) \nn\\
\hspace*{-5mm} & &           
       +  \biggl[
            \frac{1}{6(1-x)}
          - \frac{22}{3(1+x)^4}
          + \frac{44}{3(1+x)^3}
          - \frac{23}{3(1+x)^2} \nn\\
\hspace*{-5mm} & &    \hspace*{5mm} 
          + \frac{1}{6(1+x)}
          \biggr] H(0,0;x)  \nn\\
\hspace*{-5mm} & &       
       + \frac{1}{3}  \biggl[
            \frac{1}{(1+x)}
          - \frac{1}{(1-x)}
          \biggr] H(-1,0;x)  \nn\\
\hspace*{-5mm} & &       
       -  \biggl[
            \frac{3}{8(1-x)}
          - \frac{6}{(1+x)^5}
          + \frac{15}{(1+x)^4}
          - \frac{21}{2(1+x)^3}
          + \frac{3}{4(1+x)^2} \nn\\
\hspace*{-5mm} & &    \hspace*{5mm}    
          + \frac{3}{8(1+x)}
          \biggr] H(0,0,0;x)  \nn\\
\hspace*{-5mm} & + & {\mathcal O} (D-4) \, , 
\label{ex2} \\
\hspace*{-5mm} {\mathcal F}^{(2l,{\tt g})}_{3}(D,q^2) & = & 0 \, .
\label{ex3}
\eea

Note the structure of the contribution to the third {\it 
unrenormalized} form factor ${\mathcal F}^{(2l)}_{3}(D,x)$ of the seven 
diagrams given above: the contributions of the {\it Ladder}, {\it Cross} 
and {\it Vacuum-Polarization-insertion} vertices, 
the diagrams (a), (b) and (g) of Fig. \ref{fig1}, vanish 
separately, while those of the 
{\it Corner} and {\it Self-energy-insertion} diagrams, 
Fig. \ref{fig1} (c), (d), (e) and (f), cancel pairwise. 
That leads to the vanishing of the third form factor 
\be
{\mathcal F}^{(2l)}_{3}(D,q^2) = 
\sum_{{\tt graph}} {\mathcal F}^{(2l,{\tt graph})}_{3}(D,x) = 0 \ ,
\label{EFFE3}
\ee
as expected (and already repeatedly anticipated) 
from the conservation of the electromagnetic current.

\section{Renormalization subtractions \label{renormalization}}

\bfig
\bc
\subfigure[]{
\begin{fmfgraph*}(25,25) 
\fmfleft{i1,i2}
\fmfright{o}
\fmfforce{0.2w,0.93h}{v2}
\fmfforce{0.2w,0.07h}{v1}
\fmfforce{0.8w,0.5h}{v5}
\fmfforce{0.64w,0.5h}{v55}
\fmf{plain}{i1,v1}
\fmf{plain}{i2,v2}
\fmfv{label=$\otimes$}{v55}
\fmfv{l=$Z_{1}^{(1l)}$,l.a=75,l.d=.15w}{v5}
\fmf{photon}{v5,o}
\fmf{plain}{v2,v5}
\fmf{photon}{v1,v2}
\fmf{plain}{v1,v5}
\end{fmfgraph*} }
%
%
\subfigure[]{
\begin{fmfgraph*}(25,25) 
\fmfleft{i1,i2}
\fmfright{o}
\fmfforce{0.2w,0.93h}{v2}
\fmfforce{0.31w,0.79h}{v22}
\fmfforce{0.2w,0.07h}{v1}
\fmfforce{0.8w,0.5h}{v5}
\fmf{plain}{i1,v1}
\fmf{plain}{i2,v2}
\fmfv{label=$\otimes$}{v22}
\fmfv{l=$Z_{1}^{(1l)}$,l.a=60,l.d=.1w}{v2}
\fmf{photon}{v5,o}
\fmf{plain}{v2,v5}
\fmf{photon}{v1,v2}
\fmf{plain}{v1,v5}
\end{fmfgraph*} }
%
%
\subfigure[]{
\begin{fmfgraph*}(25,25) 
\fmfleft{i1,i2}
\fmfright{o}
\fmfforce{0.2w,0.93h}{v2}
\fmfforce{0.31w,0.21h}{v22}
\fmfforce{0.2w,0.07h}{v1}
\fmfforce{0.8w,0.5h}{v5}
\fmf{plain}{i1,v1}
\fmf{plain}{i2,v2}
\fmfv{label=$\otimes$}{v22}
\fmfv{l=$Z_{1}^{(1l)}$,l.a=-50,l.d=.08w}{v1}
\fmf{photon}{v5,o}
\fmf{plain}{v2,v5}
\fmf{photon}{v1,v2}
\fmf{plain}{v1,v5}
\end{fmfgraph*} }
%
%
\subfigure[]{
\begin{fmfgraph*}(25,25) 
\fmfleft{i1,i2}
\fmfright{o}
\fmfforce{0.2w,0.93h}{v2}
\fmfforce{0.2w,0.07h}{v1}
\fmfforce{0.5w,0.7h}{v3}
\fmfforce{0.52w,0.55h}{v33}
\fmfforce{0.8w,0.5h}{v5}
\fmf{plain}{i1,v1}
\fmf{plain}{i2,v2}
\fmfv{label=$\otimes$}{v33}
\fmfv{l=$\delta m^{(1l)}$,l.a=60,l.d=.15w}{v3}
\fmf{photon}{v5,o}
\fmf{plain}{v2,v5}
\fmf{photon}{v1,v2}
\fmf{plain}{v1,v5}
\end{fmfgraph*} } \\
%
%
\subfigure[]{
\begin{fmfgraph*}(25,25) 
\fmfleft{i1,i2}
\fmfright{o}
\fmfforce{0.2w,0.93h}{v2}
\fmfforce{0.2w,0.07h}{v1}
\fmfforce{0.5w,0.3h}{v3}
\fmfforce{0.52w,0.45h}{v33}
\fmfforce{0.8w,0.5h}{v5}
\fmf{plain}{i1,v1}
\fmf{plain}{i2,v2}
\fmfv{label=$\otimes$}{v33}
\fmfv{l=$\delta m^{(1l)}$,l.a=-60,l.d=.1w}{v3}
\fmf{photon}{v5,o}
\fmf{plain}{v2,v5}
\fmf{photon}{v1,v2}
\fmf{plain}{v1,v5}
\end{fmfgraph*} }
%
%
\subfigure[]{
\begin{fmfgraph*}(25,25) 
\fmfleft{i1,i2}
\fmfright{o}
\fmfforce{0.2w,0.93h}{v2}
\fmfforce{0.2w,0.07h}{v1}
\fmfforce{0.5w,0.7h}{v3}
\fmfforce{0.52w,0.55h}{v33}
\fmfforce{0.8w,0.5h}{v5}
\fmf{plain}{i1,v1}
\fmf{plain}{i2,v2}
\fmfv{label=$\otimes$}{v33}
\fmfv{l=$Z_{2}^{(1l)}$,l.a=50,l.d=.15w}{v3}
\fmf{photon}{v5,o}
\fmf{plain}{v2,v5}
\fmf{photon}{v1,v2}
\fmf{plain}{v1,v5}
\end{fmfgraph*} }
%
%
\subfigure[]{
\begin{fmfgraph*}(25,25) 
\fmfleft{i1,i2}
\fmfright{o}
\fmfforce{0.2w,0.93h}{v2}
\fmfforce{0.2w,0.07h}{v1}
\fmfforce{0.5w,0.3h}{v3}
\fmfforce{0.52w,0.45h}{v33}
\fmfforce{0.8w,0.5h}{v5}
\fmf{plain}{i1,v1}
\fmf{plain}{i2,v2}
\fmfv{label=$\otimes$}{v33}
\fmfv{l=$Z_{2}^{(1l)}$,l.a=-50,l.d=.12w}{v3}
\fmf{photon}{v5,o}
\fmf{plain}{v2,v5}
\fmf{photon}{v1,v2}
\fmf{plain}{v1,v5}
\end{fmfgraph*} }
%
%
\hspace{5mm}
\subfigure[]{
\begin{fmfgraph*}(25,25) 
\fmfleft{i1,i2}
\fmfright{o}
\fmfforce{0.2w,0.93h}{v2}
\fmfforce{0.2w,0.07h}{v1}
\fmfforce{0.2w,0.5h}{v3}
\fmfforce{0.35w,0.5h}{v33}
\fmfforce{0.8w,0.5h}{v5}
\fmf{plain}{i1,v1}
\fmf{plain}{i2,v2}
\fmfv{label=$\otimes$}{v33}
\fmfv{l=$Z_{3}^{(1l)}$,l.a=150,l.d=.1w}{v3}
\fmf{photon}{v5,o}
\fmf{plain}{v2,v5}
\fmf{photon}{v1,v2}
\fmf{plain}{v1,v5}
\end{fmfgraph*} } \\
%
%
\subfigure[]{
\begin{fmfgraph*}(25,25) 
\fmfleft{i1,i2}
\fmfright{o}
\fmfforce{0.2w,0.93h}{v2}
\fmfforce{0.2w,0.07h}{v1}
\fmfforce{0.8w,0.5h}{v5}
\fmfforce{0.64w,0.5h}{v55}
\fmf{plain}{i1,v1}
\fmf{plain}{i2,v2}
\fmfv{label=$\otimes$}{v55}
\fmfv{l=$Z_{1}^{(2l)}$,l.a=75,l.d=.15w}{v5}
\fmf{photon}{v5,o}
\fmf{plain}{v2,v5}
\fmf{phantom}{v1,v2}
\fmf{plain}{v1,v5}
\end{fmfgraph*} }
%
%
\vspace*{8mm}
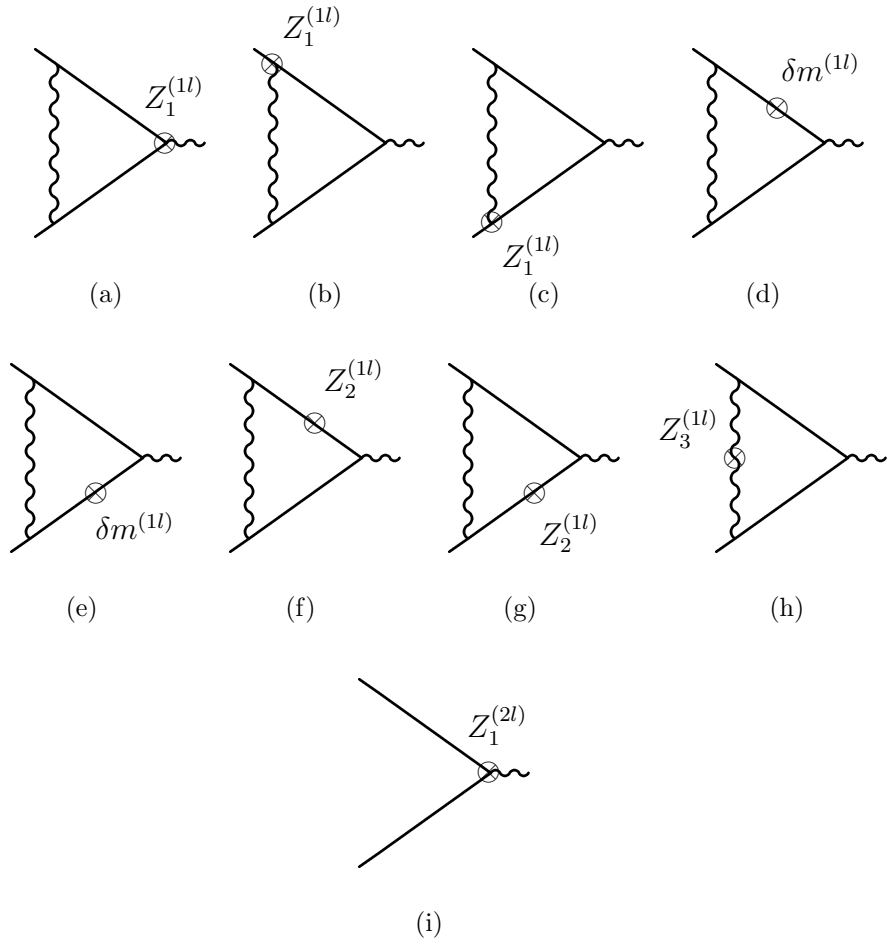
\caption{\label{fig2} Subtraction terms for the renormalization at 
2 loops. } 
\ec
\efig
As a next step we have to renormalize the 1-loop insertions in the 
above 2-loop graphs, when present (i.e. in all the graphs, with the 
exception of the {\it Cross} graph {\tt b} of Fig. \ref{fig1}). 
That will be done by subtracting from each graphs suitable 
contributions, which will be specified in the next subsection, 
proportional to the 1-loop renormalization constants 
$Z_1^{(1l)}(D)$ (charge), $\ Z_2^{(1l)}(D)$ (electron wave function; 
in QED one has $Z_2^{(1l)}(D) = - Z_1^{(1l)}(D)$, due to 
the Ward identity), $\ Z_3^{(1l)}(D)$ (photon wave function) and 
$\delta m^{(1l)}(D,m)$ (electron mass). 

Those subtractions are sufficient to obtain the 2-loop renormalized form 
factor $F_2^{(2l)}(D,q^2)$. To obtain also the renormalized value 
of $F_1^{(2l)}(D,q^2)$, we have to subtract from the unrenormalized 
charge form factor (obtained after the subtractions due to the renormalization 
of the 1-loop insertions) its value at $q^2=0$, 
$Z_{1}^{(2l)}(D) = {\mathcal{F}}_1^{(2l)}(D,q^2=0)$.

\subsection{1-loop renormalization constants times 1-loop subdiagrams
\label{1Lcnt}}

We give here the values of the subtractions to the 2-loop graphs due 
to the renormalization of the 1-loop insertions shown in 
Fig. \ref{fig2} {\tt(a-h)}. 

The 1-loop renormalization constants, $Z_{1}^{(1l)}(D)$, $\ Z_{2}^{(1l)}(D)$, 
$\ Z_{3}^{(1l)}(D)$ and $\delta m^{(1l)}(D,m)$, calculated according to the 
{\it On-Shell} renormalization prescription, have the following 
expressions exact in $D$: 
\bea
Z_{1}^{(1l)}(D) & = & 
- \frac{1}{2} \frac{(D-1)}{(D-3)(D-4)} 
\label{c001} 
 \ , \\
 & & \nonumber \\
Z_{2}^{(1l)}(D)  &=& - Z_{1}^{(1l)}(D) \qquad (\mbox{\rm Ward identity}) \ ,\\ 
 & & \nonumber \\
Z_{3}^{(1l)}(D) & = & 
\frac{2}{3(D-4)}  
\label{c002} 
 \ , \\
\delta m^{(1l)}(D,m) & = &  
m \ \frac{1}{2} \frac{(D-1)}{(D-3)(D-4)} \ ,  
\label{c003} 
\eea

The subtractions graphs of Fig. \ref{fig2} {\tt(a-h)} are defined by the
following relations:

$\bullet$ graph (a) in Fig. (\ref{fig2})

\be
\parbox{20mm}{\begin{fmfgraph*}(15,15) 
\fmfleft{i1,i2}
\fmfright{o}
\fmfforce{0.2w,0.93h}{v2}
\fmfforce{0.2w,0.07h}{v1}
\fmfforce{0.8w,0.5h}{v5}
\fmfforce{0.55w,0.5h}{v55}
\fmf{plain}{i1,v1}
\fmf{plain}{i2,v2}
\fmfv{label=$\otimes$}{v55}
\fmfv{l=$Z_{1}^{(1l)}$,l.a=55,l.d=.1w}{v5}
\fmf{photon}{v5,o}
\fmf{plain}{v2,v5}
\fmf{photon}{v1,v2}
\fmf{plain}{v1,v5}
\end{fmfgraph*} }  \stackrel{def}{=} 
 \ Z_{1}^{(1l)}(D) \ \times 
\parbox{20mm}{\begin{fmfgraph*}(15,15) 
\fmfleft{i1,i2}
\fmfright{o}
\fmfforce{0.2w,0.93h}{v2}
\fmfforce{0.2w,0.07h}{v1}
\fmfforce{0.8w,0.5h}{v5}
\fmf{plain}{i1,v1}
\fmf{plain}{i2,v2}
\fmf{photon}{v5,o}
\fmf{plain}{v2,v5}
\fmf{photon}{v1,v2}
\fmf{plain}{v1,v5}
\end{fmfgraph*} }  \, ;
\label{c1} 
\ee

$\bullet$ graphs (b) and (c) in Fig. (\ref{fig2})

\vspace*{3mm}

\be
\parbox{20mm}{\begin{fmfgraph*}(15,15) 
\fmfleft{i1,i2}
\fmfright{o}
\fmfforce{0.2w,0.93h}{v2}
\fmfforce{0.39w,0.71h}{v22}
\fmfforce{0.2w,0.07h}{v1}
\fmfforce{0.8w,0.5h}{v5}
\fmf{plain}{i1,v1}
\fmf{plain}{i2,v2}
\fmfv{label=$\otimes$}{v22}
\fmfv{l=$Z_{1}^{(1l)}$,l.a=40,l.d=.1w}{v2}
\fmf{photon}{v5,o}
\fmf{plain}{v2,v5}
\fmf{photon}{v1,v2}
\fmf{plain}{v1,v5}
\end{fmfgraph*} }  
 = 
\parbox{20mm}{\begin{fmfgraph*}(15,15) 
\fmfleft{i1,i2}
\fmfright{o}
\fmfforce{0.2w,0.93h}{v2}
\fmfforce{0.38w,0.28h}{v22}
\fmfforce{0.2w,0.07h}{v1}
\fmfforce{0.8w,0.5h}{v5}
\fmf{plain}{i1,v1}
\fmf{plain}{i2,v2}
\fmfv{label=$\otimes$}{v22}
\fmfv{l=$Z_{1}^{(1l)}$,l.a=-45,l.d=.08w}{v1}
\fmf{photon}{v5,o}
\fmf{plain}{v2,v5}
\fmf{photon}{v1,v2}
\fmf{plain}{v1,v5}
\end{fmfgraph*} }   \stackrel{def}{=} 
 \ Z_{1}^{(1l)}(D) \ \times 
\parbox{20mm}{\begin{fmfgraph*}(15,15) 
\fmfleft{i1,i2}
\fmfright{o}
\fmfforce{0.2w,0.93h}{v2}
\fmfforce{0.2w,0.07h}{v1}
\fmfforce{0.8w,0.5h}{v5}
\fmf{plain}{i1,v1}
\fmf{plain}{i2,v2}
\fmf{photon}{v5,o}
\fmf{plain}{v2,v5}
\fmf{photon}{v1,v2}
\fmf{plain}{v1,v5}
\end{fmfgraph*} }  \, ;
\label{c2} 
\ee

\vspace*{5mm}

equal to the graph (a) of Fig. (\ref{fig2}).

$\bullet$ graph (d) in Fig. (\ref{fig2})

\be
\parbox{20mm}{\begin{fmfgraph*}(15,15)
\fmfleft{i1,i2}
\fmfright{o}
\fmfforce{0.2w,0.93h}{v2}
\fmfforce{0.2w,0.07h}{v1}
\fmfforce{0.5w,0.7h}{v3}
\fmfforce{0.5w,0.51h}{v33}
\fmfforce{0.8w,0.5h}{v5}
\fmf{plain}{i1,v1}
\fmf{plain}{i2,v2}
\fmfv{label=$\otimes$}{v33}
\fmfv{l=$\delta m^{(1l)}$,l.a=60,l.d=.15w}{v3}
\fmf{photon}{v5,o}
\fmf{plain}{v2,v5}
\fmf{photon}{v1,v2}
\fmf{plain}{v1,v5}
\end{fmfgraph*} }   \stackrel{def}{=} 
\ \frac{\delta m^{(1l)}(D,m)}{m} \ \times \left( m 
\parbox{20mm}{\begin{fmfgraph*}(15,15)
\fmfleft{i1,i2}
\fmfright{o}
\fmfforce{0.2w,0.93h}{v2}
\fmfforce{0.2w,0.07h}{v1}
\fmfforce{0.5w,0.7h}{v3}
\fmfforce{0.5w,0.51h}{v33}
\fmfforce{0.8w,0.5h}{v5}
\fmf{plain}{i1,v1}
\fmf{plain}{i2,v2}
\fmfv{label=$\otimes$}{v33}
\fmf{photon}{v5,o}
\fmf{plain}{v2,v5}
\fmf{photon}{v1,v2}
\fmf{plain}{v1,v5}
\end{fmfgraph*} } \right) ;
\label{c3} 
\ee

$\bullet$ graph (e) in Fig. (\ref{fig2})

\be
\parbox{20mm}{\begin{fmfgraph*}(15,15)
\fmfleft{i1,i2}
\fmfright{o}
\fmfforce{0.2w,0.93h}{v2}
\fmfforce{0.2w,0.07h}{v1}
\fmfforce{0.5w,0.3h}{v3}
\fmfforce{0.5w,0.49h}{v33}
\fmfforce{0.8w,0.5h}{v5}
\fmf{plain}{i1,v1}
\fmf{plain}{i2,v2}
\fmfv{label=$\otimes$}{v33}
\fmfv{l=$\delta m^{(1l)}$,l.a=-60,l.d=.15w}{v3}
\fmf{photon}{v5,o}
\fmf{plain}{v2,v5}
\fmf{photon}{v1,v2}
\fmf{plain}{v1,v5}
\end{fmfgraph*} }   \stackrel{def}{=} 
\ \frac{\delta m^{(1l)}(D,m)}{m} \ \times \left( m 
\parbox{20mm}{\begin{fmfgraph*}(15,15)
\fmfleft{i1,i2}
\fmfright{o}
\fmfforce{0.2w,0.93h}{v2}
\fmfforce{0.2w,0.07h}{v1}
\fmfforce{0.5w,0.3h}{v3}
\fmfforce{0.5w,0.49h}{v33}
\fmfforce{0.8w,0.5h}{v5}
\fmf{plain}{i1,v1}
\fmf{plain}{i2,v2}
\fmfv{label=$\otimes$}{v33}
\fmf{photon}{v5,o}
\fmf{plain}{v2,v5}
\fmf{photon}{v1,v2}
\fmf{plain}{v1,v5}
\end{fmfgraph*} } \right) ;
\label{c4} 
\ee

\vspace*{5mm}

$\bullet$ graphs (f) and (g) in Fig. (\ref{fig2})

\vspace*{2mm}

\be
\parbox{20mm}{\begin{fmfgraph*}(15,15)
\fmfleft{i1,i2}
\fmfright{o}
\fmfforce{0.2w,0.93h}{v2}
\fmfforce{0.2w,0.07h}{v1}
\fmfforce{0.5w,0.7h}{v3}
\fmfforce{0.5w,0.51h}{v33}
\fmfforce{0.8w,0.5h}{v5}
\fmf{plain}{i1,v1}
\fmf{plain}{i2,v2}
\fmfv{label=$\otimes$}{v33}
\fmfv{l=$Z_{2}^{(1l)}$,l.a=60,l.d=.12w}{v3}
\fmf{photon}{v5,o}
\fmf{plain}{v2,v5}
\fmf{photon}{v1,v2}
\fmf{plain}{v1,v5}
\end{fmfgraph*} }  =
\parbox{20mm}{\begin{fmfgraph*}(15,15)
\fmfleft{i1,i2}
\fmfright{o}
\fmfforce{0.2w,0.93h}{v2}
\fmfforce{0.2w,0.07h}{v1}
\fmfforce{0.5w,0.3h}{v3}
\fmfforce{0.5w,0.49h}{v33}
\fmfforce{0.8w,0.5h}{v5}
\fmf{plain}{i1,v1}
\fmf{plain}{i2,v2}
\fmfv{label=$\otimes$}{v33}
\fmfv{l=$Z_{2}^{(1l)}$,l.a=-60,l.d=.12w}{v3}
\fmf{photon}{v5,o}
\fmf{plain}{v2,v5}
\fmf{photon}{v1,v2}
\fmf{plain}{v1,v5}
\end{fmfgraph*} } 
\stackrel{def}{=} 
\ Z_{2}^{(1l)}(D) \ \times 
\parbox{20mm}{\begin{fmfgraph*}(15,15) 
\fmfleft{i1,i2}
\fmfright{o}
\fmfforce{0.2w,0.93h}{v2}
\fmfforce{0.2w,0.07h}{v1}
\fmfforce{0.8w,0.5h}{v5}
\fmf{plain}{i1,v1}
\fmf{plain}{i2,v2}
\fmf{photon}{v5,o}
\fmf{plain}{v2,v5}
\fmf{photon}{v1,v2}
\fmf{plain}{v1,v5}
\end{fmfgraph*} }  \, ;
\label{c5} 
\ee

\vspace*{2mm}

$\bullet$ graph (h) in Fig. (\ref{fig2})

\be
\parbox{20mm}{\begin{fmfgraph*}(15,15)
\fmfleft{i1,i2}
\fmfright{o}
\fmfforce{0.2w,0.93h}{v2}
\fmfforce{0.2w,0.07h}{v1}
\fmfforce{0.2w,0.55h}{v3}
\fmfforce{0.42w,0.5h}{v33}
\fmfforce{0.8w,0.5h}{v5}
\fmf{plain}{i1,v1}
\fmf{plain}{i2,v2}
\fmfv{label=$\otimes$}{v33}
\fmfv{l=$Z_{3}^{(1l)}$,l.a=180,l.d=.15w}{v3}
\fmf{photon}{v5,o}
\fmf{plain}{v2,v5}
\fmf{photon}{v1,v2}
\fmf{plain}{v1,v5}
\end{fmfgraph*} }  
\stackrel{def}{=} 
\ Z_{3}^{(1l)}(D) \ \times 
\parbox{20mm}{\begin{fmfgraph*}(15,15)
\fmfleft{i1,i2}
\fmfright{o}
\fmfforce{0.2w,0.93h}{v2}
\fmfforce{0.2w,0.07h}{v1}
\fmfforce{0.2w,0.5h}{v3}
\fmfforce{0.42w,0.5h}{v33}
\fmfforce{0.8w,0.5h}{v5}
\fmf{plain}{i1,v1}
\fmf{plain}{i2,v2}
\fmf{photon}{v5,o}
\fmf{plain}{v2,v5}
\fmf{photon}{v1,v2}
\fmf{plain}{v1,v5}
\end{fmfgraph*} }  \, .
\label{c6} 
\ee

\vspace*{4mm}

From the {\it r.h.s} of Eqs. (\ref{c1}-\ref{c6}), it turns out that only
three 1-loop vertex subdiagrams appear in the calculation, two of them 
being in fact equal for symmetry reasons. Each of the subdiagrams can be 
written in terms of its vertex form factors by using an expression analogous 
to Eq. (\ref{b0002}) and Eq. (\ref{eq:unrenFF}). 

The first subdiagram, occurring in Eqs. 
(\ref{c1},\ref{c2},\ref{c5},\ref{c6}), is exactly the 1-loop QED vertex:
 
\be
\parbox{20mm}{\begin{fmfgraph*}(15,15)
\fmfleft{i1,i2}
\fmfright{o}
\fmfforce{0.2w,0.93h}{v2}
\fmfforce{0.2w,0.07h}{v1}
\fmfforce{0.5w,0.7h}{v3}
\fmfforce{0.5w,0.51h}{v33}
\fmfforce{0.8w,0.5h}{v5}
\fmf{plain}{i1,v1}
\fmf{plain}{i2,v2}
\fmf{photon}{v5,o}
\fmf{plain}{v2,v5}
\fmf{photon}{v1,v2}
\fmf{plain}{v1,v5}
\end{fmfgraph*} } =
\int {\mathfrak{D}}^Dk_1 \ 
\frac{{\mathcal N}^{\mu}}
{{\mathcal D}_{1} {\mathcal D}_{9} {\mathcal D}_{10} } \ , 
\label{c_c2} 
\ee
where 
\bea
\! \! \! \! \! \! \! \! \! \! {\mathcal N}^{\mu} & = &  
\bar{v}(p_2)  \gamma_{\sigma} 
[i ( \not{\! p_{2}} + \! \! \not{\! k_{1}}) \! +\! m] 
\gamma^{\mu} 
[-i ( \not{\! p_{1}} - \! \! \not{\! k_{1}}) \! +\! m] 
[-i ( \not{\! p_{1}} - \! \! \not{\! k_{1}}) \! +\! m] 
\gamma_{\sigma} u(p_1) \ . 
\eea
The corresponding form factors are 
\bea
\hspace*{-5mm} {\mathcal F}_1^{(1l)}(D,q^2) & = &  \frac{1}{(D-4)} \Biggl\{ 
          - \frac{1}{2}
       - \biggl[
            1
          - \frac{1}{(1+x)}
          - \frac{1}{(1-x)}
          \biggr] H(0;x) 
\Biggr\}   \nn\\
\hspace*{-5mm} & & 
       + \frac{1}{4} \biggl[
            1
          - \frac{2}{(1-x)} \biggr] H(0;x) \nn\\
\hspace*{-5mm} & & 
       - \frac{1}{2} \biggl[
            1
          - \frac{1}{(1+x)}
          - \frac{1}{(1-x)}
          \biggr] \bigl[ \zeta(2)
   - H(0;x) 
   - H(0,0;x) \nn\\
\hspace*{-5mm} & & \hspace*{5mm} 
   + 2 H(-1,0;x) \bigr]  \nn\\
\hspace*{-5mm} & + & 
(D-4) \Biggl\{ \frac{1}{8} \biggl[
            1
          - \frac{2}{(1-x)} \biggr] \bigl[ \zeta(2)
   - H(0,0;x) 
   + 2 H(-1,0;x) \bigr] \nn\\
\hspace*{-5mm} & & \hspace*{18mm} 
        + \frac{1}{4} \biggl[
            1
          - \frac{1}{(1+x)}
          - \frac{1}{(1-x)}
          \biggr] \bigl[ \zeta(2)
   + 2 \zeta(3) \nn\\
\hspace*{-5mm} & & \hspace*{25mm} 
          - (4 \! -  \! \zeta(2)) H(0;x)  \! 
   - \!  2 \zeta(2) H(-1;x) \! 
   - \!  H(0,0;x) \nn\\
\hspace*{-5mm} & & \hspace*{25mm} 
   + 2 H(-1,0;x)
   - H(0,0,0;x)
   + 2 H(-1,0,0;x) \nn\\
\hspace*{-5mm} & & \hspace*{25mm} 
   + 2 H(0,-1,0;x)
   - 4 H(-1,-1,0;x) \bigr] \Biggr\} \nn\\
\hspace*{-5mm} & + & {\mathcal O} \left( (D-4)^2 \right) \, , 
\label{1loopF1} \\
\hspace*{-5mm} {\mathcal F}_2^{(1l)}(D,q^2) & = & - \frac{1}{2} 
          \biggl[ 
     \frac{1}{(1-x)}
   - \frac{1}{(1+x)} \biggr] H(0;x) \nn\\
\hspace*{-5mm} & - & 
(D-4) \Biggl\{ \frac{1}{4} 
          \biggl[ 
     \frac{1}{(1-x)}
   - \frac{1}{(1+x)} \biggr] \bigl[ \zeta(2)
   - 4 H(0;x)
          - H(0,0;x) \nn\\
\hspace*{-5mm} & & \hspace*{18mm} 
   + 2 H(-1,0;x) \bigr] \Biggr\} \nn\\
\hspace*{-5mm} & + & {\mathcal O} \left( (D-4)^2 \right) \, , 
\label{1loopF2}\\
\hspace*{-5mm} {\mathcal F}_3^{(1l)}(D,q^2) & = & 0 
\label{c7} \, .
\eea
Let us observe that these are exactly the 1-loop {\it unrenormalized} 
form factors, ${\mathcal F}_i^{(1l)}(D,q^2)$ ($i=1,2,3$). 
As in the 2-loop counterterms they are multiplied by the 1-loop 
renormalization constants Eqs. (\ref{c001}-\ref{c003}), which behave 
for $D\to4$ as $1/(D-4)$, 
the 1-loop form factors must be evaluated up to the first order 
term in $(D-4)$ included; similarly, as the 1-loop unrenormalized form 
factors can develop poles in $(D-4)$ (that is the case of the charge 
form factor), the 1-loop renormalization constants are also needed 
up to first order in $(D-4)$ (a requirement trivially fulfilled by 
Eqs. (\ref{c001},\ref{c002},\ref{c003}), which are exact in $D$). 

The second subdiagram occurring in the {\it r.h.s.} of Eq. (\ref{c3}) 
is:
\be
m \parbox{20mm} {\begin{fmfgraph*}(15,15)
\fmfleft{i1,i2}
\fmfright{o}
\fmfforce{0.2w,0.93h}{v2}
\fmfforce{0.2w,0.07h}{v1}
\fmfforce{0.5w,0.7h}{v3}
\fmfforce{0.5w,0.51h}{v33}
\fmfforce{0.8w,0.5h}{v5}
\fmf{plain}{i1,v1}
\fmf{plain}{i2,v2}
\fmfv{label=$\otimes$}{v33}
\fmf{photon}{v5,o}
\fmf{plain}{v2,v5}
\fmf{photon}{v1,v2}
\fmf{plain}{v1,v5}
\end{fmfgraph*} } =
m \int {\mathfrak{D}}^Dk_1 \ 
\frac{{\mathcal U}^{\mu}}
{{\mathcal D}_{1} {\mathcal D}^{2}_{9} {\mathcal D}_{10} } \ , 
\label{c_c1} 
\ee
where 
\bea
\! \! \! \! \! \! \! \! \! \! {\mathcal U}^{\mu} & = &  
\bar{v}(p_2)  \gamma_{\sigma} 
[i ( \not{\! p_{2}} + \! \! \not{\! k_{1}}) \! +\! m] 
\gamma^{\mu} 
[-i ( \not{\! p_{1}} - \! \! \not{\! k_{1}}) \! +\! m] i
[-i ( \not{\! p_{1}} - \! \! \not{\! k_{1}}) \! +\! m] 
\gamma_{\sigma} u(p_1) \ ; 
\eea
the corresponding (dimensionless) form factors are:
\bea
\hspace*{-5mm} {\mathcal F}_1^{(\otimes,up)}(D,q^2) & = &  \frac{1}{(D-4)} 
\Biggl\{ 
            1
          - \frac{2}{(1+x)}
          - \frac{2}{(1+x)^2}
 - \biggl[ 
            \frac{1}{(1-x)}
          - \frac{2}{(1+x)} \nn\\
\hspace*{-5mm} & & \hspace*{18mm} 
          + \frac{3}{(1+x)^2}
          - \frac{2}{(1+x)^3}
          \biggr] H(0;x)
\Biggr\}   \nn\\
\hspace*{-5mm} & & 
          + 1 
   - \biggl[ \frac{1}{2(1+x)}
   - \frac{3}{2(1+x)^2}
   + \frac{1}{(1+x)^3} \biggr] H(0;x) \nn\\
\hspace*{-5mm} & & 
      - \frac{1}{2} \biggl[ 
            \frac{1}{(1-x)} \! 
          -  \! \frac{2}{(1 \! + \! x)}  \! 
          +  \! \frac{3}{(1 \! + \! x)^2} \! 
          -  \! \frac{2}{(1 \! + \! x)^3} \biggr]  \! 
   \bigl[ \zeta(2) \! 
   - H(0;x)  \nn\\
\hspace*{-5mm} & & \hspace*{10mm} 
   - H(0,0;x)
   + 2 H(-1,0;x) \bigr] \nn\\
\hspace*{-5mm} & + & 
(D-4) \Biggl\{ 
     1
          - \frac{3}{(1+x)}
          + \frac{3}{(1+x)^2}   \nn\\
\hspace*{-5mm} & & \hspace*{18mm} 
       + \!  \biggl[ \frac{1}{4} \biggl(
            \frac{1}{(1-x)} \! 
          -  \! \frac{2}{(1 \! + \! x)}  \! 
          +  \! \frac{3}{(1 \! + \! x)^2} \! 
          -  \! \frac{2}{(1 \! + \! x)^3} \biggr) \zeta(2)   \nn\\
\hspace*{-5mm} & & \hspace*{23mm}  
       -  \frac{1}{2(1-x)}
          + \frac{3}{(1+x)} 
          - \frac{9}{2(1+x)^2}  \nn\\
\hspace*{-5mm} & & \hspace*{23mm}  
          + \frac{1}{(1+x)^3} \biggr] H(0;x)  \nn\\
\hspace*{-5mm} & & \hspace*{18mm}  
       + \!  \biggl[ 
            \frac{1}{2(1-x)} \! 
          -  \! \frac{1}{(1 \! + \! x)}  \! 
          +  \! \frac{3}{2(1 \! + \! x)^2} \! 
          -  \! \frac{1}{(1 \! + \! x)^3} \biggr] \bigl[ \zeta(3)  \nn\\
\hspace*{-5mm} & & \hspace*{27mm}  
          - H(-1;x) \bigr] \nn\\
\hspace*{-5mm} & & \hspace*{18mm}  
       +  \frac{1}{4} \biggl[ 
            \frac{1}{(1-x)} \! 
          -  \! \frac{3}{(1 \! + \! x)}  \! 
          +  \! \frac{6}{(1 \! + \! x)^2} \! 
          -  \! \frac{4}{(1 \! + \! x)^3} \biggr]  \bigl[ \zeta(2) \nn\\
\hspace*{-5mm} & & \hspace*{27mm}  
          - H(0,0;x)
          + 2 H(-1,0;x) \bigr] \nn\\
\hspace*{-5mm} & & \hspace*{18mm}  
       -  \frac{1}{4} \biggl[ 
            \frac{1}{(1-x)} \! 
          -  \! \frac{2}{(1+x)}  \! 
          +  \! \frac{3}{(1+x)^2}  \! 
          -  \! \frac{2}{(1+x)^3} \biggr] \times  \nn\\
\hspace*{-5mm} & & \hspace*{27mm} 
        \times \bigl[ 
     H(0,0,0;x)
   + 4 H(-1,-1,0;x) \nn\\
\hspace*{-5mm} & & \hspace*{27mm}  
   - 2 H(-1,0,0;x)
   - 2 H(0,-1,0;x) \bigr]
\Biggr\}   \nn\\
\hspace*{-5mm} & + & {\mathcal O} \left( (D-4)^2 \right) \, , \\
\hspace*{-5mm} {\mathcal F}_2^{(\otimes,up)}(D,q^2) & = &  \frac{1}{(D-4)} 
\Biggl\{ 
            \frac{2}{(1+x)} \biggl[
     1
          - \frac{1}{(1+x)} \biggr]
 + \frac{1}{2} \biggl[ 
            \frac{1}{(1-x)}
          - \frac{3}{(1+x)} \nn\\
\hspace*{-5mm} & & \hspace*{18mm} 
          + \frac{6}{(1+x)^2}
          - \frac{4}{(1+x)^3}
          \biggr] H(0;x)
\Biggr\}   \nn\\
\hspace*{-5mm} & & 
       - \frac{2}{(1+x)} \biggl[
     1
          - \frac{1}{(1+x)} \biggr]\nn\\
\hspace*{-5mm} & & 
       + \frac{2}{(1+x)} \biggl[ 
            1
   - \frac{3}{(1+x)}
   + \frac{2}{(1+x)^2} \biggr] H(0;x) \nn\\
\hspace*{-5mm} & & 
      + \frac{1}{4} \biggl[ 
            \frac{1}{(1-x)}
          - \frac{3}{(1+x)} 
          + \frac{6}{(1+x)^2}
          - \frac{4}{(1+x)^3} \biggr] 
   \bigl[ \zeta(2) \nn\\
\hspace*{-5mm} & & \hspace*{10mm} 
   - H(0,0;x)
   + 2 H(-1,0;x) \bigr] \nn\\
\hspace*{-5mm} & + & 
(D-4) \Biggl\{ \frac{1}{(1+x)} \biggl[ 1 - \frac{1}{(1+x)} \biggr] \nn\\
\hspace*{-5mm} & & \hspace*{18mm}
       + \frac{1}{2(1+x)} \biggl[ 
             1 
   - \frac{3}{(1+x)}
   + \frac{2}{(1+x)} \biggr] 
   \bigl[ 2 \zeta(2) \nn\\
\hspace*{-5mm} & & \hspace*{23mm}
          - 3 H(0;x)
   - 2 H(0,0;x) 
   + 4 H(-1,0;x) \bigr] \nn\\
\hspace*{-5mm} & & \hspace*{18mm}
       - \frac{1}{8} \biggl[ 
            \frac{1}{(1 \! - \! x)} \! 
          -  \! \frac{3}{(1 \! + \! x)}  \! 
          +  \! \frac{6}{(1 \! + \! x)^2} \! 
          -  \! \frac{4}{(1 \! + \! x)^3} \biggr] 
   \bigl[ 2 \zeta(3) \nn\\
\hspace*{-5mm} & & \hspace*{23mm}
          + \zeta(2) ( H(0;x)
   - 2 H(-1;x) ) 
   - H(0,0,0;x)  \nn\\
\hspace*{-5mm} & & \hspace*{23mm}
         - 4 H(-1,-1,0;x)
  + 2 H(-1,0,0;x) \nn\\
\hspace*{-5mm} & & \hspace*{23mm}
  + 2 H(0,-1,0;x)\bigr]  
\Biggr\}  \nn\\
\hspace*{-5mm} & + & {\mathcal O} \left( (D-4)^2 \right) \, , \\
\hspace*{-5mm} {\mathcal F}_3^{(\otimes,up)}(D,q^2) & = &  \frac{1}{(D-4)} 
\Biggl\{ 
         - \frac{1}{(1-x)} \biggl[
     1
          - \frac{1}{(1-x)} \biggr]
 - \frac{1}{4} \biggl[ 
            \frac{1}{(1+x)}
          - \frac{3}{(1-x)} \nn\\
\hspace*{-5mm} & & \hspace*{18mm} 
          + \frac{6}{(1-x)^2}
          - \frac{4}{(1-x)^3}
          \biggr] H(0;x)
\Biggr\}   \nn\\
\hspace*{-5mm} & & 
       - \frac{1}{(1-x)} \biggl[
     1
          - \frac{1}{(1-x)} \biggr]\nn\\
\hspace*{-5mm} & & 
       + \frac{1}{4} \biggl[ 
            \frac{1}{(1+x)}
          + \frac{1}{(1-x)}
   - \frac{6}{(1-x)^2}
   + \frac{4}{(1-x)^3} \biggr] H(0;x) \nn\\
\hspace*{-5mm} & & 
      - \frac{1}{8} \biggl[ 
            \frac{1}{(1-x)}
          - \frac{3}{(1+x)} 
          + \frac{6}{(1+x)^2}
          - \frac{4}{(1+x)^3} \biggr] 
   \bigl[ \zeta(2) \nn\\
\hspace*{-5mm} & & \hspace*{10mm} 
   - H(0,0;x)
   + 2 H(-1,0;x) \bigr] \nn\\
\hspace*{-5mm} & + & 
(D-4) \Biggl\{ \frac{1}{2(1-x)} \biggl[ 1 - \frac{1}{(1-x)} \biggr] \nn\\
\hspace*{-5mm} & & \hspace*{18mm}
       - \frac{1}{4} \biggl[ 
            \frac{1}{(1+x)} 
   - \frac{3}{(1-x)^2}
   + \frac{2}{(1-x)^3} \biggr] H(0;x) \nn\\
\hspace*{-5mm} & & \hspace*{18mm}
       + \frac{1}{8} \biggl[ 
            \frac{1}{(1 \! + \! x)} \! 
          +  \! \frac{1}{(1 \! - \! x)}  \! 
          -  \! \frac{6}{(1 \! - \! x)^2} \! 
          +  \! \frac{4}{(1 \! - \! x)^3} \biggr] 
   \bigl[ \zeta(2) \nn\\
\hspace*{-5mm} & & \hspace*{23mm}
          - H(0,0;x)
   + 2 H(-1,0;x) \bigr]  \nn\\
\hspace*{-5mm} & & \hspace*{18mm}
       + \frac{1}{16} \biggl[ 
            \frac{1}{(1 \! + \! x)} \! 
          -  \! \frac{3}{(1 \! - \! x)}  \! 
          +  \! \frac{6}{(1 \! - \! x)^2} \! 
          -  \! \frac{4}{(1 \! - \! x)^3} \biggr] \times \nn\\
\hspace*{-5mm} & & \hspace*{23mm}
  \times  \bigl[ 2 \zeta(3)
          + \zeta(2) ( H(0;x)
   - 2 H(-1;x) )\nn\\
\hspace*{-5mm} & & \hspace*{28mm}
          - H(0,0,0;x)
         - 4 H(-1,-1,0;x) \nn\\
\hspace*{-5mm} & & \hspace*{28mm}
  + 2 H(-1,0,0;x)
  + 2 H(0,-1,0;x) \bigr]  
\Biggr\}  \nn\\
\hspace*{-5mm} & + & {\mathcal O} \left( (D-4)^2 \right)
\label{c8} \, .
\eea

The last subdiagram, occurring in the {\it r.h.s.} of Eq. (\ref{c4}) is:
\be
m \parbox{20mm}{\begin{fmfgraph*}(15,15)
\fmfleft{i1,i2}
\fmfright{o}
\fmfforce{0.2w,0.93h}{v2}
\fmfforce{0.2w,0.07h}{v1}
\fmfforce{0.5w,0.3h}{v3}
\fmfforce{0.5w,0.49h}{v33}
\fmfforce{0.8w,0.5h}{v5}
\fmf{plain}{i1,v1}
\fmf{plain}{i2,v2}
\fmfv{label=$\otimes$}{v33}
\fmf{photon}{v5,o}
\fmf{plain}{v2,v5}
\fmf{photon}{v1,v2}
\fmf{plain}{v1,v5}
\end{fmfgraph*} } =
m \int {\mathfrak{D}}^Dk_1 \ 
\frac{{\mathcal V}^{\mu}}
{{\mathcal D}_{1} {\mathcal D}_{9} {\mathcal D}^{2}_{10} } \ , 
\label{c_c3} 
\ee
where 
\bea
\! \! \! \! \! \! \! \! \! \! {\mathcal V}^{\mu} & = &  
\bar{v}(p_2)  \gamma_{\sigma} 
[i ( \not{\! p_{2}} + \! \! \not{\! k_{1}}) \! +\! m] i
[i ( \not{\! p_{2}} + \! \! \not{\! k_{1}}) \! +\! m]
\gamma^{\mu} 
[-i ( \not{\! p_{1}} - \! \! \not{\! k_{1}}) \! +\! m] 
\gamma_{\sigma} u(p_1) \ ; 
\eea
its form factors are:
\bea
\! \! \! \! \! \! {\mathcal F}_1^{(\otimes,down)}(D,q^2) & = & 
{\mathcal F}_1^{(\otimes,up)}(D,q^2)\\ 
\! \! \! \! \! \! {\mathcal F}_2^{(\otimes,down)}(D,q^2) & = & 
{\mathcal F}_2^{(\otimes,up)}(D,q^2) \\
\! \! \! \! \! \! {\mathcal F}_3^{(\otimes,down)}(D,q^2) & = & 
- {\mathcal F}_3^{(\otimes,up)}(D,q^2) 
\label{c9} \, .
\eea

We will write the contributions to the 2-loop form factors from the 
subtraction graphs with the 1-loop renormalization counter-terms 
of Fig. \ref{fig2} as 
${\mathcal F}_i^{(C,{\tt cnt})}(D,q^2)$ ($i=1,2,3$), where the $C$ in 
the superscript 
stands for ``counter-term'' and the ${\tt cnt}$ (${\tt cnt} \in \{
{\tt a,...,e} \}$) refers to the corresponding graphs of Fig. \ref{fig2}.
According to Eqs. (\ref{c1}-\ref{c5}), we have:
\bea
 & {\mathcal F}_i^{(C,{\tt a})}(D,q^2) \, = & 
Z_{1}^{(1l)}(D) \times {\mathcal F}_i^{(1l)}(D,q^2) \, , \\
 & {\mathcal F}_i^{(C,{\tt b})}(D,q^2)  \,= &  
{\mathcal F}_i^{(C,{\tt c})}(D,q^2)  \,  = \, Z_{1}^{(1l)}(D) 
\times {\mathcal F}_i^{(1l)}(D,q^2) \, , \\
& {\mathcal F}_i^{(C,{\tt d})}(D,q^2)  \, = & \, 
                     \frac{1}{m}\ \delta m^{(1l)}(D,m) 
\times {\mathcal F}_i^{(\otimes,up)}(D,q^2) \, , \\
 & {\mathcal F}_i^{(C,{\tt e})}(D,q^2)  \, = & \, 
                     \frac{1}{m}\ \delta m^{(1l)}(D,m) 
\times {\mathcal F}_i^{(\otimes,down)}(D,q^2) \, , \\
 & {\mathcal F}_i^{(C,{\tt f})}(D,q^2)  \, = &  
{\mathcal F}_i^{(C,{\tt g})}(D,q^2)  \, = \, Z_{2}^{(1l)}(D) \times 
{\mathcal F}_i^{(1l)}(D,q^2) \, , \\
& {\mathcal F}_i^{(C,{\tt h})}(D,q^2)  \, = & \, Z_{3}^{(1l)}(D) 
\times {\mathcal F}_i^{(1l)}(D,q^2) \, , 
\eea
where $i=1,2,3$.

As can be seen from Eqs. (\ref{c7},\ref{c8},\ref{c9}), the 
total contribution to the third form factor vanishes:
\be
{\mathcal F}^{(C)}_{3}(D,q^2) = 
\sum_{{\tt cnt}} {\mathcal F}^{(C,{\tt cnt})}_{3}(D,q^2) = 0 \ .
\label{EFFE3CT}
\ee

\subsection{2-loop charge renormalization constant. \label{2Lcnt}}

The only subtraction from the 2-loop renormalization counter-terms, in
our case, is given by the product of the charge renormalization constant
at 2-loop times the tree-level vertex, corresponding to the diagram
{\tt(i)} of Fig. \ref{fig2}:
\be
\parbox{20mm}{\begin{fmfgraph*}(15,15) 
\fmfleft{i1,i2}
\fmfright{o}
\fmfforce{0.2w,0.93h}{v2}
\fmfforce{0.2w,0.07h}{v1}
\fmfforce{0.8w,0.5h}{v5}
\fmfforce{0.56w,0.5h}{v55}
\fmf{plain}{i1,v1}
\fmf{plain}{i2,v2}
\fmfv{label=$\otimes$}{v55}
\fmfv{l=$Z_{1}^{(2l)}$,l.a=75,l.d=.15w}{v5}
\fmf{photon}{v5,o}
\fmf{plain}{v2,v5}
\fmf{phantom}{v1,v2}
\fmf{plain}{v1,v5}
\end{fmfgraph*} }  \stackrel{def}{=}  Z_{1}^{(2l)}(D) \, \times \,
\parbox{20mm}{\begin{fmfgraph*}(15,15) 
\fmfleft{i1,i2}
\fmfright{o}
\fmfforce{0.2w,0.93h}{v2}
\fmfforce{0.2w,0.07h}{v1}
\fmfforce{0.8w,0.5h}{v5}
\fmf{plain}{i1,v1}
\fmf{plain}{i2,v2}
\fmf{photon}{v5,o}
\fmf{plain}{v2,v5}
\fmf{phantom}{v1,v2}
\fmf{plain}{v1,v5}
\end{fmfgraph*} } \, .
\ee

The 2-loop charge renormalization constant $Z_1^{(2l)}(D)$ is given 
by the value at $q^2=0$ of the unrenormalized charge form factor 
at two loops, 
\begin{equation} 
{\mathcal F}^{(2l)}_{1}(D,q^2) = 
   \sum_{{\tt graph}} {\mathcal F}^{(2l,{\tt graph})}_{1}(D,q^2) 
   - \sum_{{\tt cnt}} {\mathcal F}^{(C,{\tt cnt})}_{1}(D,q^2) \, ,
\end{equation} 
where ${\tt graph} \in \{ {\tt a,...,g} \}$ runs over the 
diagrams of Fig. \ref{fig1} and ${\tt cnt} \in \{ {\tt a,...,h} \}$ runs
over the subtraction graphs of Fig. \ref{fig2}. 

As the evaluation of the value at $q^2=0$ is somewhat simpler than the 
value for arbitrary $q^2$, we give here its expression exact in $D$: 
\bea
Z_{1}^{(2l)}(D) & = & {\mathcal{F}}^{(2l)}_{1}(D,q^2=0) \nonumber\\ 
    &=&  \frac{(D-6)}{4(D-5)(D-4)(D-3)} 
            \Big( 360-650 D+470 D^2              \nn \\
    & & \hspace*{35mm} - 167 D^3 + 29 D^4 - 2 D^5 \Big) 
\left( \frac{1}{m^2} \, \parbox{20mm}{\begin{fmfgraph*}(15,15)
\fmfleft{i}
\fmfright{o}
\fmf{plain}{i,v1}
\fmf{plain}{v2,o}
\fmf{photon,tension=.15,left}{v1,v2}
\fmf{plain,tension=.15}{v1,v2}
\fmf{photon,tension=.15,right}{v1,v2}
\end{fmfgraph*} }  \hspace*{-3mm} \right)
\nn \\
& + &  \frac{(D + 4)}{8 (D - 6) (D - 4)^2} 
           \Big( 736 - 1348 D + 800 D^2 
            \nn \\
& & \hspace*{35mm} 
            - 207 D^3 + 24 D^4 - D^5 \Big) 
\left( \frac{1}{m^2} \, \parbox{20mm}{\begin{fmfgraph*}(15,15)
\fmfleft{i}
\fmfright{o}
\fmf{plain}{i,v1}
\fmf{plain}{v2,o}
\fmf{plain,tension=.15,left}{v1,v2}
\fmf{plain,tension=.15}{v1,v2}
\fmf{plain,tension=.15,right}{v1,v2}
\end{fmfgraph*} } \hspace*{-3mm} \right)
\nn \\
& + & \frac{(D-2)}{48(D-7)(D-6)(D-5)^2(D-4)^2(D-3)^2} 
           \times \nn \\
& & \hspace*{15mm}
           \Big(  7131744 - 9801144 D + 1271956 D^2 + 5512286 D^3 
             \nn \\
& & \hspace*{18mm}
             - 4884843 D^4 + 2058126 D^5 - 514065 D^6 + 79836 D^7 
             \nn \\
& & \hspace*{18mm}
             - 7567 D^8 + 400 D^9 - 9 D^{10} \Big) 
 \left( \frac{1}{m^4} \! \! 
\parbox{15mm}{\begin{fmfgraph*}(15,15)
\fmfleft{i}
\fmfright{o}
\fmf{phantom}{i,v1}
\fmf{phantom}{v1,o}
\fmf{plain,right=45}{v1,v1}
\fmf{plain,left=45}{v1,v1}
\end{fmfgraph*} } \hspace*{-3mm} \right) \, , 
\eea
where the MIs depicted in the r.h.s. are those of Fig. 7 of \cite{Bon1}.

The corresponding expansion in $(D-4)$ is:
\bea
Z_{1}^{(2l)}(D) & = &
       - \frac{9}{8}   \frac{1}{(D-4)^2} 
       + \frac{55}{32} \frac{1}{(D-4)}  
       - \frac{7685}{1152}
       \nn \\
       & &
       + \frac{55}{8} \zeta(2)
       - 6 \zeta(2) \ln2
       + \frac{3}{2} \zeta(3)
       + {\mathcal O} (D-4) \, .
\eea
Let us recall once more that this counter-term is required for the 
renormalization of the {\it charge} form factor 
${\mathcal F}^{(2l)}_{1}(D,q^2)$ only.

\section{The renormalized form factors \label{fullyren}}

As a first step the renormalization procedure requires the proper 
subtraction of the 8 graphs with the counter-terms at one loop of Fig. 
\ref{fig2} 
from the 7 unrenormalized 2-loop graphs of section \ref{diagrams}. 
That is carried out according to the following scheme: 
\bea
\left. \parbox{20mm}{\begin{fmfgraph*}(15,15)
\fmfleft{i1,i2}
\fmfright{o}
\fmf{plain}{i1,v1}
\fmf{plain}{i2,v2}
\fmf{photon}{v5,o}
\fmf{plain,tension=.3}{v2,v3}
\fmf{plain,tension=.3}{v3,v5}
\fmf{plain,tension=.3}{v1,v4}
\fmf{plain,tension=.3}{v4,v5}
\fmf{photon,tension=0}{v2,v1}
\fmf{photon,tension=0}{v4,v3}
\end{fmfgraph*} } \right| _{ren} & = & 
\parbox{20mm}{\begin{fmfgraph*}(15,15)
\fmfleft{i1,i2}
\fmfright{o}
\fmf{plain}{i1,v1}
\fmf{plain}{i2,v2}
\fmf{photon}{v5,o}
\fmf{plain,tension=.3}{v2,v3}
\fmf{plain,tension=.3}{v3,v5}
\fmf{plain,tension=.3}{v1,v4}
\fmf{plain,tension=.3}{v4,v5}
\fmf{photon,tension=0}{v2,v1}
\fmf{photon,tension=0}{v4,v3}
\end{fmfgraph*} } - 
\parbox{20mm}{\begin{fmfgraph*}(15,15) 
\fmfleft{i1,i2}
\fmfright{o}
\fmfforce{0.2w,0.93h}{v2}
\fmfforce{0.2w,0.07h}{v1}
\fmfforce{0.8w,0.5h}{v5}
\fmfforce{0.55w,0.5h}{v55}
\fmf{plain}{i1,v1}
\fmf{plain}{i2,v2}
\fmfv{label=$\otimes$}{v55}
\fmfv{l=$Z_{1}$,l.a=55,l.d=.1w}{v5}
\fmf{photon}{v5,o}
\fmf{plain}{v2,v5}
\fmf{photon}{v1,v2}
\fmf{plain}{v1,v5}
\end{fmfgraph*} } \, , \\
 & & \nonumber \\
\left. \parbox{20mm}{\begin{fmfgraph*}(15,15)
\fmfleft{i1,i2}
\fmfright{o}
\fmf{plain}{i1,v1}
\fmf{plain}{i2,v2}
\fmf{photon}{v5,o}
\fmf{plain,tension=.3}{v2,v3}
\fmf{plain,tension=.3}{v3,v5}
\fmf{plain,tension=.3}{v1,v4}
\fmf{plain,tension=.3}{v4,v5}
\fmf{photon,tension=0}{v2,v4}
\fmf{photon,tension=0}{v1,v3}
\end{fmfgraph*} 
}  \right| _{ren} & = & 
\parbox{20mm}{\begin{fmfgraph*}(15,15)
\fmfleft{i1,i2}
\fmfright{o}
\fmf{plain}{i1,v1}
\fmf{plain}{i2,v2}
\fmf{photon}{v5,o}
\fmf{plain,tension=.3}{v2,v3}
\fmf{plain,tension=.3}{v3,v5}
\fmf{plain,tension=.3}{v1,v4}
\fmf{plain,tension=.3}{v4,v5}
\fmf{photon,tension=0}{v2,v4}
\fmf{photon,tension=0}{v1,v3}
\end{fmfgraph*} 
} \, , \\
 & & \nonumber \\
\left. \parbox{20mm}{\begin{fmfgraph*}(15,15)
\fmfleft{i1,i2}
\fmfright{o}
\fmfforce{0.2w,0.93h}{v1}
\fmfforce{0.2w,0.07h}{v2}
\fmfforce{0.2w,0.5h}{v3}
\fmfforce{0.8w,0.5h}{v5}
\fmf{plain}{i1,v2}
\fmf{plain}{i2,v1}
\fmf{photon}{v5,o}
\fmf{plain,tension=0}{v2,v5}
\fmf{plain,tension=0}{v3,v4}
\fmf{photon,tension=.4}{v1,v4}
\fmf{plain,tension=.4}{v4,v5}
\fmf{plain,tension=0}{v1,v3}
\fmf{photon,tension=0}{v2,v3}
\end{fmfgraph*} }    \right| _{ren} & = & 
\parbox{20mm}{\begin{fmfgraph*}(15,15)
\fmfleft{i1,i2}
\fmfright{o}
\fmfforce{0.2w,0.93h}{v1}
\fmfforce{0.2w,0.07h}{v2}
\fmfforce{0.2w,0.5h}{v3}
\fmfforce{0.8w,0.5h}{v5}
\fmf{plain}{i1,v2}
\fmf{plain}{i2,v1}
\fmf{photon}{v5,o}
\fmf{plain,tension=0}{v2,v5}
\fmf{plain,tension=0}{v3,v4}
\fmf{photon,tension=.4}{v1,v4}
\fmf{plain,tension=.4}{v4,v5}
\fmf{plain,tension=0}{v1,v3}
\fmf{photon,tension=0}{v2,v3}
\end{fmfgraph*} }  - 
\parbox{20mm}
{\begin{fmfgraph*}(15,15) 
\fmfleft{i1,i2}
\fmfright{o}
\fmfforce{0.2w,0.93h}{v2}
\fmfforce{0.4w,0.70h}{v22}
\fmfforce{0.2w,0.07h}{v1}
\fmfforce{0.8w,0.5h}{v5}
\fmf{plain}{i1,v1}
\fmf{plain}{i2,v2}
\fmfv{label=$\otimes$}{v22}
\fmfv{l=$Z_{1}$,l.a=60,l.d=.1w}{v2}
\fmf{photon}{v5,o}
\fmf{plain}{v2,v5}
\fmf{photon}{v1,v2}
\fmf{plain}{v1,v5}
\end{fmfgraph*} } \, , \\
 & & \nonumber \\
\left. \parbox{20mm}{\begin{fmfgraph*}(15,15)
\fmfleft{i1,i2}
\fmfright{o}
\fmfforce{0.2w,0.93h}{v2}
\fmfforce{0.2w,0.07h}{v1}
\fmfforce{0.2w,0.5h}{v3}
\fmfforce{0.8w,0.5h}{v5}
\fmf{plain}{i1,v1}
\fmf{plain}{i2,v2}
\fmf{photon}{v5,o}
\fmf{plain,tension=0}{v2,v5}
\fmf{plain,tension=0}{v3,v4}
\fmf{photon,tension=.4}{v1,v4}
\fmf{plain,tension=.4}{v4,v5}
\fmf{plain,tension=0}{v1,v3}
\fmf{photon,tension=0}{v2,v3}
\end{fmfgraph*} }   \right| _{ren} & = & 
\parbox{20mm}{\begin{fmfgraph*}(15,15)
\fmfleft{i1,i2}
\fmfright{o}
\fmfforce{0.2w,0.93h}{v2}
\fmfforce{0.2w,0.07h}{v1}
\fmfforce{0.2w,0.5h}{v3}
\fmfforce{0.8w,0.5h}{v5}
\fmf{plain}{i1,v1}
\fmf{plain}{i2,v2}
\fmf{photon}{v5,o}
\fmf{plain,tension=0}{v2,v5}
\fmf{plain,tension=0}{v3,v4}
\fmf{photon,tension=.4}{v1,v4}
\fmf{plain,tension=.4}{v4,v5}
\fmf{plain,tension=0}{v1,v3}
\fmf{photon,tension=0}{v2,v3}
\end{fmfgraph*} }  - 
\parbox{20mm}{\begin{fmfgraph*}(15,15)
\fmfleft{i1,i2}
\fmfright{o}
\fmfforce{0.2w,0.93h}{v2}
\fmfforce{0.2w,0.07h}{v1}
\fmfforce{0.4w,0.30h}{v11}
\fmfforce{0.8w,0.5h}{v5}
\fmf{plain}{i1,v1}
\fmf{plain}{i2,v2}
\fmfv{label=$\otimes$}{v11}
\fmfv{l=$Z_{1}$,l.a=-10,l.d=.2w}{v1}
\fmf{photon}{v5,o}
\fmf{plain}{v2,v5}
\fmf{photon}{v1,v2}
\fmf{plain}{v1,v5}
\end{fmfgraph*} } \, , \\
 & & \nonumber \\
\left. \parbox{20mm}{\begin{fmfgraph*}(15,15)
\fmfleft{i1,i2}
\fmfright{o}
\fmfforce{0.2w,0.93h}{v2}
\fmfforce{0.2w,0.07h}{v1}
\fmfforce{0.8w,0.5h}{v5}
\fmf{plain}{i1,v1}
\fmf{plain}{i2,v2}
\fmf{photon}{v5,o}
\fmf{plain}{v2,v3}
\fmf{photon,tension=.25,right}{v3,v4}
\fmf{plain,tension=.25}{v3,v4}
\fmf{plain}{v4,v5}
\fmf{plain}{v1,v5}
\fmf{photon}{v1,v2}
\end{fmfgraph*} }   \right| _{ren} & = & 
\parbox{20mm}{\begin{fmfgraph*}(15,15)
\fmfleft{i1,i2}
\fmfright{o}
\fmfforce{0.2w,0.93h}{v2}
\fmfforce{0.2w,0.07h}{v1}
\fmfforce{0.8w,0.5h}{v5}
\fmf{plain}{i1,v1}
\fmf{plain}{i2,v2}
\fmf{photon}{v5,o}
\fmf{plain}{v2,v3}
\fmf{photon,tension=.25,right}{v3,v4}
\fmf{plain,tension=.25}{v3,v4}
\fmf{plain}{v4,v5}
\fmf{plain}{v1,v5}
\fmf{photon}{v1,v2}
\end{fmfgraph*} }  
- \parbox{20mm}
{\begin{fmfgraph*}(15,15)
\fmfleft{i1,i2}
\fmfright{o}
\fmfforce{0.2w,0.93h}{v2}
\fmfforce{0.2w,0.07h}{v1}
\fmfforce{0.5w,0.7h}{v3}
\fmfforce{0.5w,0.51h}{v33}
\fmfforce{0.8w,0.5h}{v5}
\fmf{plain}{i1,v1}
\fmf{plain}{i2,v2}
\fmfv{label=$\otimes$}{v33}
\fmfv{l=$\delta m$,l.a=60,l.d=.15w}{v3}
\fmf{photon}{v5,o}
\fmf{plain}{v2,v5}
\fmf{photon}{v1,v2}
\fmf{plain}{v1,v5}
\end{fmfgraph*} }  
- \parbox{20mm}
{\begin{fmfgraph*}(15,15)
\fmfleft{i1,i2}
\fmfright{o}
\fmfforce{0.2w,0.93h}{v2}
\fmfforce{0.2w,0.07h}{v1}
\fmfforce{0.5w,0.7h}{v3}
\fmfforce{0.5w,0.51h}{v33}
\fmfforce{0.8w,0.5h}{v5}
\fmf{plain}{i1,v1}
\fmf{plain}{i2,v2}
\fmfv{label=$\otimes$}{v33}
\fmfv{l=$Z_{2}$,l.a=80,l.d=.15w}{v3}
\fmf{photon}{v5,o}
\fmf{plain}{v2,v5}
\fmf{photon}{v1,v2}
\fmf{plain}{v1,v5}
\end{fmfgraph*} } \, , \\
 & & \nonumber \\
\left. \parbox{20mm}{\begin{fmfgraph*}(15,15)
\fmfleft{i1,i2}
\fmfright{o}
\fmfforce{0.2w,0.93h}{v1}
\fmfforce{0.2w,0.07h}{v2}
\fmfforce{0.8w,0.5h}{v5}
\fmf{plain}{i1,v2}
\fmf{plain}{i2,v1}
\fmf{photon}{v5,o}
\fmf{plain}{v2,v3}
\fmf{photon,tension=.25,left}{v3,v4}
\fmf{plain,tension=.25}{v3,v4}
\fmf{plain}{v4,v5}
\fmf{plain}{v1,v5}
\fmf{photon}{v1,v2}
\end{fmfgraph*} }  \right| _{ren}  & = & \parbox{20mm}{\begin{fmfgraph*}(15,15)
\fmfleft{i1,i2}
\fmfright{o}
\fmfforce{0.2w,0.93h}{v1}
\fmfforce{0.2w,0.07h}{v2}
\fmfforce{0.8w,0.5h}{v5}
\fmf{plain}{i1,v2}
\fmf{plain}{i2,v1}
\fmf{photon}{v5,o}
\fmf{plain}{v2,v3}
\fmf{photon,tension=.25,left}{v3,v4}
\fmf{plain,tension=.25}{v3,v4}
\fmf{plain}{v4,v5}
\fmf{plain}{v1,v5}
\fmf{photon}{v1,v2}
\end{fmfgraph*} } 
- \parbox{20mm}
{\begin{fmfgraph*}(15,15)
\fmfleft{i1,i2}
\fmfright{o}
\fmfforce{0.2w,0.93h}{v2}
\fmfforce{0.2w,0.07h}{v1}
\fmfforce{0.5w,0.7h}{v3}
\fmfforce{0.5w,0.49h}{v33}
\fmfforce{0.8w,0.5h}{v5}
\fmf{plain}{i1,v1}
\fmf{plain}{i2,v2}
\fmfv{label=$\otimes$}{v33}
\fmfv{l=$\delta m$,l.a=-60,l.d=.4w}{v3}
\fmf{photon}{v5,o}
\fmf{plain}{v2,v5}
\fmf{photon}{v1,v2}
\fmf{plain}{v1,v5}
\end{fmfgraph*} }  
- \parbox{20mm}
{\begin{fmfgraph*}(15,15)
\fmfleft{i1,i2}
\fmfright{o}
\fmfforce{0.2w,0.93h}{v2}
\fmfforce{0.2w,0.07h}{v1}
\fmfforce{0.5w,0.7h}{v3}
\fmfforce{0.5w,0.49h}{v33}
\fmfforce{0.8w,0.5h}{v5}
\fmf{plain}{i1,v1}
\fmf{plain}{i2,v2}
\fmfv{label=$\otimes$}{v33}
\fmfv{l=$Z_{2}$,l.a=-60,l.d=.4w}{v3}
\fmf{photon}{v5,o}
\fmf{plain}{v2,v5}
\fmf{photon}{v1,v2}
\fmf{plain}{v1,v5}
\end{fmfgraph*} } \, , \\
 & & \nonumber \\
\left. \parbox{20mm}{\begin{fmfgraph*}(15,15)
\fmfleft{i1,i2}
\fmfright{o}
\fmfforce{0.2w,0.93h}{v2}
\fmfforce{0.2w,0.07h}{v1}
\fmfforce{0.2w,0.3h}{v3}
\fmfforce{0.2w,0.7h}{v4}
\fmfforce{0.8w,0.5h}{v5}
\fmf{plain}{i1,v1}
\fmf{plain}{i2,v2}
\fmf{photon}{v5,o}
\fmf{plain}{v2,v5}
\fmf{photon}{v1,v3}
\fmf{photon}{v2,v4}
\fmf{plain}{v1,v5}
\fmf{plain,left}{v3,v4}
\fmf{plain,right}{v3,v4}
\end{fmfgraph*} } \right| _{ren}  & = & 
\parbox{20mm}{\begin{fmfgraph*}(15,15)
\fmfleft{i1,i2}
\fmfright{o}
\fmfforce{0.2w,0.93h}{v2}
\fmfforce{0.2w,0.07h}{v1}
\fmfforce{0.2w,0.3h}{v3}
\fmfforce{0.2w,0.7h}{v4}
\fmfforce{0.8w,0.5h}{v5}
\fmf{plain}{i1,v1}
\fmf{plain}{i2,v2}
\fmf{photon}{v5,o}
\fmf{plain}{v2,v5}
\fmf{photon}{v1,v3}
\fmf{photon}{v2,v4}
\fmf{plain}{v1,v5}
\fmf{plain,left}{v3,v4}
\fmf{plain,right}{v3,v4}
\end{fmfgraph*} }
- \hspace*{0.5cm}
\parbox{20mm}
{\begin{fmfgraph*}(15,15)
\fmfleft{i1,i2}
\fmfright{o}
\fmfforce{0.2w,0.93h}{v2}
\fmfforce{0.2w,0.07h}{v1}
\fmfforce{0.2w,0.5h}{v3}
\fmfforce{0.42w,0.5h}{v33}
\fmfforce{0.8w,0.5h}{v5}
\fmf{plain}{i1,v1}
\fmf{plain}{i2,v2}
\fmfv{label=$\otimes$}{v33}
\fmfv{l=$Z_{3}$,l.a=180,l.d=.15w}{v3}
\fmf{photon}{v5,o}
\fmf{plain}{v2,v5}
\fmf{photon}{v1,v2}
\fmf{plain}{v1,v5}
\end{fmfgraph*} } \, .
\eea
As already anticipated in the previous section, we will indicate by 
${\mathcal F}_i^{(2l)}(D,q^2)$ the sum of the 
contributions to the form factors from all the above graphs: 
\begin{equation} 
{\mathcal F}^{(2l)}_{i}(D,q^2) = 
   \sum_{{\tt graph}} {\mathcal F}^{(2l,{\tt graph})}_{i}(D,q^2) 
   - \sum_{{\tt cnt}} {\mathcal F}^{(C,{\tt cnt})}_{i}(D,q^2) \, ,
\end{equation} 
where ${\tt graph} \in \{ {\tt a,...,g} \}$ runs over the 
diagrams of Fig. \ref{fig1} and ${\tt cnt} \in \{ {\tt a,...,h} \}$ runs
over the counter-terms of Fig. \ref{fig2}. 
Note that, as $ Z_2^{(1l)}(D) = - Z_1^{(1l)}(D) $ (the Ward identity), 
all the terms with $ Z_2^{(1l)}(D) $ are canceled by corresponding 
terms with $ Z_1^{(1l)}(D) $, so that 
only a single term proportional to $ Z_1^{(1l)}(D) $ remains in 
the sum of the counter-terms. 

For $i=3$ the renormalized form factor vanishes, as expected 
\be 
  F^{(2l)}_{3}(D,q^2) = 0 \, . 
\ee 

To obtain the 2-loop fully renormalized form factors $ F^{(2l)}_{i}(D,q^2) $ 
we have to subtract from the first form factor 
${\mathcal F}^{(2l)}_{1}(D,q^2)$ its value at $q^2=0$ (2-loop charge 
renormalization); no renormalization is needed for the second 
form factor, and we have finally 
\begin{eqnarray} 
 F^{(2l)}_{1}(D,q^2) &=& {\mathcal F}^{(2l)}_{1}(D,q^2) 
                      -  {\mathcal F}^{(2l)}_{1}(D,0)
   \label{2lFiren} \nn\\ 
                     &=& {\mathcal F}^{(2l)}_{1}(D,q^2) 
                      -   Z_{1}^{(2l)}(D) \, , \\
 F^{(2l)}_{2}(D,q^2) &=& {\mathcal F}^{(2l)}_{2}(D,q^2) \ .
\end{eqnarray} 

We report now the analytic expression of the UV-renormalized form factors 
$ F^{(2l)}_{i}(D,q^2) $ in the space-like region $-s = q^2> 0$, 
in terms of HPLs of the variable $x$ already introduced in Eq. (\ref{b00015}), 
$$ x = \frac{\sqrt{q^2+4} - \sqrt{q^2} }{\sqrt{q^2+4} + \sqrt{q^2} } \ , $$ 
up to the finite term in the expansion in $(D-4)$: 
\bea
\hspace*{-5mm} F^{(2l)}_{1}(D,q^2) & = & 
       \frac{1}{(D-4)^2} \Bigg\{
          \frac{1}{2}
       -  \Bigg[
            1
          - \frac{1}{(1-x)}
          - \frac{1}{(1+x)}
          \Bigg] H(0;x) \nn \\
\hspace*{-5mm} & & \hspace*{20mm}
       +  \Bigg[
            1
          + \frac{1}{(1-x)^2}
          - \frac{1}{(1-x)}
          + \frac{1}{(1+x)^2} \nn \\
\hspace*{-5mm} & & \hspace*{25mm}
          - \frac{1}{(1+x)}
          \Bigg] H(0,0;x) 
       \Bigg\} \nn \\
\hspace*{-5mm} & - &
        \frac{1}{(D-4)} \Bigg\{
            1
          + \frac{1}{2} \Bigg[
                    1  
                  - \frac{1}{(1-x)}
                  - \frac{1}{(1+x)}
             \Bigg] \bigl[ \zeta(2)
                           + 2 H(-1,0;x) \bigr] \nn \\
\hspace*{-5mm} & & \hspace*{18mm}
       -  \Bigg[
            \frac{7}{4}
          - \frac{2}{(1-x)}
          - \frac{3}{2(1+x)}
          \Bigg] H(0;x) \nn \\
       \hspace*{-5mm} & & \hspace*{18mm}
       + \Bigg[
            1
          + \frac{2}{(1-x)^2}
          - \frac{3}{2(1-x)}
          + \frac{1}{(1+x)^2} \nn \\
       \hspace*{-5mm} & & \hspace*{23mm}
          - \frac{1}{2(1+x)}
         \Bigg] H(0,0;x) \nn \\
       \hspace*{-5mm} & & \hspace*{18mm}
       -  \frac{1}{2}  \Bigg[
            1 \! 
          +  \! \frac{1}{(1-x)^2} \! 
          -  \! \frac{1}{(1-x)} \! 
          +  \! \frac{1}{(1+x)^2} \! 
          -  \! \frac{1}{(1+x)} \! 
          \Bigg] \times \nn \\
       \hspace*{-5mm} & & \hspace*{23mm}
          \times \bigl[  H(0;x) \zeta(2) 
                 + 4 H(-1,0,0;x)
                 + H(0,-1,0;x)  \nn \\
       \hspace*{-5mm} & & \hspace*{28mm}
                 - 3 H(0,0,0;x)
          \bigr]
      \Bigg\} \nn \\
  \hspace*{-5mm} & &
          +  \! \frac{1387}{216} \! 
          - \frac{49}{9(1+x)} \biggl[ 
     1 \! 
   - \frac{1}{(1+x)} \biggr] \! 
       + \!  \Bigg[
            \frac{51}{16}  \! 
          +  \! \frac{1}{2(1-x)} \! 
          - \frac{82}{3(1+x)^4}  \nn \\
       \hspace*{-5mm} & &
          + \frac{200}{3(1+x)^3}
          - \frac{221}{6(1+x)^2}
          - \frac{33}{8(1+x)}
            \Bigg] \zeta(2)
       - \Bigg[
            3 
          + \frac{18}{(1+x)^2}  \nn \\
       \hspace*{-5mm} & &
          - \frac{18}{(1+x)}
            \Bigg] \zeta(2) \ln2
       - \Bigg[
            \frac{7}{4} 
          +  \frac{1}{(1-x)^2}
          - \frac{1}{2(1-x)}
          - \frac{42}{(1+x)^4} \nn \\
       \hspace*{-5mm} & &
          + \frac{84}{(1+x)^3}
          - \frac{95}{2(1+x)^2}
          + \frac{6}{(1+x)}
            \Bigg] \zeta(3)
      + \Bigg[
            \frac{181}{40}
          + \frac{61}{40(1-x)^2} \nn \\
       \hspace*{-5mm} & &
          -  \! \frac{1219}{320(1-x)} \! 
          -  \! \frac{171}{4(1+x)^5} \! 
          +  \! \frac{855}{8(1 \! + \! x)^4} \! 
          -  \! \frac{6867}{80(1 \! + \! x)^3} \! 
          +  \! \frac{749}{32(1 \! + \! x)^2} \! 
          \nn \\
          \hspace*{-5mm} & &
          - \frac{1731}{320(1+x)}
            \Bigg] \zeta^2(2) \nn \\
 \hspace*{-5mm} & &
       -  \Bigg[
            \frac{3355}{864} \! 
          -  \! \frac{985}{216(1-x)} \! 
          -  \! \frac{89}{9(1+x)^3} \! 
          + \!  \frac{89}{9(1 \! + \! x)^2} \! 
          -  \! \frac{3521}{432(1 \! + \! x)} \! 
          \nn \\
\hspace*{-5mm} & & \hspace*{5mm}
      + \Bigg(
            \frac{53}{12} 
          + \frac{2}{(1-x)^2}
          - \frac{19}{24(1-x)}
          + \frac{21}{(1+x)^5}
          - \frac{54}{(1+x)^4}
          \nn \\
\hspace*{-5mm} & & \hspace*{5mm}
          + \frac{45}{(1+x)^3}
          - \frac{3}{(1+x)^2}
          - \frac{349}{24(1+x)}
             \Bigg) \zeta(2)
      - \Bigg(
            2 
          - \frac{7}{4(1-x)}  \nn\\
\hspace*{-5mm} & & \hspace*{5mm}
          + \frac{42}{(1+x)^5}
          - \frac{105}{(1+x)^4}
          + \frac{91}{(1+x)^3}
          - \frac{63}{2(1+x)^2}  \nn\\
\hspace*{-5mm} & & \hspace*{5mm}
          + \frac{5}{4(1+x)}
             \Bigg) \zeta(3)
          \Bigg] H(0;x)  \nn \\
 \hspace*{-5mm} & &
       +  \Bigg[
            \frac{5}{2} 
          + \frac{1}{2(1-x)}
          + \frac{45}{(1+x)^4}
          - \frac{90}{(1+x)^3}
          + \frac{135}{2(1+x)^2} \nn \\
 \hspace*{-5mm} & & \hspace*{5mm} 
          - \frac{22}{(1+x)}
          \Bigg]  \zeta(2) H(-1;x) \nn \\
 \hspace*{-5mm} & &
       +  \Bigg[
            \frac{2137}{144} \! 
          +  \! \frac{4}{(1 \! - \! x)^2} \! 
          -  \! \frac{21}{2(1 \! - \! x)} \! 
          +  \! \frac{494}{9(1 \! + \! x)^4} \! 
          -  \! \frac{1258}{9(1 \! + \! x)^3} \! 
          -  \! \frac{1258}{9(1 \! + \! x)^3} \nn \\
\hspace*{-5mm} & & \hspace*{5mm} 
          + \frac{1130}{9(1+x)^2}
          - \frac{1081}{24(1+x)}
      + \Bigg(
            \frac{13}{4}
          + \frac{5}{4(1-x)^2}
          - \frac{105}{32(1-x)} \nn \\
\hspace*{-5mm} & & \hspace*{5mm} 
          - \frac{33}{2(1+x)^5}
          + \frac{165}{4(1+x)^4}
          - \frac{281}{8(1+x)^3}
          + \frac{203}{16(1+x)^2}
          \nn \\
\hspace*{-5mm} & & \hspace*{5mm} 
          - \frac{137}{32(1+x)}
            \Bigg) \zeta(2)
          \Bigg] H(0,0;x) \nn \\
 \hspace*{-5mm} & &
       -  \Bigg[
            \frac{55}{8} \! 
          -  \! \frac{9}{(1-x)} \! 
          -  \! \frac{48}{(1+x)^3} \! 
          +  \! \frac{72}{(1+x)^2} \! 
          -  \! \frac{115}{4(1+x)} \! 
          \Bigg] H(-1,0;x) \nn \\
 \hspace*{-5mm} & &
       -  \Bigg[
            \frac{5}{2}  \! 
          -  \! \frac{1}{2(1-x)^2} \! 
          -  \! \frac{19}{16(1-x)} \! 
          -  \! \frac{45}{(1 \! + \! x)^5} \! 
          +  \! \frac{225}{2(1 \! + \! x)^4} \! 
          -  \! \frac{363}{4(1 \! + \! x)^3} \nn \\
\hspace*{-5mm} & & \hspace*{5mm} 
          + \frac{185}{8(1+x)^2}
          - \frac{67}{16(1+x)}
          \Bigg] \zeta(2) H(0,-1;x) \nn \\
 \hspace*{-5mm} & &
       + \Bigg[
            4 \! 
          -  \! \frac{4}{(1-x)} \! 
          -  \! \frac{12}{(1+x)^3} \! 
          +  \! \frac{18}{(1+x)^2} \! 
          -  \! \frac{10}{(1+x)} \! 
       + \!  \Bigg(
            1 \! 
          +  \!  \frac{1}{(1-x)^2} \nn\\
\hspace*{-5mm} & & \hspace*{5mm}
          - \frac{3}{2(1-x)}
          - \frac{36}{(1+x)^5}
          + \frac{90}{(1+x)^4}
          - \frac{78}{(1+x)^3}
          + \frac{28}{(1+x)^2} \nn \\
\hspace*{-5mm} & & \hspace*{5mm}
          - \frac{7}{2(1+x)}
            \Bigg)  \zeta(2)
         \Bigg] H(1,0;x) \nn \\
\hspace*{-5mm} & &
       - \Bigg[
            1
          - \frac{1}{(1-x)}
          - \frac{1}{(1+x)}
         \Bigg]  H(-1,-1,0;x) \nn \\
\hspace*{-5mm} & & 
       - \Bigg[
            \frac{3}{2} \! 
          +  \! \frac{4}{(1-x)^2} \! 
          -  \! \frac{7}{2(1-x)} \! 
          +  \! \frac{63}{(1+x)^4} \! 
          -  \! \frac{126}{(1+x)^3} \! 
          +  \! \frac{135}{2(1+x)^2}  \nn \\
\hspace*{-5mm} & & \hspace*{5mm}
          - \frac{4}{(1+x)}
         \Bigg] H(-1,0,0;x) \nn \\
\hspace*{-5mm} & &
       + \Bigg[
            4
          - \frac{1}{2(1-x)}
          + \frac{96}{(1+x)^4}
          - \frac{192}{(1+x)^3}
          + \frac{221}{2(1+x)^2} \nn \\
\hspace*{-5mm} & & \hspace*{5mm}
          - \frac{15}{(1+x)}
         \Bigg] H(0,-1,0;x) \nn \\
\hspace*{-5mm} & &
       - \Bigg[
            \frac{8}{3} \! 
          -  \! \frac{1}{(1-x)^2} \! 
          +  \! \frac{41}{24(1-x)} \! 
          +  \! \frac{21}{(1+x)^5} \! 
          -  \! \frac{69}{(1+x)^4} \! 
          +  \! \frac{75}{(1+x)^3} \nn \\
\hspace*{-5mm} & & \hspace*{5mm}
          - \frac{95}{4(1+x)^2}
          - \frac{223}{24(1+x)}
         \Bigg] H(0,0,0;x) \nn \\
 \hspace*{-5mm} & &
       - \Bigg[
            \frac{5}{2}
          + \frac{2}{(1-x)^2}
          - \frac{2}{(1-x)}
          + \frac{24}{(1+x)^4}
          - \frac{49}{(1+x)^3}
          + \frac{29}{(1+x)^2} \nn\\
\hspace*{-5mm} & & \hspace*{5mm}
          - \frac{5}{(1+x)}
         \Bigg] H(0,1,0;x) \nn \\
 \hspace*{-5mm} & &
       + \Bigg[
            4 \! 
          +  \! \frac{90}{(1+x)^4} \! 
          -  \! \frac{180}{(1+x)^3} \! 
          +  \! \frac{197}{2(1+x)^2} \! 
          -  \! \frac{17}{2(1+x)} \! 
         \Bigg] H(1,0,0;x) \nn \\
 \hspace*{-5mm} & &
       - \Bigg[
            \frac{7}{2} \! 
          +  \! \frac{5}{2(1-x)^2} \! 
          -  \! \frac{67}{16(1-x)} \! 
          +  \! \frac{63}{(1 \! + \! x)^5} \! 
          -  \! \frac{315}{2(1 \! + \! x)^4} \! 
          +  \! \frac{517}{4(1 \! + \! x)^3} \nn \\
\hspace*{-5mm} & & \hspace*{5mm}
          - \frac{271}{8(1+x)^2} 
          - \frac{19}{16(1+x)}
         \Bigg] H(0,-1,0,0;x) \nn \\
 \hspace*{-5mm} & &
       - \Bigg[
            \frac{11}{2} \! 
          +  \! \frac{5}{2(1-x)^2} \! 
          -  \! \frac{21}{8(1-x)} \! 
          -  \! \frac{96}{(1 \! + \! x)^5} \! 
          +  \! \frac{240}{(1 \! + \! x)^4} \! 
          -  \! \frac{373}{2(1 \! + \! x)^3} \nn \\
\hspace*{-5mm} & & \hspace*{5mm}
          + \frac{169}{4(1+x)^2}
          - \frac{45}{8(1+x)}
         \Bigg] H(0,0,-1,0;x)  \nn \\
 \hspace*{-5mm} & &
       + \Bigg[
            \frac{27}{4} \! 
          +  \! \frac{17}{4(1 \! - \! x)^2} \! 
          -  \! \frac{217}{32(1 \! - \! x)} \! 
          -  \! \frac{3}{2(1 \! + \! x)^5} \! 
          +  \! \frac{15}{4(1 \! + \! x)^4} \! 
          -  \! \frac{17}{8(1 \! + \! x)^3} \nn \\
\hspace*{-5mm} & & \hspace*{5mm}
          + \frac{59}{16(1+x)^2}
          - \frac{201}{32(1+x)}
         \Bigg] H(0,0,0,0;x) \nn \\
 \hspace*{-5mm} & &
       + \Bigg[
            3
          + \frac{1}{(1-x)^2}
          - \frac{2}{(1-x)}
          - \frac{24}{(1+x)^5}
          + \frac{60}{(1+x)^4}
          - \frac{44}{(1+x)^3}  \nn \\
\hspace*{-5mm} & & \hspace*{5mm}
          + \frac{7}{(1+x)^2}
          - \frac{2}{(1+x)}
         \Bigg] H(0,0,1,0;x) \nn \\
 \hspace*{-5mm} & &
       - \Bigg[
            1 \! 
          +  \! \frac{7}{4(1-x)} \! 
          -  \! \frac{90}{(1+x)^5} \! 
          +  \! \frac{225}{(1+x)^4} \! 
          -  \! \frac{175}{(1+x)^3} \! 
          +  \! \frac{75}{2(1+x)^2} \nn \\
\hspace*{-5mm} & & \hspace*{5mm}
          - \frac{5}{4(1+x)}
         \Bigg] H(0,1,0,0;x) \nn \\
 \hspace*{-5mm} & &
       + \Bigg[
            2
          + \frac{2}{(1-x)^2}
          - \frac{5}{2(1-x)}
          - \frac{36}{(1+x)^5}
          + \frac{90}{(1+x)^4}
          - \frac{78}{(1+x)^3}  \nn \\
\hspace*{-5mm} & & \hspace*{5mm}
          + \frac{29}{(1+x)^2}
          - \frac{9}{2(1+x)}
         \Bigg] H(1,0,0,0;x)  \nn \\
 \hspace*{-5mm} & &
       +  \Bigg[
            1
          + \frac{1}{(1-x)^2}
          - \frac{1}{(1-x)}
          + \frac{1}{(1+x)^2}
          - \frac{1}{(1+x)}
          \Bigg]  \big[ \zeta(3) H(1;x)  \nn\\
\hspace*{-5mm} & & \hspace*{5mm}
                        +  \! \zeta(2) H(-1,0;x) \! 
                        +  \! 4 H( \! -1, \! -1,0,0;x) \! 
                        +  \! 2 H( \! -1,0, \! -1,0;x) \nn \\
\hspace*{-5mm} & & \hspace*{5mm}
                        -  \! 3 H(-1,0,0,0;x)  \! 
                        +  \! H(0,-1, \! -1,0;x) \! 
                        -  \! 2 H(1,0, \! -1,0;x) \nn \\
\hspace*{-5mm} & & \hspace*{5mm}
                        + 2 H(1,0,1,0;x)
                 \big] \nn \\
\hspace*{-5mm} & + & {\mathcal O}(D-4)
\label{FF1fr} \ , \\
\hspace*{-5mm} F^{(2l)}_{2}(D,q^2) & = &  
        \frac{1}{(D-4)}  \Bigg\{
          \frac{1}{2} \left[ \frac{1}{(1+x)} - \frac{1}{(1-x)} 
                     \right] H(0;x) 
       + \Bigg[
             \frac{1}{(1+x)^2}
             \nn \\
\hspace*{-5mm} & & \hspace*{18mm}
          -  \frac{1}{(1+x)}
          -  \frac{1}{(1-x)^2}
          +  \frac{1}{(1-x)}
          \Bigg] H(0,0;x) 
          \Bigg\} \nn \\
 \hspace*{-5mm} & &
       - \Bigg[
            \frac{17}{8(1-x)} \! 
          -   \!  \frac{26}{(1+x)^4} \! 
          +  \!   \frac{64}{(1+x)^3} \! 
          -   \!  \frac{43}{(1+x)^2} \! 
          +  \! \frac{23}{8(1+x)} \! 
         \Bigg] \zeta(2)  \nn \\
\hspace*{-5mm} & &
       + \Bigg[
            \frac{12}{(1+x)^2}
          - \frac{12}{(1+x)}
         \Bigg] \zeta(2) \ln2
       - \Bigg[
            \frac{69}{80(1-x)^3}
          - \frac{207}{160(1-x)^2} \nn \\
\hspace*{-5mm} & &
          + \frac{327}{320(1-x)}
          - \frac{171}{4(1+x)^5}
          + \frac{855}{8(1+x)^4}
          - \frac{3291}{40(1+x)^3} \nn \\
\hspace*{-5mm} & &
          + \frac{1323}{80(1+x)^2}
          + \frac{327}{320(1+x)}
         \Bigg] \zeta^2(2)
       - \Bigg[
            \frac{5}{2(1-x)^2}
          - \frac{5}{2(1-x)} \nn \\
\hspace*{-5mm} & &
          +   \frac{42}{(1+x)^4}
          -   \frac{84}{(1+x)^3}
          + \frac{91}{2(1+x)^2}
          - \frac{7}{2(1+x)}
         \Bigg] \zeta(3) \nn \\
\hspace*{-5mm} & &
          + \frac{17}{3(1+x)} \biggl[ 
     1
   - \frac{1}{(1+x)} \biggr] \nn \\
 \hspace*{-5mm} & & 
       -  \Bigg[
             \Bigg(
            \frac{19}{8(1-x)^2} \! 
          -  \! \frac{75}{16(1-x)} \! 
          -  \! \frac{21}{(1+x)^5} \! 
          +  \! \frac{54}{(1+x)^4} \! 
          -  \! \frac{173}{4(1+x)^3}  \nn \\
\hspace*{-5mm} & &  \hspace*{5mm}
          +  \! \frac{13}{4(1 \! + \! x)^2} \! 
          +  \! \frac{149}{16(1 \! + \! x)} \! 
             \Bigg) \zeta(2) \! 
      +  \! \Bigg(
            \frac{7}{8(1-x)} \! 
          +  \!   \frac{42}{(1 \! + \! x)^5} \! 
          -  \!  \frac{105}{(1 \! + \! x)^4} \nn \\
\hspace*{-5mm} & & \hspace*{5mm}
          + \frac{175}{2(1+x)^3} 
          - \frac{105}{4(1+x)^2}
          + \frac{7}{8(1+x)}
             \Bigg) \zeta(3)
          - \frac{7}{144(1-x)}  \nn \\
\hspace*{-5mm} & & \hspace*{5mm}
          + \frac{31}{3(1+x)^3}
          - \frac{31}{2(1+x)^2}
          + \frac{751}{144(1+x)}
          \Bigg] H(0;x) \nn \\
 \hspace*{-5mm} & &
       +  \Bigg[
            \frac{9}{4(1-x)^2}
          - \frac{9}{4(1-x)}
          - \frac{45}{(1+x)^4}
          + \frac{90}{(1+x)^3}
          - \frac{231}{4(1+x)^2} \nn \\
\hspace*{-5mm} & &  \hspace*{5mm}
          + \frac{51}{4(1+x)}
          \Bigg] \zeta(2) H(-1;x) \nn \\
 \hspace*{-5mm} & &
       -  \Bigg[
            \frac{9}{4(1-x)}
          + \frac{48}{(1+x)^3}
          - \frac{72}{(1+x)^2}
          + \frac{87}{4(1+x)}
          \Bigg]  H(-1,0;x) \nn \\
\hspace*{-5mm} & &
       +  \Bigg[
            \frac{9}{4(1-x)^3} \! 
          - \frac{27}{8(1-x)^2} \! 
          + \frac{21}{16(1-x)} \! 
          - \frac{45}{(1+x)^5} \! 
          + \frac{225}{2(1+x)^4} \nn \\
\hspace*{-5mm} & &  \hspace*{5mm}
          - \frac{87}{(1+x)^3}
          + \frac{18}{(1+x)^2}
          + \frac{21}{16(1+x)}
          \Bigg] \zeta(2) H(0,-1;x) \nn \\
 \hspace*{-5mm} & &
       -  \Bigg[
             \Bigg(
            \frac{7}{8(1 \! - \! x)^3} \! 
          -  \! \frac{21}{16(1 \! - \! x)^2} \! 
          +  \! \frac{1}{32(1 \! - \! x)} \! 
          -  \! \frac{33}{2(1 \! + \! x)^5} \! 
          +  \! \frac{165}{4(1 \! + \! x)^4}  \nn \\
\hspace*{-5mm} & &  \hspace*{5mm}
          - \frac{135}{4(1+x)^3}
          + \frac{75}{8(1+x)^2}
          + \frac{1}{32(1+x)}
             \Bigg) \zeta(2)
          - \frac{1}{2(1-x)^2} \nn \\
\hspace*{-5mm} & &  \hspace*{5mm}
          - \frac{1}{8(1-x)}
          + \frac{166}{3(1+x)^4}
          - \frac{422}{3(1+x)^3}
          + \frac{709}{6(1+x)^2} \nn \\
\hspace*{-5mm} & &  \hspace*{5mm}
          - \frac{773}{24(1+x)}
          \Bigg] H(0,0;x) \nn \\
\hspace*{-5mm} & &
       +  \Bigg[
             \Bigg(
            \frac{3}{4(1-x)}
          + \frac{36}{(1+x)^5}
          - \frac{90}{(1+x)^4}
          + \frac{75}{(1+x)^3}
          - \frac{45}{2(1+x)^2} \nn \\
\hspace*{-5mm} & &  \hspace*{5mm}
          + \frac{3}{4(1+x)}
             \Bigg) \zeta(2)
          - \frac{1}{(1-x)}
          + \frac{12}{(1+x)^3}
          - \frac{18}{(1+x)^2} \nn \\
\hspace*{-5mm} & &  \hspace*{5mm}
          + \frac{7}{(1+x)} 
          \Bigg]  H(1,0;x) \nn \\
 \hspace*{-5mm} & &
       +  \Bigg[
            \frac{1}{4(1-x)^2}
          - \frac{1}{4(1-x)}
          + \frac{63}{(1+x)^4}
          - \frac{126}{(1+x)^3}
          + \frac{257}{4(1+x)^2}  \nn \\
       \hspace*{-5mm} & &  \hspace*{5mm}
          - \frac{5}{4(1+x)}
          \Bigg] H(-1,0,0;x) \nn \\
 \hspace*{-5mm} & &
       -  \Bigg[
            \frac{1}{2(1-x)^2}
          - \frac{1}{2(1-x)}
          +  \frac{96}{(1+x)^4}
          -  \frac{192}{(1+x)^3}
          + \frac{197}{2(1+x)^2}  \nn \\
\hspace*{-5mm} & &  \hspace*{5mm}
          - \frac{5}{2(1+x)}
          \Bigg] H(0,-1,0;x) \nn \\
 \hspace*{-5mm} & &
       - \Bigg[
            \frac{7}{8(1-x)^2}
          - \frac{51}{16(1-x)}
          -  \frac{21}{(1+x)^5}
          +  \frac{69}{(1+x)^4} 
          - \frac{293}{4(1+x)^3}\nn \\
\hspace*{-5mm} & &  \hspace*{5mm}
          + \frac{91}{4(1+x)^2}
          + \frac{77}{16(1+x)}
          \Bigg] H(0,0,0;x) \nn \\
 \hspace*{-5mm} & &
       - \Bigg[
            \frac{1}{(1-x)^2}
          - \frac{1}{(1-x)}
          -  \frac{24}{(1+x)^4}
          +  \frac{48}{(1+x)^3}
          -  \frac{25}{(1+x)^2} \nn \\
\hspace*{-5mm} & &  \hspace*{5mm}
          + \frac{1}{(1+x)}
         \Bigg] H(0,1,0;x) \nn \\
 \hspace*{-5mm} & &
       - \Bigg[
            \frac{1}{(1-x)^2}
          - \frac{1}{(1-x)}
          +  \frac{90}{(1+x)^4}
          -  \frac{180}{(1+x)^3}
          + \frac{89}{(1+x)^2} \nn \\
\hspace*{-5mm} & &  \hspace*{5mm}
          + \frac{1}{(1+x)}
         \Bigg] H(1,0,0;x) \nn \\
 \hspace*{-5mm} & &
       + \Bigg[
            \frac{1}{4(1-x)^3} \! 
          - \frac{3}{8(1-x)^2} \! 
          - \frac{7}{16(1-x)} \! 
          +  \! \frac{63}{(1+x)^5} \! 
          - \frac{315}{2(1+x)^4}  \nn \\
\hspace*{-5mm} & &  \hspace*{5mm}
          + \frac{124}{(1+x)^3}
          - \frac{57}{2(1+x)^2}
          - \frac{7}{16(1+x)}
         \Bigg] H(0,-1,0,0;x) \nn \\
 \hspace*{-5mm} & &
       + \Bigg[
            \frac{1}{2(1-x)^3}
          - \frac{3}{4(1-x)^2}
          + \frac{7}{2(1-x)}
          - \frac{96}{(1+x)^5}
          + \frac{240}{(1+x)^4}  \nn \\
\hspace*{-5mm} & &  \hspace*{5mm}
          - \frac{357}{2(1+x)^3}
          + \frac{111}{4(1+x)^2} 
          + \frac{7}{2(1+x)}
         \Bigg] H(0,0,-1,0;x) \nn \\
 \hspace*{-5mm} & &
       - \Bigg[
            \frac{3}{8(1-x)^3} \! 
          -  \! \frac{9}{16(1-x)^2} \! 
          +  \! \frac{11}{32(1-x)} \! 
          -  \! \frac{3}{2(1 \! + \! x)^5}  \! 
          +  \! \frac{15}{4(1 \! + \! x)^4} \nn \\
\hspace*{-5mm} & &  \hspace*{5mm}
          - \frac{2}{(1+x)^3}
          - \frac{3}{4(1+x)^2}
          + \frac{11}{32(1+x)}
         \Bigg] H(0,0,0,0;x) \nn \\
 \hspace*{-5mm} & &
       - \Bigg[
            \frac{3}{2(1-x)}
          - \frac{24}{(1+x)^5}
          + \frac{60}{(1+x)^4}
          - \frac{42}{(1+x)^3}
          + \frac{3}{(1+x)^2} \nn \\
\hspace*{-5mm} & &  \hspace*{5mm}
          + \frac{3}{2(1+x)}
         \Bigg] H(0,0,1,0;x) \nn \\
 \hspace*{-5mm} & &
       + \Bigg[
            \frac{25}{8(1-x)}
          -  \frac{90}{(1+x)^5}
          +  \frac{225}{(1+x)^4}
          - \frac{335}{2(1+x)^3}
          + \frac{105}{4(1+x)^2} \nn \\
\hspace*{-5mm} & &  \hspace*{5mm}
          + \frac{25}{8(1+x)}
         \Bigg] H(0,1,0,0;x) \nn \\
 \hspace*{-5mm} & &
       + \Bigg[
            \frac{3}{4(1-x)}
          +  \frac{36}{(1+x)^5}
          -  \frac{90}{(1+x)^4}
          +  \frac{75}{(1+x)^3}
          - \frac{45}{2(1+x)^2}  \nn \\
\hspace*{-5mm} & &  \hspace*{5mm}
          + \frac{3}{4(1+x)}
         \Bigg] H(1,0,0,0;x)  \nn \\
\hspace*{-5mm} & + & {\mathcal O}(D-4) \ .
\label{FF2fr}
\eea

Even after the full renormalization has been carried out, the on-shell 
renormalized form factors still develop polar singularities in $(D-4)$, 
due to soft IR divergences. These divergences are not physical and are 
removed in any physical process by the corresponding divergences due 
to soft real emission.

As it is easy to check explicitly, the IR divergences are the same as 
in~\cite{2loop1,Pie}, provided the following formal replacement is done 
\be
\log \left( \frac{\lambda}{m} \right) = - \frac{1}{(D-4)} \, , 
\ee 
so that they follow the general structure already pointed out in~\cite{YFS}. 
The finite parts are however different in the $D$-dimensional 
and in the $\lambda$-mass regularization schemes.

\subsection{Continuation to time-like momentum transfer and imaginary 
parts \label{im}}

The expressions of the UV-renormalized form factors given in the
previous section, Eqs. (\ref{FF1fr},\ref{FF2fr}), can be analytically 
continued to the time-like region, $S = -Q^2 > 0$, and in particular 
above the physical threshold $S>4m^2$, where the form factors 
develop an imaginary part. 

The analytic continuation to $S+i\epsilon$ for $S>4m^2$ is performed 
with the substitution 
\be
x = - y + i \epsilon \, ,
\label{analyticY}
\ee
where now:
\be
y = \frac{\sqrt{s} - \sqrt{s-4}    }{\sqrt{s} + \sqrt{s-4} } \,
  = \frac{\sqrt{S} - \sqrt{S-4m^2} }{\sqrt{S} + \sqrt{S-4m^2} } \, , 
\label{Yins}
\ee 
with $S=m^2 s$ according to Eq. (\ref{defq2ands}). 

The imaginary part of the form factors is originated by the imaginary parts
developed by the polylogarithms with rightmost index equal to 0 (for
details see \cite{Pie,Polylog,Polylog3}).

In the kinematical region above threshold the form factors can be written as 
\bea
F^{(2l)}_{1}(D,-s-i\epsilon) & = & \Re \, F^{(2l)}_{1}(D,-s) 
                               + i \pi \, \Im \, F^{(2l)}_{1}(D,-s) \, , \\
F^{(2l)}_{2}(D,-s-i\epsilon) & = & \Re \, F^{(2l)}_{2}(D,-s) 
                               + i \pi \, \Im \, F^{(2l)}_{2}(D,-s) \, ; 
\eea
for short, we give the explicit expressions of the imaginary parts only: 
\bea
\hspace*{-5mm} \Im \, F^{(2l)}_{1}(D,-s) & = & 
     \frac{1}{(D-4)^2} 
       \Bigg\{
          - 1
          + \frac{1}{(1-y)}
          + \frac{1}{(1+y)} 
          + \Bigg[
              1
            + \frac{1}{(1-y)^2}
          \nn \\ 
\hspace*{-5mm} & & \hspace*{20mm} 
            - \frac{1}{(1-y)}
            + \frac{1}{(1+y)^2}
            - \frac{1}{(1+y)}
          \Bigg] H(0;y)
       \Bigg\} \nn \\
\hspace*{-5mm} & + &
       \frac{1}{(D-4)}
       \Bigg\{
            \frac{7}{4}
          - \frac{3}{2(1-y)}
          - \frac{2}{(1+y)}
          - \Bigg[
            1
          + \frac{1}{(1-y)^2}
          \nn \\ 
\hspace*{-5mm} & & \hspace*{18mm} 
          - \frac{1}{2(1-y)}
          + \frac{2}{(1+y)^2}
          - \frac{3}{2(1+y)}
          \Bigg] H(0;y)
          \nn \\ 
\hspace*{-5mm} & & \hspace*{18mm} 
          + \Bigg[
            1
          - \frac{1}{(1-y)}
          - \frac{1}{(1+y)}
          \Bigg] H(1;y)
          \nn \\ 
\hspace*{-5mm} & & \hspace*{18mm} 
          + \frac{1}{2} \Bigg[
            1 \! 
          +  \! \frac{1}{(1 \! - \! y)^2} \! 
          -  \! \frac{1}{(1 \! - \! y)} \! 
          +  \! \frac{1}{(1 \! + \! y)^2} \! 
          -  \! \frac{1}{(1 \! + \! y)} \! 
          \Bigg]  \! \times \nn \\ 
\hspace*{-5mm} & & \hspace*{25mm} 
           \times \big[
            4 \zeta(2)
          -  3 H(0,0;y)
          -  2 H(0,1;y) \nn \\ 
\hspace*{-5mm} & & \hspace*{25mm} 
          -  4 H(1,0;y)
          \big]
       \Bigg\} \nn \\
\hspace*{-5mm} & &
          -  \! \frac{3355}{864} \! 
          +  \! \frac{89}{9(1-y)^3} \! 
          -  \! \frac{89}{6(1-y)^2} \! 
          +  \! \frac{3521}{432(1-y)} \! 
          +  \! \frac{985}{216(1 \! + \! y)} \nn \\ 
\hspace*{-5mm} & & 
       - \Bigg(
            \frac{7}{4}  \! 
          +  \!  \frac{15}{(1-y)^4} \! 
          -  \!  \frac{30}{(1-y)^3} \! 
          +  \! \frac{83}{4(1-y)^2} \! 
          -  \! \frac{21}{4(1-y)} \! 
          +  \! \frac{3}{(1 \! + \! y)^2} \nn \\ 
\hspace*{-5mm} & & 
          - \frac{5}{2(1+y)}
            \Bigg) \zeta(2) 
       + \Bigg(
            2 
          +  \frac{42}{(1-y)^5}
          -  \frac{105}{(1-y)^4}
          +  \frac{91}{(1-y)^3} \nn \\ 
\hspace*{-5mm} & & 
          - \frac{63}{2(1-y)^2}
          + \frac{5}{4(1-y)}
          - \frac{7}{4(1+y)}
            \Bigg) \zeta(3) \nn \\ 
\hspace*{-5mm} & & 
       + \Bigg[
            \frac{2137}{144}
          + \frac{494}{9(1-y)^4}
          - \frac{1258}{9(1-y)^3}
          + \frac{1130}{9(1-y)^2}
          - \frac{1081}{24(1-y)}\nn \\
\hspace*{-5mm} & & \hspace*{5mm}
          + \frac{4}{(1+y)^2} 
          - \frac{21}{2(1+y)}
          - \Bigg(
            \frac{7}{2} 
          + \frac{15}{(1-y)^5}
          - \frac{75}{2(1-y)^4} \nn \\
\hspace*{-5mm} & & \hspace*{5mm}
          + \frac{33}{(1-y)^3}
          - \frac{9}{(1-y)^2}
          - \frac{2}{(1-y)}
          + \frac{3}{(1+y)^2}  \nn \\
\hspace*{-5mm} & & \hspace*{5mm}
          - \frac{7}{2(1+y)}
            \Bigg) \zeta(2)
          \Bigg] H(0;y) \nn \\
\hspace*{-5mm} & &
       - \Bigg[
            4
          - \frac{12}{(1-y)^3}
          + \frac{18}{(1-y)^2}
          - \frac{10}{(1-y)}
          - \frac{4}{(1+y)}
         \Bigg] H(-1;y) \nn \\
\hspace*{-5mm} & &
       + \Bigg[
            \frac{55}{8} \! 
          -  \frac{48}{(1-y)^3} \! 
          +  \!  \frac{72}{(1-y)^2} \! 
          - \frac{115}{4(1-y)} \! 
          - \frac{9}{(1+y)} \! 
         \Bigg] H(1;y) \nn \\
\hspace*{-5mm} & &
       - \Bigg[
            4 \! 
          +  \!  \frac{90}{(1-y)^4} \! 
          -  \!  \frac{180}{(1-y)^3} \! 
          +  \! \frac{197}{2(1 \! - \! y)^2} \! 
          -  \! \frac{17}{2(1 \! - \! y)} \! 
         \Bigg] H(-1,0;y) \nn \\
\hspace*{-5mm} & &
       + \Bigg[
            \frac{5}{2} \! 
          +  \!  \frac{24}{(1-y)^4} \! 
          -  \!  \frac{48}{(1-y)^3} \! 
          +  \!  \frac{29}{(1-y)^2} \! 
          -  \!  \frac{5}{(1-y)} \! 
          +  \!  \frac{2}{(1+y)^2} \nn \\
\hspace*{-5mm} & & \hspace*{5mm}
          -  \frac{2}{(1+y)}
         \Bigg]  H(0,-1;y) \nn \\
\hspace*{-5mm} & & 
       - \Bigg[
            \frac{8}{3} \! 
          +  \!  \frac{21}{(1 \! - \! y)^5} \! 
          -  \!  \frac{69}{(1 \! - \! y)^4} \! 
          +  \!  \frac{75}{(1 \! - \! y)^3} \! 
          -  \! \frac{95}{4(1 \! - \! y)^2} \! 
          -  \! \frac{223}{24(1 \! - \! y)} \nn \\
\hspace*{-5mm} & & \hspace*{5mm}
          - \frac{1}{(1+y)^2} 
          + \frac{41}{24(1+y)}
         \Bigg] H(0,0;y) \nn \\
\hspace*{-5mm} & &
       - \Bigg[
            4
          +  \frac{96}{(1-y)^4}
          -  \frac{192}{(1-y)^3}
          + \frac{221}{2(1-y)^2} 
          -  \frac{15}{(1-y)} \nn \\
\hspace*{-5mm} & & \hspace*{5mm}
          - \frac{1}{2(1+y)}
         \Bigg] H(0,1;y) \nn \\
\hspace*{-5mm} & &
       + \Bigg[
            \frac{3}{2} \! 
          +   \! \frac{63}{(1-y)^4} \! 
          -  \!  \frac{126}{(1-y)^3} \! 
          +  \! \frac{135}{2(1-y)^2} \! 
          -  \! \frac{4}{(1-y)} \! 
          +  \! \frac{4}{(1+y)^2} \nn \\
\hspace*{-5mm} & & \hspace*{5mm}
          - \frac{7}{2(1+y)}
         \Bigg] H(1,0;y) \nn \\
\hspace*{-5mm} & &
       - \Bigg[
            1
          - \frac{1}{(1-y)}
          - \frac{1}{(1+y)}
         \Bigg] H(1,1;y) \nn \\
\hspace*{-5mm} & & 
       - \Bigg[
            2 \! 
          -  \!  \frac{36}{(1-y)^5} \! 
          +  \!  \frac{90}{(1-y)^4} \! 
          -  \!  \frac{78}{(1-y)^3}  \! 
          +  \! \frac{29}{(1-y)^2} \! 
          -  \! \frac{9}{2(1-y)}  \nn \\
\hspace*{-5mm} & & \hspace*{5mm}
          + \frac{2}{(1+y)^2}
          - \frac{5}{2(1+y)}
         \Bigg]  H(-1,0,0;y) \nn \\
\hspace*{-5mm} & &
       + \Bigg[
            1
          -  \frac{90}{(1-y)^5}
          +  \frac{225}{(1-y)^4}
          -  \frac{175}{(1-y)^3}
          + \frac{75}{2(1-y)^2} \nn \\
\hspace*{-5mm} & & \hspace*{5mm}
          - \frac{5}{4(1-y)}
          + \frac{7}{4(1+y)}
         \Bigg] H(0,-1,0;y) \nn \\
\hspace*{-5mm} & &
       - \Bigg[
            3 \! 
          -   \! \frac{24}{(1-y)^5} \! 
          +   \! \frac{60}{(1-y)^4} \! 
          -  \!  \frac{44}{(1-y)^3}  \! 
          +  \! \frac{7}{(1-y)^2} \! 
          -  \! \frac{2}{(1-y)} \nn \\
\hspace*{-5mm} & & \hspace*{5mm}
          + \frac{1}{(1+y)^2} 
          - \frac{2}{(1+y)}
         \Bigg] H(0,0,-1;y) \nn \\
\hspace*{-5mm} & &
       + \Bigg[
            \frac{27}{4}
          - \frac{3}{2(1-y)^5}
          + \frac{15}{4(1-y)^4}
          - \frac{17}{8(1-y)^3} 
          + \frac{59}{16(1-y)^2} \nn \\
\hspace*{-5mm} & & \hspace*{5mm}
          - \frac{201}{32(1-y)}
          + \frac{17}{4(1+y)^2} 
          - \frac{217}{32(1+y)}
         \Bigg] H(0,0,0;y) \nn \\
\hspace*{-5mm} & &
       + \Bigg[
            \frac{11}{2}
          - \frac{96}{(1-y)^5}
          + \frac{240}{(1-y)^4}
          - \frac{373}{2(1-y)^3}
          + \frac{169}{4(1-y)^2} \nn \\
\hspace*{-5mm} & & \hspace*{5mm}
          - \frac{45}{8(1-y)}
          + \frac{5}{2(1+y)^2}
          - \frac{21}{8(1+y)}
         \Bigg] H(0,0,1;y) \nn \\
\hspace*{-5mm} & &
       + \Bigg[
            \frac{7}{2}
          + \frac{63}{(1-y)^5}
          - \frac{315}{2(1-y)^4}
          + \frac{517}{4(1-y)^3} 
          - \frac{271}{8(1-y)^2} \nn \\
\hspace*{-5mm} & & \hspace*{5mm}
          - \frac{19}{16(1-y)}
          + \frac{5}{2(1+y)^2} 
          - \frac{67}{16(1+y)}
         \Bigg] H(0,1,0;y) \nn \\
\hspace*{-5mm} & &
       + \Bigg[
            1 
          + \frac{1}{(1-y)^2}
          - \frac{1}{(1-y)}
          + \frac{1}{(1+y)^2}
          - \frac{1}{(1+y)}
         \Bigg] \times \nn \\
\hspace*{-5mm} & & \hspace*{5mm}
         \times \big[ 
             \zeta(2) H(-1;y)
           - 4 \zeta(2) H(1;y)
           + 2 H(-1,0,-1;y)    \nn \\
\hspace*{-5mm} & & \hspace*{10mm}
           - 2 H(-1,0,1;y)
           + H(0,1,1;y)
           + 3 H(1,0,0;y) \nn \\
\hspace*{-5mm} & & \hspace*{10mm}
           + 2 H(1,0,1;y)
           + 4 H(1,1,0;y)
         \big] \nn \\
\hspace*{-5mm} & + & {\mathcal O}(D-4) \, , \\
\hspace*{-5mm} \Im \, F^{(2l)}_{2}(D,-s) & = & 
      \frac{1}{(D-4)}
       \Bigg\{
            \frac{1}{2(1-y)}
          - \frac{1}{2(1+y)}
          + \Bigg[
            \frac{1}{(1-y)^2}
          - \frac{1}{(1-y)} \nn \\
\hspace*{-5mm} & & \hspace*{18mm}
          - \frac{1}{(1+y)^2}
          + \frac{1}{(1+y)}
            \Bigg] H(0;y) 
       \Bigg\} \nn \\
\hspace*{-5mm} & &     
          - \frac{31}{3(1-y)^3}
          + \frac{31}{2(1-y)^2}
          - \frac{751}{144(1-y)}
          + \frac{7}{144(1+y)} \nn \\
\hspace*{-5mm} & & 
         + \Bigg[
            \frac{15}{(1-y)^4} \! 
          -  \! \frac{30}{(1-y)^3} \! 
          +  \! \frac{39}{2(1-y)^2} \! 
          -  \! \frac{9}{2(1-y)} \! 
          -  \!  \frac{3}{2(1+y)^2} \nn \\
\hspace*{-5mm} & &
          +  \frac{3}{2(1+y)}
           \Bigg] \zeta(2)
         - \Bigg[
             \frac{42}{(1-y)^5}
          -  \frac{105}{(1-y)^4}
          + \frac{175}{2(1-y)^3}  \nn \\
\hspace*{-5mm} & & 
          - \frac{105}{4(1-y)^2} 
          + \frac{7}{8(1-y)}
          + \frac{7}{8(1+y)}
           \Bigg] \zeta(3) \nn \\
\hspace*{-5mm} & &
       - \Bigg[
            \frac{12}{(1 \! - \! y)^3} \! 
          - \frac{18}{(1 \! - \! y)^2} \! 
          + \frac{7}{(1 \! - \! y)} \! 
          - \frac{1}{(1 \! + \! y)} \! 
         \Bigg] H(-1;y) \nn \\
\hspace*{-5mm} & &
       - \Bigg[
            \frac{166}{3(1 \! - \! y)^4} \! 
          - \!  \frac{422}{3(1 \! - \! y)^3} \! 
          +  \! \frac{709}{6(1 \! - \! y)^2} \! 
          -  \! \frac{773}{24(1 \! - \! y)} \! 
          -  \! \frac{1}{2(1 \! + \! y)^2} \nn \\
\hspace*{-5mm} & & \hspace*{5mm}
          - \frac{1}{8(1+y)}
          - \Bigg(
            \frac{15}{(1-y)^5}
          - \frac{75}{2(1-y)^4}
          + \frac{127}{4(1-y)^3} \nn \\
\hspace*{-5mm} & & \hspace*{5mm}
          - \frac{81}{8(1-y)^2} 
          + \frac{5}{16(1-y)}
          - \frac{1}{2(1+y)^3}
          + \frac{3}{4(1+y)^2} \nn \\
\hspace*{-5mm} & & \hspace*{5mm}
          + \frac{5}{16(1+y)}
            \Bigg) \zeta(2)
         \Bigg] H(0;y) \nn \\
\hspace*{-5mm} & &
       + \Bigg[
            \frac{48}{(1-y)^3}
          - \frac{72}{(1-y)^2}
          + \frac{87}{4(1-y)} 
          + \frac{9}{4(1+y)}
         \Bigg] H(1;y) \nn \\
\hspace*{-5mm} & &
       + \Bigg[
             \frac{90}{(1-y)^4}
          -  \frac{180}{(1-y)^3}
          +  \frac{89}{(1-y)^2}
          + \frac{1}{(1-y)}
          + \frac{1}{(1+y)^2} \nn \\
\hspace*{-5mm} & & \hspace*{5mm}
          - \frac{1}{(1+y)}
         \Bigg]  H(-1,0;y) \nn \\
\hspace*{-5mm} & &
       - \Bigg[
             \frac{24}{(1-y)^4}
          -  \frac{48}{(1-y)^3}
          +  \frac{25}{(1-y)^2}
          - \frac{1}{(1-y)}
          - \frac{1}{(1+y)^2} \nn \\
\hspace*{-5mm} & & \hspace*{5mm}
          + \frac{1}{(1+y)}
         \Bigg] H(0,-1;y) \nn \\
\hspace*{-5mm} & &
       + \Bigg[ 
            \frac{21}{(1-y)^5} \! 
          -  \! \frac{69}{(1-y)^4} \! 
          +  \! \frac{293}{4(1-y)^3} \! 
          -  \! \frac{91}{4(1-y)^2} \! 
          -  \! \frac{77}{16(1-y)} \nn \\
\hspace*{-5mm} & & \hspace*{5mm}
          - \frac{7}{8(1+y)^2}
          + \frac{51}{16(1+y)}
         \Bigg] H(0,0;y) \nn \\
\hspace*{-5mm} & &
       + \Bigg[ 
            \frac{96}{(1-y)^4} \! 
          -  \! \frac{192}{(1-y)^3} \! 
          +  \! \frac{197}{2(1-y)^2} \! 
          -  \! \frac{5}{2(1-y)} \! 
          +  \! \frac{1}{2(1+y)^2} \nn \\
\hspace*{-5mm} & & \hspace*{5mm}
          - \frac{1}{2(1+y)}
         \Bigg] H(0,1;y) \nn \\
\hspace*{-5mm} & &
       - \Bigg[
            \frac{63}{(1-y)^4} \! 
          -  \! \frac{126}{(1-y)^3} \! 
          +  \! \frac{257}{4(1-y)^2} \! 
          -  \! \frac{5}{4(1-y)} \! 
          +  \! \frac{1}{4(1+y)^2} \nn \\
\hspace*{-5mm} & & \hspace*{5mm}
          - \frac{1}{4(1+y)}
         \Bigg] H(1,0;y) \nn \\
\hspace*{-5mm} & &
       - \Bigg[
             \frac{36}{(1-y)^5} \! 
          -  \frac{90}{(1-y)^4} \! 
          +  \frac{75}{(1-y)^3} \! 
          - \frac{45}{2(1-y)^2} \! 
          + \frac{3}{4(1-y)} \nn \\
\hspace*{-5mm} & & \hspace*{5mm}
          + \frac{3}{4(1+y)}
         \Bigg] H(-1,0,0;y) \nn \\
\hspace*{-5mm} & &
       + \Bigg[ 
             \frac{90}{(1-y)^5} \! 
          -   \! \frac{225}{(1-y)^4} \! 
          +  \! \frac{335}{2(1-y)^3} \! 
          -  \! \frac{105}{4(1-y)^2} \! 
          -  \! \frac{25}{8(1-y)} \nn \\
\hspace*{-5mm} & & \hspace*{5mm}
          - \frac{25}{8(1+y)}
         \Bigg] H(0,-1,0;y) \nn \\
\hspace*{-5mm} & &
       - \Bigg[ 
             \frac{24}{(1-y)^5} \! 
          -  \frac{60}{(1-y)^4} \! 
          +   \! \frac{42}{(1-y)^3} \! 
          -  \frac{3}{(1-y)^2} \! 
          - \frac{3}{2(1-y)} \nn \\
\hspace*{-5mm} & & \hspace*{5mm}
          - \frac{3}{2(1+y)}
         \Bigg]  H(0,0,-1;y) \nn \\
\hspace*{-5mm} & &
       + \Bigg[ 
            \frac{3}{2(1-y)^5} \! 
          -  \! \frac{15}{4(1-y)^4} \! 
          +  \! \frac{2}{(1-y)^3} \! 
          +  \! \frac{3}{4(1-y)^2} \! 
          -  \! \frac{11}{32(1-y)} \nn \\
\hspace*{-5mm} & & \hspace*{5mm}
          - \frac{3}{8(1+y)^3} \! 
          +  \! \frac{9}{16(1+y)^2} \! 
          - \frac{11}{32(1+y)} \! 
         \Bigg] H(0,0,0;y) \nn \\
\hspace*{-5mm} & &
       + \Bigg[
            \frac{96}{(1-y)^5} \! 
          -  \! \frac{240}{(1-y)^4} \! 
          +  \! \frac{357}{2(1-y)^3} \! 
          -  \! \frac{111}{4(1-y)^2} \! 
          -  \! \frac{7}{2(1-y)} \nn \\
\hspace*{-5mm} & & \hspace*{5mm}
          - \frac{1}{2(1+y)^3}
          + \frac{3}{4(1+y)^2}
          - \frac{7}{2(1+y)}
         \Bigg] H(0,0,1;y) \nn \\
\hspace*{-5mm} & &
       - \Bigg[
            \frac{63}{(1-y)^5} \! 
          -  \! \frac{315}{2(1-y)^4} \! 
          +  \! \frac{124}{(1-y)^3}  \! 
          -  \! \frac{57}{2(1-y)^2} \! 
          -  \! \frac{7}{16(1-y)} \nn \\
\hspace*{-5mm} & & \hspace*{5mm}
          + \frac{1}{4(1+y)^3}
          - \frac{3}{8(1+y)^2}
          - \frac{7}{16(1+y)}
         \Bigg] H(0,1,0;y) \nn \\
\hspace*{-5mm} & + & {\mathcal O}(D-4) \, .
\eea

\section{Expansion for $Q^2 \gg m^2$ \label{Q2>>a}}

We present in this section the asymptotic limit $Q^2 \gg m^2$ ($x \to
0$) of the space-like UV-renormalized form factors, given in section 
\ref{fullyren}, putting for brevity $L = \log (Q^2/m^2) = \log (q^2)$.
The analytic continuation for large time-like $S=-Q^2$ can be carried
out with the replacement $Q^2=-(S+i \epsilon)$, generating an
imaginary parts in the logarithm $L$ (which becomes 
$L = \log(-s-i\epsilon) = \log{(s)} - i \pi\ ,$). 
\bea
F^{2l}_{1}(D,q^2) & = &  
         \frac{1}{(D-4)^2} \Bigg\{
            \frac{1}{2}
          - L
          + \frac{1}{2} L^2 
       + \frac{1}{q^2}   \Bigg(
          - 2
          + 2 L
          \Bigg)
       \nn \\
       & & \qquad \qquad \quad
       +  \frac{1}{q^4}   \Bigg(
            5
          - 5 L
          + 2 L^2
          \Bigg)
       \nn \\
       & & \qquad \qquad \quad
       +  \frac{1}{q^6}   \Bigg(
          - \frac{50}{3}
          + \frac{68}{3} L
          - 8 L^2
          \Bigg)
       \Bigg\} \nn \\
 & &
       + \frac{1}{(D-4)} \Bigg\{ 
          - 1
          + \frac{1}{2} \zeta(2)
          + \frac{7}{4} L
          - \frac{1}{2} \zeta(2) L 
          - L^2
          + \frac{1}{4} L^3
       \nn \\
       & & \qquad \qquad \quad
       +  \frac{1}{q^2}   \Bigg(
            \frac{5}{2}
          - \zeta(2)
          - \frac{7}{2} L
          + 2 L^2
          \Bigg)
       \nn \\
       & & \qquad \qquad \quad
       +  \frac{1}{q^4}   \Bigg(
          - 6
          + \frac{5}{2} \zeta(2)
          + \frac{55}{4} L
          - 2 \zeta(2) L 
       \nn \\
       & & \qquad \qquad \quad
          - \frac{29}{4} L^2
          + L^3
          \Bigg)
       +  \frac{1}{q^6}   \Bigg(
            \frac{212}{9}
          - \frac{34}{3} \zeta(2)
       \nn \\
       & & \qquad \qquad \quad
          - \frac{500}{9} L
          + 8 \zeta(2) L 
          + \frac{97}{3} L^2
          - 4 L^3
          \Bigg)
       \Bigg\} \nn \\
 & &
          + \frac{1387}{216}
          + \frac{33}{16} \zeta(2)
          - 3  \zeta(2) \ln2
          - \frac{11}{4} \zeta(3)
          - \frac{59}{40} \zeta^2(2)
          \nn \\
          & &
          - \frac{3355}{864} L
          + \frac{1}{12} \zeta(2) L 
          + 2 \zeta(3) L 
          + \frac{571}{288} L^2
          - \frac{3}{8} \zeta(2) L^2 
           \nn \\
          & &
         - \frac{4}{9} L^3
          + \frac{7}{96} L^4
          \nn \\
 & &
       + \frac{1}{q^2}   \Bigg(
          - \frac{6301}{432}
          - \frac{173}{24} \zeta(2)
          + 18  \zeta(2) \ln2
          - \frac{3}{2} \zeta(3)
          + \frac{8}{5} \zeta^2(2)
           \nn \\
          & & \qquad \quad
          + \frac{487}{48} L
          - \frac{15}{4} \zeta(2) L 
          + 3 \zeta(3) L 
          - \frac{97}{16} L^2
          + \frac{1}{2} \zeta(2) L^2 
           \nn \\
          & & \qquad \quad
          + \frac{7}{6} L^3
          - \frac{1}{48} L^4
          \Bigg) \nn \\
 & &
       + \frac{1}{q^4}   \Bigg(
            \frac{12871}{144}
          - \frac{109}{12} \zeta(2)
          - 72  \zeta(2) \ln2
          + \frac{123}{2} \zeta(3)
           \nn \\
          & & \qquad \quad
          - \frac{503}{20} \zeta^2(2)
          - \frac{7487}{216} L
          + \frac{401}{12} \zeta(2) L 
          - 37 \zeta(3) L 
           \nn \\
          & & \qquad \quad
          + \frac{1261}{36} L^2
          - \frac{27}{4} \zeta(2) L^2 
          - \frac{295}{72} L^3
          + \frac{7}{12} L^4
          \Bigg) \nn \\
 & &
       + \frac{1}{q^6}   \Bigg(
          - \frac{94525}{144}
          + \frac{5317}{18} \zeta(2)
          + 288  \zeta(2) \ln2
          - \frac{1447}{3} \zeta(3)
           \nn \\
          & & \qquad \quad
          + 227 \zeta^2(2)
          + \frac{11401}{216} L
          - \frac{1175}{6} \zeta(2) L 
          + 296 \zeta(3) L 
           \nn \\
          & & \qquad \quad
          - \frac{4451}{24} L^2
          + 55 \zeta(2) L^2 
          + \frac{403}{18} L^3
          - \frac{9}{4} L^4
          \Bigg) \nn \\
 & + & {\mathcal O} \Bigg(\frac{1}{q^8} \Bigg) \ , \\
F^{2l}_{2}(D,q^2) & = & 
         \frac{1}{(D-4)} \Bigg\{
         \frac{1}{q^2}   \Bigg(
            L
          - L^2
          \Bigg)
       + \frac{1}{q^4}   \Bigg(
            2
          - 6 L
          + 2 L^2
          \Bigg)
       \nn \\
       & & \qquad \qquad \quad
       + \frac{1}{q^6}   \Bigg(
          - 11
          + 20 L
          - 8 L^2
          \Bigg)
       \Bigg\} \nn \\
 & & 
       + \frac{1}{q^2}   \Bigg(
            \frac{17}{3}
          + \frac{11}{4} \zeta(2)
          - 12  \zeta(2) \ln2
           \nn \\
          & & \qquad \quad
          + \zeta(3)
          - \frac{379}{72} L
          + 3 \zeta(2) L 
          + \frac{55}{24} L^2
          - \frac{1}{2} L^3
          \Bigg) \nn \\
 & &
       + \frac{1}{q^4}   \Bigg(
          - \frac{1717}{36}
          + \frac{31}{2} \zeta(2)
          + 48  \zeta(2) \ln2
          - 56 \zeta(3)
           \nn \\
          & & \qquad \quad
          + \zeta(3)
          + \frac{66}{5} \zeta^2(2)
          + \frac{931}{36} L
          - 19 \zeta(2) L 
          + 28 \zeta(3) L 
           \nn \\
          & & \qquad \quad
          - \frac{289}{12} L^2
          + 4 \zeta(2) L^2 
          + \frac{7}{3} L^3
          - \frac{1}{12} L^4
          \Bigg) \nn \\
 & &
       + \frac{1}{q^6}   \Bigg(
          + \frac{34471}{72}
          - 254 \zeta(2)
          - 192 \zeta(2) \ln2 
          + 448 \zeta(3)
           \nn \\
          & & \qquad \quad
          - \frac{858}{5} \zeta^2(2)
          - \frac{295}{6} L
          + 129 \zeta(2) L 
          - 252 \zeta(3) L 
           \nn \\
          & & \qquad \quad
          + 147 L^2
          - 44 \zeta(2) L^2 
          - \frac{83}{6} L^3
          + \frac{1}{4} L^4
          \Bigg) \nn \\
& + & {\mathcal O}\Bigg(\frac{1}{q^8} \Bigg) \, .
\eea

\section{Expansion for $Q^2 \ll m^2$ \label{Q2<<a}}

We present in this section the expansion in $Q^2=m^2q^2$ around $Q^2=0$, 
valid for $Q^2 \ll m^2$ ($x \to 1$), of the form factors. Since we are 
in the analyticity region, the expansions are valid for space-like as 
well as time-like values of $Q^2$. The expansion reads 
\bea
F^{(2l)}_{1}(D,q^2) & = & 
            \frac{q^4}{(D-4)^{2}} \Biggl\{ \frac{1}{18} 
                          - q^2   \biggl( \frac{1}{60} \biggr)
                                  \Biggr\} \nn\\
& - & \frac{q^4}{(D-4)} \Biggl\{ \frac{1}{24} 
                          - q^2   \biggl( \frac{31}{1440} \biggr) 
                        \Biggr\} \nn\\
&  & +  \,  q^2  \, \biggl(
                             \frac{4819}{5184}
                           + \frac{49}{72} \zeta(2)
                           - 3 \zeta(2) \ln{2}
                           + \frac{3}{4} \zeta(3)
                    \biggr) \nn\\
& &  - \, q^4   \,  \biggl(
            \frac{1349}{6480}
          + \frac{8731}{28800} \zeta(2)
          - \frac{11}{10} \zeta(2) \ln{2}
          + \frac{11}{40} \zeta(3)
          \biggr) \nn\\
& &        + \, q^6   \,  \biggl(
            \frac{109069}{1814400} \! 
          +  \! \frac{723901}{5644800} \zeta(2) \! 
          -  \! \frac{113}{280} \zeta(2) \ln{2} \! 
          +  \! \frac{113}{1120} \zeta(3) \! 
          \biggr) \nn\\
& + & {\mathcal O} \bigl( q^8 
\bigr) \, , \\
F^{(2l)}_{2}(D,q^2) & = & - \frac{q^2}{(D-4)} \Biggl\{ \frac{1}{6} 
       -  \, q^2  \,  \biggl(
            \frac{19}{360}
          \biggr)
       +  \, q^4   \,  \biggl(
            \frac{73}{5040}
          \biggr)
\Biggr\} \nn\\
& &       +   \frac{197}{144}  
          +   \frac{1}{2} \zeta(2)  
          -   3 \zeta(2) \ln{2}  
          +   \frac{3}{4} \zeta(3)  \nn\\
& &        -  \, q^2   \,  \biggl(
            \frac{1031}{2160}  
          + \frac{13}{20} \zeta(2)
          -   \frac{23}{10} \zeta(2) \ln{2}
          + \frac{23}{40} \zeta(3)
          \biggr) \nn\\
& &        +   \,  q^4   \,  \biggl(
            \frac{2551}{15120}  
          + \frac{187}{525} \zeta(2)
          - \frac{15}{14} \zeta(2) \ln{2}  
          + \frac{15}{56} \zeta(3)
          \biggr) \nn\\
& &        -  \, q^6   \,  \biggl(
            \frac{69019}{1209600}
          + \frac{140951}{940800} \zeta(2)
          - \frac{29}{70} \zeta(2) \ln{2}
          + \frac{29}{280} \zeta(3)
          \biggr) \nn\\
& + & {\mathcal O} \bigl( q^8 \bigr) \, .
\eea

It is to be noted that the absence of the first order term in $q^2$ 
from the (infrared) singular part of $F_1^{(2l)}(D,q^2)$ and the absence
of the zeroth order term from the (infrared) singular part of
$F_2^{(2l)}(D,q^2)$ guarantee the finiteness of the 2-loop value for the 
charge slope of the electron ($-F_{1}^{(2l)}$$'(0)$ in the current 
notation) and of the magnetic anomaly $F_{2}^{(2l)}(0)$. The corresponding 
values are of course in agreement with the well known results in the 
literature (see for example \cite{2loop1}).

\section{Summary}

We carried out the analytic calculation of the renormalized on shell 
form factors of the electron vertex at two loops in perturbative QED. 
We gave the full results, for arbitrary momentum 
transfer $S$ and on-shell fermionic external lines with finite mass 
$m$, in the space-like region, $S < 0$, and the imaginary parts in the 
time-like region above the threshold, $S > 4m^2$. 
The results are expressed in term of 1-dimensional HPLs 
of maximum weight 4.

We gave also the expansions of the form factors in the two kinematical 
regions of large momentum transfer, $S \to \infty$, and small momentum 
transfer, $S \to 0$, recovering in particular the known results for the 
charge slope and the anomalous magnetic moment of the electron. 

For the calculation we used the MIs calculated in a previous paper,
regularized in the dimensional regularization scheme. Both UV and soft
IR divergences are regularized with the parameter $D$, dimension of the
space-time. In the final results, after the renormalization of the UV
divergences, the form factors still possess polar singularities in $(D-4)$, 
because of the soft IR divergences (the poles in $1/(D-4)$ cancel out 
from any physical observable when soft real photons are also accounted 
for).

\section{Acknowledgment}
We are grateful to J. Vermaseren for his kind assistance in the use
of the algebra manipulating program {\tt FORM}~\cite{FORM}, by which
all our calculations were carried out.

We wish to thank A. Ferroglia for useful discussions. 

\appendix

\section{Propagators \label{app1}}

We list here the denominators of the integral expressions appeared in
the paper.
\bea
{\mathcal D}_{1} & = & k_{1}^{2} \, , \\
{\mathcal D}_{2} & = & k_{2}^{2} \, , \\
{\mathcal D}_{3} & = & (k_{1}-k_{2})^{2} \, , \\
{\mathcal D}_{4} & = & (p_{1}-k_{1})^{2} \, , \\
{\mathcal D}_{5} & = & (p_{2}-k_{2})^{2} \, , \\
{\mathcal D}_{6} & = & [k_{1}^{2}+m^2] \, , \\
{\mathcal D}_{7} & = & [k_{2}^{2}+m^2] \, , \\
{\mathcal D}_{8} & = & [(k_{1}+k_{2})^{2}+m^2] \, , \\
{\mathcal D}_{9} & = & [(p_{1}-k_{1})^{2}+m^2] \, , \\
{\mathcal D}_{10} & = & [(p_{2}+k_{1})^{2}+m^2] \, , \\
{\mathcal D}_{11} & = & [(p_{2}-k_{2})^{2}+m^2] \, , \\
{\mathcal D}_{12} & = & [(p_{1}-k_{1}+k_{2})^{2}+m^2] \, , \\
{\mathcal D}_{13} & = & [(p_{2}+k_{1}-k_{2})^{2}+m^2] \, .
\eea

\section{One-loop results \label{app2}}

For completeness we give in this appendix the 1-loop form factors, 
corresponding to the first order in $\alpha / \pi$ in 
Eqs. (\ref{b00020}-\ref{b00022}), expanded up to the first order 
in $(D-4)$. 

The Feynman diagrams involved are those in Fig. \ref{fig3}, where the
order $\alpha / \pi$ Feynman diagram and the relative subtraction for
the renormalization of the charge form factor are shown.

\bfig
\bc
\subfigure[]{
\begin{fmfgraph*}(30,30)
\fmfleft{i1,i2}
\fmfright{o}
\fmfforce{0.2w,0.93h}{v2}
\fmfforce{0.2w,0.07h}{v1}
\fmfforce{0.8w,0.5h}{v5}
\fmf{fermion}{i1,v1}
\fmf{fermion}{i2,v2}
\fmf{photon}{v5,o}
\fmflabel{$p_2$}{i1}
\fmflabel{$p_1$}{i2}
\fmf{fermion,tension=.3}{v2,v5}
\fmf{fermion,tension=.3}{v1,v5}
\fmf{photon,tension=0}{v2,v1}
\end{fmfgraph*}}
%
%
\hspace{10mm}
\subfigure[]{
\begin{fmfgraph*}(30,30)
\fmfleft{i1,i2}
\fmfright{o}
\fmfforce{0.2w,0.93h}{v2}
\fmfforce{0.2w,0.07h}{v1}
\fmfforce{0.8w,0.5h}{v5}
\fmfforce{0.66w,0.5h}{v55}
\fmf{plain}{i1,v1}
\fmf{plain}{i2,v2}
\fmf{photon}{v5,o}
\fmflabel{$p_2$}{i1}
\fmflabel{$p_1$}{i2}
\fmfv{label=$\otimes$}{v55}
\fmfv{l=$Z_{1}^{(1l)}$,l.a=70,l.d=.12w}{v5}
\fmf{fermion,tension=.3}{v2,v5}
\fmf{fermion,tension=.3}{v1,v5}
\end{fmfgraph*}}
%
%
\vspace*{8mm}
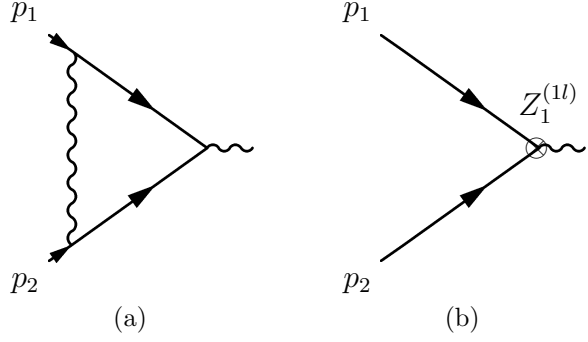
\caption{\label{fig3} 1-loop vertex diagrams for the QED form factor.}
\ec
\efig

The unrenormalized form factors ${\mathcal F}_1^{(1l)}(D,q^2)$, 
${\mathcal F}_2^{(1l)}(D,q^2)$ and ${\mathcal F}_3^{(1l)}(D,q^2)$ for the 
diagram {\tt(a)} in Fig. \ref{fig3} were already given in 
Eqs. (\ref{1loopF1}--\ref{c7}), while 
$Z^{(1l)}_1(D)$ was given in Eq. (\ref{c001}). The counter-term in Fig. 
\ref{fig3} {\tt(b)} contributes only to the renormalization of the {\it
charge} form factor, being the product of the renormalization constant
$Z_{1}^{(1l)}(D)$ at 1-loop level, Eq. (\ref{c001}), and the 
tree-level vertex.

The UV-renormalized form factors at 1-loop level, in the space-like
region $-S=Q^2=m^2q^2 > 0$, up to the first order in $(D-4)$, are the 
following: 
\bea
F^{(1l)}_{1}(D,q^2) & = &   \frac{1}{(D-4)} \Biggl\{ 
1 - \biggl[ 1 - \frac{1}{(1-x)} - \frac{1}{(1+x)} \biggr] H(0;x)
\Biggr\} 
\nn\\
& &       - 1 
          + \frac{1}{2} \biggl[ 
	    \frac{1}{2} \! 
	  - \! \frac{1}{(1\! +\! x)}
          \biggr] H(0;x) \nn\\
& & - \! \frac{1}{2} \biggl[ 1 \! - \! 
   \frac{1}{(1\! -\! x)} \! - \! \frac{1}{(1\! +\! x)} \biggr] \bigl[ 4 
   \zeta(2)  - 2 H(0;x) - H(0,0;x)  \nn\\
& &   - 2 H(1,0;x) \bigr] \nn\\
& + & (D-4) \Biggl\{ 1 
       - \frac{1}{8} \biggl[ 
	    1 \! 
	  -  \! \frac{2}{(1 \! + \! x)} \biggr]
         \bigl[ \zeta(2)  \! 
	  -  \! H(0,0;x)  \! 
	  -  \! 2 H( \! -1,0;x) \bigr] \nn\\
& & \hspace*{18mm}
       + \frac{1}{4} \biggl[ 
            1 
	  - \frac{1}{(1-x)} 
	  - \frac{1}{(1+x)} \biggr] 
	 \bigl[
            2( \zeta(2) + \zeta(3) )  \nn\\
& & \hspace*{23mm}
	  + ( \zeta(2)  \! -  \! 4 ) H(0;x)  \! 
	  -  \! 2 \zeta(2) H( \! 1;x) \! 
          -  \! 2 H(0, \! 0;x)   \nn\\
& & \hspace*{23mm}
	  +  \! 2 H( \! -1,0;x)  \! 
	  -  \! H(0,0,0;x)  \! 
	  -  \! 4 H( \! -1, \! -1,0;x)   \nn\\
& & \hspace*{23mm}
	  +  \! 2 H(-1,0,0;x) \! 
          +  \! 2 H(0,-1,0;x) \bigr] \Biggr\}  \nn\\
& + &  {\mathcal O} \left( (D-4)^2 \right)
\label{appb1}
\, , \\
F^{(1l)}_{2}(D,q^2) & = &  -  \frac{1}{2} \left[ \frac{1}{(1-x)} -
\frac{1}{(1+x)} \right] H(0;x) \nn\\
& - &   (D-4) \Biggl\{ \frac{1}{4} \left[ \frac{1}{(1-x)} -
\frac{1}{(1+x)} \right] \bigl[ \zeta(2) 
           - 4 H(0;x)   \nn\\
& & \hspace*{23mm}
           - 4 H(0,0;x) 
	   + 2 H(-1,0;x)
\bigr] \Biggr\}   \nn\\
& + &  {\mathcal O} \left( (D-4)^2 \right)
\label{appb2} \, , \\
F^{(1l)}_{3}(D,q^2) & = &  0 \, .
\eea 
where $F_{2}^{(2l)}(D,q^2)$ is exactly equal to 
${\mathcal{F}}_{2}(D,q^2)$, given in Eq. (\ref{1loopF2}) and 
$F^{(1l)}_{1}(D,q^2)$ is obtained by subtracting $Z^{(1l)}_1$ of
Eq. (\ref{c001}) from ${\mathcal{F}}_{1}(D,q^2)$ given in 
Eq. (\ref{1loopF1}). 

Eqs. (\ref{appb1},\ref{appb2}) can be analytically continued in the
time-like region $S=-Q^2>0$ and in particular above the physical
threshold $S>4m^2$, where an imaginary part appears. Using the
substitution of Eq. (\ref{analyticY}) and writing ($s=S/m^2$) 
\bea
F^{(1l)}_{1}(D,-s-i\epsilon) & = & \Re \, F^{(1l)}_{1}(D,-s) 
                         + i \pi \, \Im \, F^{(1l)}_{1}(D,-s) \, , \\ 
F^{(1l)}_{2}(D,-s-i\epsilon) & = & \Re \, F^{(1l)}_{2}(D,-s) 
                         + i \pi \, \Im \, F^{(1l)}_{2}(D,-s) \, ,
\eea
the imaginary parts have the following expressions:
\bea
\Im \, F^{(1l)}_{1}(D,-s) & = & - \frac{1}{(D-4)} \Biggl\{ 
1 - \frac{1}{(1-y)} - \frac{1}{(1+y)} 
\Biggr\}  \nn\\
& &  +\frac{1}{4} - \frac{1}{2(1+y)} + \frac{1}{2} \biggl[ 1 - 
\frac{1}{(1-y)} - \frac{1}{(1+y)} \biggr] \bigl[ 1 + H(0;y) \nn\\
& & + 2 H(1;y)
\bigr] \nn\\
& - & (D-4) \Biggl\{ \frac{1}{8} \biggl[ 
          1 
	- \frac{2}{(1+y)} \biggr] \bigl[
	  H(0;y)
	+ 2 H(1;y) \bigr] \nn\\
& & \hspace*{18mm}
      + \frac{1}{4} \biggl[ 
          1 \! 
	-  \! \frac{1}{(1-y)}  \! 
	-  \! \frac{1}{(1+y)} \biggr] \bigl[ 
          4 \! 
	-  \! 2 \zeta(2) \! 
	+  \! H(0;y) \nn\\
& & \hspace*{23mm}
	+ 2 H(1;y)
	+ H(0,0;y)
	+ 2 H(0,1;y) \nn\\
& & \hspace*{23mm}
	+ 2 H(1,0;y)
	+ 4 H(1,1;y) \bigr] \Biggr\}   \nn\\
& + &  {\mathcal O} \left( (D-4)^2 \right) \, , 
\label{appb3} \\
\Im \, F^{(1l)}_{2}(D,-s) & = & \frac{1}{2} \left[ \frac{1}{(1-y)} -
\frac{1}{(1+y)} \right] \nn\\
& - & (D-4) \Biggl\{  \frac{1}{4} \left[ \frac{1}{(1-y)} -
\frac{1}{(1+y)} \right] \bigl[ 
          \zeta(2) 
	- 4 H(0;y) \nn\\
& & \hspace*{23mm}
        - H(0,0;y) 
	+ 2 H(-1,0;y) \bigr]
\Biggr\}  \nn\\
& + &  {\mathcal O} \left( (D-4)^2 \right) \, .
\label{appb4}
\eea
The presence of $1/(D-4)$ singularities even in the on-shell renormalized 
form factors is due to the fact that soft IR divergences are still present. 
As already recalled, in any physical quantity they
will cancel against similar divergences due to soft real photons.

\end{fmffile}
\end{document}